\documentclass[prb,twocolumn,showpacs,groupedaddress,showkeys,preprintnumbers,amsmath,amssymb,floatfix]{revtex4}
\usepackage{graphicx,epsfig}% Include figure files
\usepackage{dcolumn}% Align table columns on decimal point
\usepackage{bm,texdraw}% bold math

\begin{document}

\title{Boundary critical behavior at $\bm{m}$-axial Lifshitz points
  for a boundary plane parallel to the modulation axes}
\author{H.~W. Diehl}
\author{A. Gerwinski}
\author{S. Rutkevich}
\altaffiliation{%
On leave from: Institute of Solid State and Semiconductor Physics,
220072 Minsk, Belarus.}%
\affiliation{%
Fachbereich Physik, Universit{\"a}t Duisburg-Essen, 45117 Essen, Germany}

\date{\today}

\begin{abstract}
  The critical behavior of semi-infinite $d$-dimensional systems with
  $n$-component order parameter $\bm{\phi}$ and short-range
  interactions is investigated at an $m$-axial bulk Lifshitz point
  whose wave-vector instability is isotropic in an $m$-dimensional
  subspace of $\mathbb{R}^d$. The associated $m$ modulation axes are
  presumed to be parallel to the surface, where $0\le m\le d-1$. An
  appropriate semi-infinite $|\bm{\phi}|^4$ model representing the
  corresponding universality classes of surface critical behavior is
  introduced. It is shown that the usual $O(n)$ symmetric boundary
  term $\propto \bm{\phi}^2$ of the Hamiltonian must be supplemented
  by one of the form $\mathring{\lambda}\,
  \sum_{\alpha=1}^m(\partial\bm{\phi}/\partial x_\alpha)^2$ involving
  a dimensionless (renormalized) coupling constant $\lambda$.  The
  implied boundary conditions are given, and the general form of the
  field-theoretic renormalization of the model below the upper
  critical dimension $d^*(m)=4+{m}/{2}$ is clarified.  Fixed points
  describing the ordinary, special, and extraordinary transitions are
  identified and shown to be located at a nontrivial value $\lambda^*$
  if $\epsilon\equiv d^*(m)-d>0$. The surface critical exponents of
  the ordinary transition are determined to second order in
  $\epsilon$.  Extrapolations of these $\epsilon$ expansions yield
  values of these exponents for $d=3$ in good agreement with recent
  Monte Carlo results for the case of a uniaxial ($m=1$) Lifshitz
  point. The scaling dimension of the surface energy density is shown
  to be given exactly by $d+m\,(\theta-1)$, where
  $\theta=\nu_{l4}/\nu_{l2}$ is the anisotropy exponent.
\end{abstract}

\maketitle
%\tableofcontents\newpage
\section{Introduction}
\label{sec:Intro}

As is well known, there are many physical systems whose phase diagrams
exhibit a Lifshitz point.\cite{Hor80,Sel88,Sel92,Die02} A Lifshitz
point is a multicritical point at which a disordered, a homogeneous
ordered, and a modulated ordered phase meet. It divides the critical
line that separates the disordered phase from the two ordered ones
into two sections, and marks the onset of a wave-vector instability.
In the case of an $m$-axial Lifshitz point this instability occurs in
an $m$-dimensional subspace of $\mathbb{R}^d$. Prominent examples of
systems with uniaxial ($m=1$) Lifshitz points are, on the experimental
side, the metallic compound\cite{BSOC80,YCS84,ZSK00,BBO00} MnP, and on
the theoretical side, the three-dimensional axial
next-nearest-neighbor Ising (ANNNI) model.\cite{FS80,PH01}

Although the notion of a Lifshitz point was introduced decades
ago,\cite{HLS75a} the successful application of modern field theory
tools to the study of critical behavior at such points is a fairly
recent development.\cite{MC98,MC99,DS00a,SD01,DS01a,DS02} A full
two-loop renormalization group analysis and $\epsilon$ expansion of
all critical exponents to second order in $\epsilon$ about the upper
critical dimension $d^*(m)=4+m/2$ ($0\le m\le 8$) has been
accomplished for general values of $m$ only two years
ago.\cite{DS00a,SD01}

In the present paper we wish to study the effects of surfaces on the
critical behavior at an $m$-axial Lifshitz point. Previous work on
this problem is scarce. It started in 1986 with Gumbs' investigation
\cite{Gum86} based on Landau theory. This was reconsidered, completed,
and partly corrected by Binder and Frisch.\cite{BF99} In either one of
these papers, only the uniaxial case $m=1$ was considered, and the
special axis along which ferromagnetic nearest-neighbor and
antiferromagnetic next-nearest-neighbor interactions compete was
chosen perpendicular to the surface plane.  The complementary case in
which the special axis of the considered semi-infinite ANNNI model is
parallel to the surface was dealt with on the level of mean field
theory in subsequent work by Frisch \emph{et al}.\cite{FKB00} Even
more recently, Pleimling\cite{Ple02} presented results of Monte Carlo
simulations of semi-infinite ANNNI models for both types of surface
orientations.

A characteristic feature of critical behavior at Lifshitz points is
\emph{anisotropic scale invariance}. Let $\Delta x_\alpha$ and $\Delta
x_\beta$ with $1\le\alpha\le m$ and $m<\beta\le d$ denote
displacements along the corresponding Euclidean axes and suppose that
the $\Delta x_\beta$ are rescaled by a factor $\ell$.  Anisotropic
scale invariance means that the $\Delta x_\alpha$ must be rescaled by
a nontrivial power $\ell^\theta$ in order that the system looks
self-similar on different scales.\cite{com:ASIgen}

An obvious consequence of this property is that two distinct cases of
surface orientations must be distinguished when dealing with surface
critical behavior at bulk Lifshitz points: a parallel one for which
the surface normal $\bm{n}$ is along a $\beta$-direction (all
$\alpha$-directions being parallel to the surface plane), and a
perpendicular one for which $\bm{n}$ is along an $\alpha$-direction.
Since the distance $z$ from the surface scales differently in these
two cases, one expects (i) that the respective values of the surface
exponents are in general different, even in mean-field theory, and
(ii) that the boundary terms which must be included in the Hamiltonian
are different as well.  Expectation (i) is borne out by both the
mean-field results of Refs.~\onlinecite{BF99} and \onlinecite{FKB00}
as well as by Pleimling's Monte Carlo results;\cite{Ple02} (ii) turns
out to be equally true.\cite{DGunpub}

In this paper\cite{DRGcom,DRG03} we will focus our attention on the
case of \emph{parallel} surface orientation, leaving the study of the
other, ``perpendicular'' case to a subsequent paper [which will
confirm the above conclusion (ii)]. To our knowledge, experimental
investigations of even bulk critical behavior at Lifshitz
points\cite{BSOC80,BBO00} have so far been restricted to the uniaxial
case $m=1$, and this is also still the only one for which Monte Carlo
simulation results on surface critical behavior at bulk Lifshitz
points are available.\cite{Ple02}

Although $m=1$ thus appears to be the case of greatest interest, we
will keep the value of $m$ as general as possible, imposing at this
stage no restriction other than the obvious one $0\le m \le
d-1$.\cite{com:mres} Just as in essentially all renormalization group
(RG) analyses of bulk critical behavior with $m>1$, we will make,
however, the simplifying assumption that the wave-vector instability
is \emph{isotropic} in the $m$-dimensional subspace of $\alpha$
directions. Let us nevertheless add a cautionary remark about this
presumed ``$m$-isotropy''. In a recent study of bulk critical
behavior,\cite{DSZ03} this assumption has been relaxed by requiring
only invariance in this subspace under the cubic, or even a lower,
symmetry group. The results indicate that the isotropic bulk fixed
point for $m\ge 2$ gets destabilized by bulk terms of second order in
$\bm{\phi}$ and fourth order in the derivatives $\partial/\partial
x_\alpha$, although the associated crossover exponent appears to be
fairly small. The inclusion of such non-isotropic terms is beyond the
scope of the present work and, of course, meaningless in the uniaxial
case $m=1$.

The remainder of this paper is organized as follows. In the next
section we specify the Hamiltonian representing the corresponding
universality classes for surface critical behavior at $m$-axial bulk
Lifshitz points. We explain which boundary terms must be included in
it and give the implied boundary conditions. In Sec.~\ref{sec:ren} we
first clarify the renormalization of the model for general values of
its parameters and derive the resulting RG equations. On the basis of
two-loop results, we then discuss the form of the RG flow in the space
of the surface interaction constants and identify the fixed points
describing the ordinary, special, and extraordinary transitions.

Section~\ref{sec:RGOT} deals specifically with the ordinary
transition. We show that the RG analysis can be simplified in much the
same way as in the $m=0$ case of a critical point\cite{Die86a} by
choosing Dirichlet boundary conditions, considering correlation
functions involving the normal derivative of the order parameter at
the boundary, and making use of the boundary operator expansion. The
required RG functions are evaluated to two-loop order; the scaling
index of the surface energy density is determined exactly.

In Sec.~\ref{sec:critexp} results to order $\epsilon^2$ are given for
the critical exponents of the ordinary transition. These $\epsilon$
expansions are exploited to estimate the values of these exponents for
the uniaxial, one-component case $m=n=1$ in three dimensions.
Section~\ref{sec:concl} contains a brief summary and concluding
remarks.  Finally, there are five appendixes, explaining details of
our calculations.

\section{The Hamiltonian, its boundary terms, and implied boundary
  conditions}
\label{sec:mod}

We consider systems with short-range interactions, assuming also that
the perturbations of these interactions induced by the presence of the
surface decay to zero within a short distance from it. With these
assumptions the Hamiltonian can be taken to be of the form
\begin{equation}
  \label{eq:Hamf}
    {\mathcal{H}}={\int_{\mathfrak{V}}}\mathcal{L}_{\text{b}}(\bm{x})\,dV+ 
    {\int_{\mathfrak{B}}}\mathcal{L}_1(\bm{x})\,dA\;, 
\end{equation}
where $\mathcal{L}_{\text{b}}(\bm{x})$ and $\mathcal{L}_1(\bm{x})$
depend on the order parameter density
$\bm{\phi}(\bm{x})=(\phi_a(\bm{x}),\,a=1,\ldots,n)$ and its spatial
derivatives up to a finite order. Here the volume and surface
integrals extend over $\mathfrak{V}$, the $d$-dimensional half-space
$\mathbb{R}^d_+=\mathbb{R}^{d-1}{\times} [0,\infty)$, and
$\mathfrak{B}$, the $d-1$ dimensional surface plane $z=0$,
respectively.  Writing
$\bm{x}=\big((x_\alpha),(x_\beta)\big)=(\bm{r},z)$, we split the
position vector $\bm{x}$ into its $m$-dimensional component
$(x_\alpha)\in\mathbb{R}^m$ and $(d-m)$-dimensional one
$(x_\beta)=\big((r_\beta),z\big)$, where $\bm{r}=\big((r_\alpha)\equiv
(x_\alpha),(r_\beta)\big)$ is the $(d-1)$-dimensional coordinate along
the surface. We choose the same bulk density as in
Refs.~\onlinecite{DS00a,SD01,DS01a,DS02}, namely
\begin{eqnarray}
  \label{eq:Lb}
  {\mathcal L}_{\text{b}}(\bm{x})&=&\frac{\mathring{\sigma}}{2}\,
  \bigg(\sum_{\alpha=1}^m\partial_\alpha^2\bm{\phi} \bigg)^2  
    +\frac{1}{2}\,
\sum_{\beta=m+1}^d{(\partial_\beta\bm{\phi})}^2
\nonumber\\&&\mbox{}
+\frac{\mathring{\rho}}{2}\,\sum_{\alpha=1}^m{(\partial_\alpha\bm{\phi})}^2
 +\frac{\mathring{\tau}}{2}\,
\bm{\phi}^2+\frac{\mathring{u}}{4!}\,|\bm{\phi} |^4\;,
\end{eqnarray}
where $\partial_\alpha$ and $\partial_\beta$ denote the spatial
derivatives $\partial/\partial x_\alpha$ and $\partial/\partial
x_\beta$ with $1\le\alpha\le m$ and $m+1\le\beta\le d$, respectively.
Had we not assumed that the wave-vector instability is isotropic in
the subspace ${\mathbb R}^m$, the term
$\big(\sum_{\alpha=1}^m\partial^2_\alpha\phi\big)^2$ would have to be
supplemented by similar, albeit less symmetric, terms involving four
derivatives, such as\cite{DSZ03}
$\sum_{\alpha=1}^m(\partial_\alpha^2\bm{\phi})^2$.

In order to decide which monomials should be included in
$\mathcal{L}_1(\bm{x})$, we use power counting. Recalling from
Ref.~\onlinecite{DS00a} the naive dimensions $[x_\beta]=[z]=\mu^{-1}$,
$[x_\alpha]=\mathring{\sigma}^{1/4}\,\mu^{-1/2}$, and
$[\phi(\bm{x})]=\mathring{\sigma}^{-m/8}\,\mu^{(d-2-m/2)/2}$ (where
$\mu$ is an arbitrary momentum scale), we include only such $O(n)$
invariant monomials whose interaction constants have non-negative
$\mu$-dimensions at the upper critical dimension $d^*(m)$. It is not
difficult to see that the choice
\begin{equation}
  \label{eq:L1}
  \mathcal{L}_1(\bm{x})=\frac{\mathring{c}}{2}\,\bm{\phi}^2 +
\frac{\mathring{\lambda}}{2}\,
\sum_{\alpha=1}^m{(\partial_\alpha\bm{\phi})}^2 
\end{equation}
is sufficient since the two additional candidates
$\bm{\phi}\partial_\alpha\partial_\alpha\bm{\phi}$ and
$\bm{\phi}\partial_n\bm{\phi}$ both may be dropped.  The integral
$\int_{\mathfrak{B}}$ of the former reduces to that of the derivative
term retained in Eq.~(\ref{eq:L1}) upon integration by parts, and the
latter can be shown to be \emph{redundant} in much the same way as in
the familiar case of the standard semi-infinite $\phi^4$
model.\cite{Die86a} One must merely note that the action defined by
Eqs.~(\ref{eq:Hamf})--(\ref{eq:L1}) leads to the boundary
condition\cite{com:bc,Die86a,DJ92,Die94b}
\begin{equation}
  \label{eq:bc}
  \partial_n\bm{\phi}(\bm{x})={\big(\mathring{c}-\mathring{\lambda}\,
  \partial_\alpha\partial_\alpha\big)}\,\bm{\phi}(\bm{x})\;, 
  \quad\bm{x}\in\mathfrak{B}\;,
\end{equation}
which in turn implies that the surface term
${\int_{\mathfrak{B}}}\,\bm{\phi}\partial_n\bm{\phi}$ is equivalent to
a linear combination of the two monomials present in $\mathcal{L}_1$.
Here a sum convention which will frequently be employed below has been
introduced: Pairs of equal $\alpha$ and $\beta$ indices are to be
summed over $\alpha=1,\ldots,m$ and $\beta=m+1,\ldots,d$,
respectively.

We need the free propagator of the disordered phase,
$G(\bm{x},\bm{x}')=[\delta^2\mathcal{H}/\delta\bm{\phi}
\delta\bm{\phi} |_{\bm{\phi}=0}]^{-1}(\bm{x},\bm{x}')$. This is a
solution to the equation
\begin{equation}
  \label{eq:freepropeq}
  {\big[\mathring{\sigma}\,{(\partial_\alpha\partial_\alpha)}^2-
      \mathring{\rho}\,\partial_\alpha\partial_\alpha
      -\partial_\beta\partial_\beta +\mathring{\tau}\big]}\,
   G(\bm{x},\bm{x}')=\delta(\bm{x}-\bm{x}')\;,
\end{equation}
subject to the boundary condition
\begin{equation}
  \label{eq:bcG}
  {\big(\partial_n-\mathring{c}+\mathring{\lambda}\,
      \partial_\alpha\partial_\alpha\big)}\,G(\bm{x},\bm{x}')=0\;,
\quad\bm{x}\in\mathfrak{B}\;,\;\;
\bm{x}'\notin\mathfrak{B}\;,
\end{equation}
where $\partial_n$ ($\equiv \partial_z$) denotes the derivative along
the inward normal $\bm{n}$.

We take periodic boundary conditions along all $d-1$ axes parallel to
the surface. Let $\bm{p}=((p_\alpha),(p_\beta)) \in
\mathbb{R}^m{\times}\mathbb{R}^{d-m-1}$ be the momentum conjugate to
$\bm{r}$, and $\hat{G}(\bm{p};z,z')$ be the corresponding Fourier
transform of $G$ with respect to the $d-1$ coordinates parallel to the
surface. In this $\bm{p}z$ representation $\hat{G}$ is easily
evaluated; one finds
\begin{eqnarray}
  \label{eq:Ghatgen}
 \lefteqn{\hat{G}(\bm{p};z,z')}&&\nonumber\\
&=&\frac{1}{2\mathring{\kappa}_{\bm{p}}}
  {\bigg[}e^{-\mathring{\kappa}_{\bm{p}}|z-z'|} 
-\frac{\mathring{c}_{\bm{p}}-
  \mathring{\kappa}_{\bm{p}}}{\mathring{c}_{\bm{p}}+\mathring{\kappa}_{\bm{p}}}\,
e^{-\mathring{\kappa}_{\bm{p}}(z+z')}{\bigg]} 
\end{eqnarray}
with
\begin{equation}
  \label{eq:cbp}
  \mathring{c}_{\bm{p}}\equiv
   \mathring{c}+ \mathring{\lambda}\,{p_\alpha}{p_\alpha}
\end{equation}
and
\begin{equation}
  \label{eq:kappap}
  \mathring{\kappa}_{\bm{p}}=\sqrt{\mathring{\tau}+\mathring{\rho}\,
    p_\alpha p_\alpha+p_\beta p_\beta+ \mathring{\sigma}\,{\big(p_\alpha
      p_\alpha\big)}^2}\;.
\end{equation}

The part of $\hat{G}$ that depends on $|z-z'|$ is its bulk analog
$\hat{G}_{\text{b}}$; the remainder is the contribution induced by the
presence of the surface. Results for the free bulk propagator
$G_{\text{b}}(\bm{x}-\bm{x}')$ in position and momentum space can be
found in Refs.~\onlinecite{DS00a} and \onlinecite{SD01}. Specifically
at the (Gaussian) Lifshitz point $\mathring{\tau}=\mathring{\rho}=0$,
one has
\begin{equation}
  \label{eq:Gbx}
  G_{\text{b}}(\bm{x})=X^{-2+\epsilon}\,\mathring{\sigma}^{-m/4}\,
   \Phi_{m,d}{\big(\mathring{\sigma}^{-1/4}\,\check{x}\,X^{-1/2}\big)} 
\end{equation}
with
\begin{equation}
  \label{eq:X}
 X\equiv\sqrt{x_\beta\,x_\beta}\;,\quad
 \check{x}\equiv\sqrt{x_\alpha\,x_\alpha} \;,
\end{equation}
and
\begin{eqnarray}
  \label{eq:Phi}
  \Phi_{m,d}(\upsilon)&=&\frac{1}{2^{2+m}\,\pi^{(6+m-2\epsilon)/4}}\,
  {\bigg[}
  \frac{\Gamma{\big(1-
      \frac{\epsilon}{2}\big)}}{\Gamma{\big(\frac{m+2}{4}\big)}}\,
\nonumber\\&&\times
\mathop{_{1\!}{F}_{2}}\nolimits{\Big(1-\frac{\epsilon}{2};\frac{1}{2},\frac{m+2}{4};
    \frac{\upsilon^4}{64}\Big)}
-\frac{\upsilon^2\,\Gamma\big(\frac{3-\epsilon}{2}\big)}{4\,
   \Gamma\big(1+\frac{m}{4}\big)}\,
\nonumber\\&&\times
\mathop{_{1\!}{F}_{2}}\nolimits{\Big(\frac{3-\epsilon}{2};\frac{3}{2},1+\frac{m}{4};
        \frac{\upsilon^4}{64}\Big)}{\bigg]}\;,
\end{eqnarray}
where $\epsilon=4+\frac{m}{2}-d$ while $_{p}F_q$ means the generalized
hypergeometric function.

In our calculations below we need, in particular, the free bulk
propagator between a point $\bm{x}=(\bm{r},z)$ and its mirror point
$\check{\bm{x}}\equiv(\bm{r},-z)=\bm{x}-2z\,\bm{n}$. With the aid of
the Taylor expansion of the scaling function $\Phi_{m,d}(\upsilon)$
given in Eqs.~(10) and (11) of Ref.~\onlinecite{SD01} one easily
obtains the result
\begin{eqnarray}
  \label{eq:Gbxxs}
  G_{\text{b}}(\bm{x}-\check{\bm{x}})&=&G_{\text{b}}(2z\,\bm{n})=
  \Phi_{m,d}(0)\,\mathring{\sigma}^{-m/4}\,(2z)^{\epsilon-2}
  \nonumber\\[\smallskipamount] 
&=&F_{m,\epsilon}\,\frac{
    \Gamma(2 - \epsilon)\,
      \sin ({\epsilon\,\pi }/{2})}
    {\epsilon\,\pi\,\mathring{\sigma}^{m/4}}\,z^{\epsilon-2}\;,
\end{eqnarray}
in which 
\begin{equation}\label{Fmeps}
F_{m,\epsilon}=
\frac{\Gamma{\left(1+{\epsilon/ 2}\right)}
\,\Gamma^2{\left(1-{\epsilon/ 2}\right)}\,
\Gamma{\left({m}/{4}\right)}}{(4\,\pi)^{({8+m-2\,\epsilon})/{4}}\,
\Gamma(2-\epsilon)\, 
\Gamma{\left({m}/{2}\right)}}
\end{equation}
is a factor introduced in Ref.~\onlinecite{SD01}; just as there, we
will absorb it in the renormalized coupling constant $u$ to be defined
below.

\section{Renormalization group}
\label{sec:ren}

\subsection{General considerations}
\label{sec:RGgencons}

Let us introduce the cumulants involving $N$ fields $\phi_{a_j}$ at
points $\bm{x}_j$ off the surface and $M$ boundary fields
$\phi^{\mathfrak{B}}_{b_k}(\bm{r}_k)\equiv\phi_{b_k}(\bm{r}_k,0)$,
namely
\begin{equation}
  \label{eq:GNMdef}
  G^{(N,M)}(\bm{x};\bm{r})
=\biggl\langle \prod_{j=1}^N\phi_{a_j}(\bm{x}_j)
\prod_{k=1}^M
\phi^{\mathfrak{B}}_{b_k}(\bm{r}_k)
\biggr\rangle^{\text{cum}}.
\end{equation}
Here $\bm{x}$ and $\bm{r}$ are convenient short-hands for the sets of
position variables $\{\bm{x}_j\}$ and $\{\bm{r}_k\}$, and the
components indices $\{a_j\}$ and $\{b_k\}$ have been suppressed on the
right-hand side for the sake of notational simplicity.

We wish to analyze the critical behavior of these functions in bulk
dimensions $d\le d^*(m)$ using a field-theoretic RG approach and the
$\epsilon=d^*(m)-d$ expansion. To regularize their ultraviolet (uv)
singularities, we employ dimensional regularization. Aside from the
bulk uv singularities induced by the bulk part of the free propagator,
additional primitive ones localized on the surface occur.  

Consider first the former ``bulk uv singularities''. In previous
investigations\cite{DS00a,SD01} of the bulk model,
$\mathring{\rho}$ and $\mathring{\tau}$ were set to their values
$\mathring{\rho}_{\text{LP}}$ and $\mathring{\tau}_{\text{LP}}$ at the
Lifshitz point in the actual calculations, or else deviations
$\delta\mathring{\tau}\equiv\mathring{\tau}-
\mathring{\tau}_{\text{LP}}\ne 0$ with
$\delta\mathring{\rho}\equiv\mathring{\rho}-\mathring{\rho}_{\text{LP}}=0$
and $\delta\mathring{\tau}\ne 0$ with $\delta\mathring{\rho}=0$ were
considered. This is sufficient to determine the two RG eigenexponents
$1/\nu_{l2}$ and $\varphi/\nu_{l2}$ associated with the corresponding
eigenoperators at the infrared (ir) stable fixed point (where
$\nu_{l2}$ is a standard correlation exponent while $\varphi$ means
the crossover exponent pertaining to $\delta\mathring{\rho}$). One
must remember, however, that in the $\rho$-dependent scaling forms of
the correlation functions derived in Refs.~\onlinecite{DS00a} and
\onlinecite{SD01} the linear scaling fields associated with these
eigenexponents ought to be replaced by nonlinear ones when considering
general deviations $\delta\mathring{\rho}$ and $\delta\mathring{\tau}$
from the Lifshitz point.

Here we wish to go beyond these previous analyses by allowing both
$\delta\mathring{\tau}$ and $\delta\mathring{\rho}$ to be nonzero and
not necessarily small. Since the naive dimension of
$\mathring{\rho}\,\mathring{\sigma}^{-1/2}$ is
$[\mathring{\rho}\,\mathring{\sigma}^{-1/2}]=\mu$, we see that
deviations $\delta\mathring{\rho}\ne 0$ may give contributions to the
renormalization of $\delta\mathring{\tau}$.  That is, counterterms of
the form $\propto\mu\rho^2\int_{\mathfrak{V}}\bm{\phi}^2$ (where
$\rho$ is the dimensionless renormalized counterpart of
$\delta\mathring{\rho}\,\mathring{\sigma}^{-1/2}$) are possible.
Combining this with the considerations made in Ref.~\onlinecite{DS00a}
and \onlinecite{SD01}, we can conclude that the bulk uv singularities
can absorbed by making the following ``bulk reparametrizations'':
\begin{eqnarray}
  \label{eq:bulkrep}
  \bm{\phi}&=&Z_\phi^{1/2}\,\bm{\phi}_{\text{ren}}\;,
  \nonumber\\[\smallskipamount]
  \mathring{\sigma}&=&Z_\sigma\,\sigma\;,
\nonumber\\[\smallskipamount]
\mathring{\tau}-\mathring{\tau}_{\text{LP}}&=&
\mu^2\,Z_\tau\,{\big[\tau+A_\tau\,\rho^2\big]}\;, 
\nonumber\\[\smallskipamount]
\left(\mathring{\rho}-\mathring{\rho}_{\text{LP}}\right)\,
{\mathring{\sigma}}^{-1/2}&=&\mu\,Z_\rho\,\rho\;,
\nonumber\\[\smallskipamount]
   \mathring{u}\,{\mathring{\sigma}}^{-m/4}\,F_{m,\epsilon}&=&
   \mu^\epsilon\,Z_u\,u\;.\\ 
\nonumber
\end{eqnarray}
Here $Z_\iota=Z_\iota(u,\epsilon)$, $\iota=\sigma, \,\rho,\,u$ are
bulk renormalization factors for which results to two-loop order were
given in Refs.~\onlinecite{DS00a} and \onlinecite{SD01} for general
values of $m$. In our perturbative approach based on dimensional
regularization and the $\epsilon$ expansion the critical values
$\mathring{\tau}_{\text{LP}}$ and $\mathring{\rho}_{\text{LP}}$
vanish, as usual. $A_\tau$ is the additional renormalization function
associated with the counterterm discussed above. Owing to the
restrictive assumptions made in Ref.~\onlinecite{DS00a} and
\onlinecite{SD01}, it was not needed --- and hence not computed ---
there. A straightforward one-loop calculation yields
\begin{equation}
  \label{eq:Atau}
  A_\tau(u,\epsilon)=-\frac{n+2}{3}\,\frac{m}{16}\,\frac{u}{\epsilon}
  +O(u^2)\;. 
\end{equation}

Next, we turn to the ``surface uv singularities'' induced by the free
propagator's ``surface part'', i.e., the part proportional to
$e^{-\mathring{\kappa}_{\bm{p}}|z-z'|}$. They require additional
counterterms, localized on the surface. To determine their form, we
use power counting in conjunction with what is generally known about
such field theories with planar boundaries.\cite{Die86a,Die97} In
comparison to the usual $({m=0})$ semi-infinite $\phi^4$ model, a
qualitative change occurs: There is a surface variable with zero
$\mu$-dimension, namely
$\mathring{\lambda}\,\mathring{\sigma}^{-1/2}$. This implies that the
surface renormalization functions do not only depend on the
renormalized bulk coupling constant $u$ but additionally on the
renormalized analog of this bare variable, which we denote as
$\lambda$. Furthermore, power counting suggests that a surface
counterterm of the form $\propto \int_{\mathfrak{B}}{dA}\,
\sum_\alpha(\partial_\alpha\bm{\phi})^2$ is needed even if
$\mathring{\lambda}$ is set to zero. If a large-momentum cutoff
$\Lambda$ were used to regularize the uv singularities of the theory,
then the associated renormalization function would diverge as
$\ln\Lambda$.  Hence it must have poles in $\epsilon$ in
renormalization schemes based on dimensional regularization such as
ours.

Finally, arguments completely analogous to those giving the
contributions $\propto \rho^2$ to the renormalization of
$\mathring{\tau}$ tell us that the renormalization of the surface
enhancement variable $\mathring{c}$ involves contributions linear in
$\rho$, and hence is not multiplicative.  The upshot of these
considerations is that the following reparametrizations of surface
quantities are needed:
\begin{eqnarray}
  \label{eq:surfrep}
  \bm{\phi}^{\mathfrak{B}}&=&(Z_\phi
  Z_1)^{1/2}\,\bm{\phi}^{\mathfrak{B}}_{\text{ren}}\;,
\nonumber\\[\smallskipamount]
  \mathring{c}-\mathring{c}_{\text{sp}}&=&
  \mu\,Z_c\,{\big[c+A_c(u,\lambda,\epsilon)\,\rho\big]}\;, 
\nonumber\\[\smallskipamount]
\mathring{\lambda}\,\mathring{\sigma}^{-1/2}&=&\lambda+
P_\lambda(u,\lambda,\epsilon)\;.
\end{eqnarray}
Here the surface renormalization factors
$Z_{1,c}=Z_{1,c}(u,\lambda,\epsilon)$ have the form
\begin{eqnarray}
  \label{eq:Zsurf}
  Z_{1,c}-1&=&\sum_{i,j=1}^\infty
  Z_{1,c}^{(i,-j)}(\lambda)\,u^i\,\epsilon^{-j}\nonumber\\ 
  &=&\sum_{i,j=1}^\infty\sum_{k=0}^\infty Z_{1,c}^{(i,-j;k)}\,
    u^i\, \epsilon^{-j}\,\lambda^k\;,
\end{eqnarray}
provided we fix them by requiring that the poles be minimally
subtracted. Likewise we have for the renormalization function
$P_\lambda$,
\begin{eqnarray}
  \label{eq:Plambda}
  P_\lambda(u,\lambda,\epsilon)&=&\sum_{i,j=1}^\infty
  P_{\lambda}^{(i,-j)}(\lambda)\,u^i\,\epsilon^{-j}\nonumber\\ 
  &=&\sum_{i,j=1}^\infty\sum_{k=0}^\infty P_\lambda^{(i,-j;k)}\,
     u^i\,\epsilon^{-j}\,\lambda^k\;,
\end{eqnarray}
and similarly for $A_c$. 

It is easy to see that the one-loop coefficient
$P_\lambda^{(1,-1)}(\lambda)$ vanishes for $\lambda=0$. This follows
from the fact that the tadpole graph \;\raisebox{-0.15em}{\includegraphics[width=1.25em]{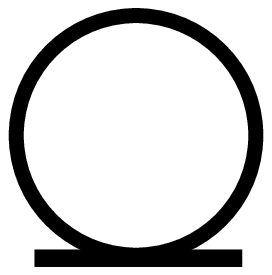}}\; is independent of
the momentum $\bm{p}$, so that no subtraction corresponding to a surface
two-point counterterm $\propto \bm{p}_\alpha^2$ is required for the
one-loop graph of $\langle\phi\,\phi^{\mathfrak{B}}\rangle$ if
$\lambda=0$. On the other hand, a subtraction of this kind \emph{is
  needed} for its two-loop subgraph
\raisebox{-0.5em}{\begin{texdraw} \drawdim pt \setunitscale 3 \linewd
    0.3 \move(-1 0)  \move(8 0) \lellip rx:5
    ry:2.5  \move(1 0) \rlvec(14 0) 
\end{texdraw}}
with $\lambda=0$ (where the crossed circle marks the external surface
point). Hence, $P_\lambda^{(2,-1)}(\lambda=0)$ \emph{does not
  vanish}.

\subsection{One-loop results for general values of $\lambda$}
\label{sec:onelgenlam}

At one-loop order, the bulk renormalization factors $Z_\phi$,
$Z_\sigma$, $Z_\tau$, $Z_\rho$, and $Z_u$ are known to be independent
of $m$ and hence equal to their $m=0$ analogs. Obviously, this
property cannot be expected to carry over to the surface
renormalization factors $Z_1(u,\lambda,\epsilon)$ and
$Z_c(u,\lambda,\epsilon)$ because of their $\lambda$~dependence. On
the other hand, it does hold for $Z_1(u,0,\epsilon)$ and
$Z_c(u,0,\epsilon)$.

To see this, note that a straightforward calculation presented in
Appendix~\ref{app:Plam} yields the one-loop results
\begin{equation}
  \label{eq:Z1onel}
  Z_1(u,\lambda,\epsilon)=1+\frac{n+2}{3}\,
  \frac{i_1(\lambda;m)\,u}{2\epsilon} + O(u^2)\;,
\end{equation}
\begin{equation}
  \label{eq:Zconel}
  Z_c(u,\lambda,\epsilon)=1+\frac{n+2}{3}\,
  \frac{{[2\,i_2(\lambda;m)-i_1(\lambda;m)]}\,u}{2\epsilon} + O(u^2)\;,
\end{equation}
and
\begin{equation}
  \label{eq:Plambdares}
  P_\lambda(u,\lambda,\epsilon)=-\frac{n+2}{3}\,
  \frac{i_1(\lambda;m)\,\lambda\,u}{2\epsilon}\,+O(u^2)\;.
\end{equation}
Here $i_1(\lambda;m)\equiv i_1(\lambda,0;m)$ and $i_2(\lambda;m)\equiv
i_2(\lambda,0;m)$ are special cases of the integrals
\begin{equation}
  \label{eq:i1}
  i_1(\lambda,\epsilon;m)={\int_0^1}\!dt\, \frac{2\,t^{(m-2)/2}\,
    {\big(1-t^2\big)}^{(2-2\epsilon-m)/4}}{B[m/4,(6-m-2\epsilon)/4]}\,
  \frac{1-\lambda\,t}{1+\lambda\,t}
\end{equation}
and
\begin{eqnarray}
  \label{eq:i2}
  \lefteqn{i_2(\lambda,\epsilon;m)}&&\nonumber\\
&=&{\int_0^1}\!dt\,
  \frac{2\,t^{(m-2)/2}\,
    {\big(1-t^2\big)}^{(2-2\epsilon-m)/4}}{B[m/4,(6-m-2\epsilon)/4]}\,
  \frac{1}{{(1+\lambda\,t)}^2}\;,\nonumber\\
\end{eqnarray}
where $B(a,b)$ is the Euler Beta function.

At $\lambda=0$ and for $\lambda\to\infty$ we have
\begin{equation}
  \label{eq:ilambda0inf}
  i_1(0;m)=i_2(0;m)=-i_1(\infty;m)=1\;,\quad i_2(\infty;m)=0\;.
\end{equation}
These values for $\lambda=0$ ensure that our one-loop results
(\ref{eq:Z1onel}) and (\ref{eq:Zconel}) for $Z_1$ and $Z_c$ reduce to
their $m=0$ analogs\cite{com:comp} when $\lambda$ is set to zero.

\subsection{RG equations}

Upon changing $\mu$ at fixed values of the bare interaction constants,
we see that the renormalized functions
$G^{(N,M)}_{\text{ren}}=Z_\phi^{-(N+M)/2}Z_1^{-M/2}\,G^{(N,M)}$
satisfy the RG equations
\begin{equation}
  \label{eq:RGE}
  {\left[ {\mathcal D}_\mu +\frac{N+M}{2}\,\eta_\phi+
      \frac{M}{2}\,\eta_1\right]}G^{(N,M)}_{\text{ren}}=0
\end{equation}
with
\begin{equation}
  \label{eq:Dmu}
  {\mathcal D}_\mu\equiv\mu\partial_\mu+\sum_{\wp
    =u,\sigma,\tau,\rho,c,\lambda}\beta_\wp\,\partial_\wp\;.
\end{equation}
Here $\eta_\phi(u)$ and $\eta_1(u,\lambda)$ denote particular ones of
the functions
\begin{equation}
  \label{eq:etasdef}
  \eta_\wp\equiv\left.\mu\partial_\mu\right|_0\ln
  Z_\wp\;,\quad\wp=\phi,u,\sigma,\tau,\rho,1,c\,, 
\end{equation}
where $\left.\mu\partial_\mu\right|_0$ stands for a $\mu$ derivative
at fixed bare interaction constants $\mathring{u}$, $\mathring{\tau}$,
$\mathring{\rho}$, $\mathring{\sigma}$, $\mathring{c}$, and
$\mathring{\lambda}$.

The beta functions appearing in Eq.~(\ref{eq:Dmu}) are defined
via
\begin{equation}
  \label{eq:betasdef}
  \beta_\wp\equiv\left.\mu\partial_\mu\right|_0\wp\;,\quad\wp
  =u,\sigma,\tau,\rho,c,\lambda\,. 
\end{equation}
They can be conveniently
expressed in terms of the $\eta_\wp$,
\begin{equation}
  \label{eq:btau}
  b_\tau(u)\equiv A_\tau\, {\big[
    \left.\mu\partial_\mu\right|_0\ln A_\tau+
    \eta_\tau-2\eta_\rho\big]}\;,  
\end{equation}
and
\begin{equation}
  \label{eq:bcfunc}
  b_c(u,\lambda)=A_c\,{\big[
    \left.\mu\partial_\mu\right|_0\ln A_c+\eta_c-\eta_\rho\big]}\;.
\end{equation}
We have
\begin{eqnarray}
  \label{eq:betauetc}
  \beta_u(u,\epsilon)&=&-u\,[\epsilon+\eta_u(u)]\;,
\nonumber\\
\beta_\sigma(u,\sigma)&=&-\sigma\,\eta_\sigma(u)\;,
\nonumber\\
\beta_\tau(u,\tau,\rho)&=&-\tau\,[2+\eta_\tau(u)]-\rho^2\,b_\tau(u) \;,
\nonumber\\
\beta_\rho(u,\rho)&=&-\rho\,[1+\eta_\rho(u)]\;,
\nonumber\\
\beta_c(u,\lambda,\rho,c)&=&
-c\,[1+\eta_c(u,\lambda)]-\rho\,b_c(u,\lambda)\;, 
\end{eqnarray}
and
\begin{equation}
  \label{eq:betalambda}
  \beta_\lambda(u,\lambda)=\frac{-\beta_u(u,\epsilon)\,\partial_u
    P_\lambda(u,\lambda,\epsilon)}{1+\partial_\lambda
    P_\lambda(u,\lambda,\epsilon)} \;. 
\end{equation}
That the functions $\eta_{u,\sigma,\tau,\rho}(u)$, $b_\tau(u)$,
$\eta_{c,1}(u,\lambda)$, $b_c(u,\lambda)$, and
$\beta_\lambda(u,\lambda)$ are independent of $\epsilon$ is due to our
use of the minimal subtraction prescription. As usual, the eta
functions can be written in terms of $u$ derivatives of the residues
of the $Z$ factors; we have
\begin{equation}
  \label{eq:minsubformeta}
  \eta_\wp(u,\lambda)=-u\partial_u\mathop{\text{Res}}_{\epsilon=0}
  Z_\wp(u,\lambda,\epsilon)\;,\;\;\wp=\phi,u,\sigma,\tau,\rho,1,c\;.
\end{equation}
The analogous results for $b_\tau$, $b_c$, and $\beta_\lambda$,
\begin{equation}
  \label{eq:btaumin}
  b_\tau(u)=-u\partial_u\mathop{\text{Res}}_{\epsilon=0}
  A_\tau(u,\epsilon)\;, 
\end{equation}
\begin{equation}
  \label{eq:bcmin}
  b_c(u,\lambda)=-u\partial_u\mathop{\text{Res}}_{\epsilon=0}
  A_c(u,\lambda,\epsilon)\;, 
\end{equation}
and
\begin{eqnarray}
  \label{eq:betalamres}
  \beta_\lambda(u,\lambda)&=&u\partial_u\mathop{\text{Res}}_{\epsilon=0}
    P_\lambda(u,\lambda,\epsilon) \nonumber\\ &=&
 \sum_{i=1}^\infty i\,P_\lambda^{(i,-1)}(\lambda)\,u^i\;,
\end{eqnarray}
can be derived from Eqs.~(\ref{eq:btau}), (\ref{eq:bcfunc}), and
(\ref{eq:betalambda}), respectively.

Upon substituting Eq.~(\ref{eq:Atau}) into Eq.~(\ref{eq:btaumin}), we
obtain
\begin{equation}
  \label{eq:btaures}
  b_\tau(u)=\frac{n+2}{3}\,\frac{m\,u}{16}+O(u^2)\;.
\end{equation}
Our perturbative result for $\beta_\lambda$ will be given and
discussed below [see Eqs.~(\ref{eq:betalambdares}) and
(\ref{eq:P21zerom2})--(\ref{eq:P2min1numval}), and
Fig.~\ref{fig:betalamonel}]. The function $b_c(u)$ will not be
computed in this paper since its explicit form is not needed for our
subsequent analysis.

\subsection{Flow equations and fixed points}
\label{sec:foweq}

To exploit the RG equations (\ref{eq:RGE}) via characteristics, we
introduce running coupling constants $\bar{\wp}(\ell)$ into which the
$\wp$ evolve under a change $\mu\to\bar{\mu}(\ell)=\mu\ell$ of the
momentum scale. They are solutions to the flow equations
\begin{equation}
  \label{eq:floweq}
  \ell\frac{d}{d\ell}\bar{\wp}(\ell)=
  \beta_\wp[\bar{u}(\ell),\ldots]\;,\quad\wp=
  u,\sigma,\tau,\rho,c,\lambda\,,
\end{equation}
satisfying the initial conditions
\begin{equation}
  \label{eq:initcond}
  \bar{\wp}(1)=\wp\;,\quad\wp=
  u,\sigma,\tau,\rho,c,\lambda\,.
\end{equation}

As is well known, for dimensions $d<d^*(m)$ the bulk critical behavior
at the Lifshitz point $\tau=\rho=0$ (with $\sigma>0$) is described by a
fixed point that is located at the nontrivial zero
$u^*=u^*(m,\epsilon)$ of the beta function $\beta_u$. According to
Eq.~(60) of Ref.~\onlinecite{SD01}, the expansion to order
$\epsilon^2$ of this root is given by
\begin{eqnarray}\label{eq:ustar}
u^*&=&\frac{2\,\epsilon}{3}\,\frac{9}{n+8}+\frac{8\,\epsilon^2}{27}\,
{\bigg[\frac{9}{n+8}\bigg]}^3\,{\bigg\{}
3\,\frac{5n+22}{27}\,J_u(m)\nonumber\\
&&\mbox{}+\frac{1}{24}\,\frac{n+2}{3}
\,{\bigg[\frac{j_{\sigma}(m)}{8\,(m+2)}-j_\phi(m)
\bigg]}
\bigg\}+O{(\epsilon^3)}\;,\nonumber\\
\end{eqnarray}
where $J_u(m)$ is one of the four single integrals $j_\phi(m)$,
$j_\sigma(m)$, $j_\rho(m)$, and $J_u(m)$ in terms of which the
two-loop results for the bulk renormalization factors $Z_\phi(u)$,
$Z_\sigma(u)$, $Z_\rho(u)$, $Z_\tau(u)$, and $Z_u(u)$ were expressed
in Ref.~\onlinecite{SD01}. It is given by
\begin{equation}\label{eq:Ju}
J_u(m)=1-\frac{C_E+\psi{\left(2-\frac{m}{4}\right)}}{2}+
j_u(m)\;,
\end{equation}
where $C_E=0.577216\ldots$ denotes Euler's constant while $\psi(x)$ is
the digamma function and $j_u(m)$ means the integral
\begin{equation}\label{eq:judef}
j_u(m)=\frac{B_m}{2^{4+m}\,\pi^{(6+m)/4}}\,
{\int_0^\infty}\!d\upsilon\,\upsilon^{m-1}\,\Phi_{m,d^*}^2(\upsilon)\,
\Theta_m(\upsilon)
\end{equation}
with
\begin{eqnarray}\label{eq:Theta}
\Theta_m(\upsilon)&=&\frac{\upsilon^4}{32}\,
\frac{1}{\Gamma(\frac{3}{2}+\frac{m}{4})}
\,\mathop{_{2}{F}_{3\!}}\nolimits{\Big(1,1;\frac{3}{2},2,\frac{3}{2}+\frac{m}{4};
\frac{\upsilon^4}{64}\Big)}
\nonumber\\&&\mbox{}
-\frac{\upsilon^2}{4}\frac{\sqrt{\pi}}{\Gamma(1+\frac{m}{4})} 
\;\mathop{_{1\!}{F}_{2}}\nolimits{\Big(\frac{1}{2};\frac{3}{2},1+\frac{m}{4};
  \frac{\upsilon^4}{64}\Big)}\quad 
\end{eqnarray}
and
\begin{equation}\label{Bm}
B_m\equiv \frac{S_{4-\frac{m}{2}}\, S_m}{F_{m,0}^2}=
{\frac{{2^{10 + m}}\,{{\pi }^{6 + {\frac{3\,m}{4}}}}\,
     \Gamma({\frac{m}{2}})}{\Gamma(
      2 - {\frac{m}{4}})\,
     {{\Gamma({\frac{m}{4}})}^2}}}\;.
\end{equation}
The quantity $S_m$ denotes the surface area $S_m\equiv
2\,\pi^{m/2}/\Gamma(m/2)$ of an $m$-dimensional unit sphere.

Since some of our analytical results to be given below involve besides
the functions $j_\phi(m)$ and $j_\sigma(m)$ also $j_\rho(m)$, let us
recall their definition here for completeness. We have
\begin{equation}\label{eq:jphidef}
j_\phi(m)\equiv B_m\,J_{0,3}(m)
%=B_m\,{\int_0^\infty}\!{d}\upsilon\,\upsilon^{m-1}\,
%\Phi_{m,d^*}^3(\upsilon)\;, 
\end{equation}
and
\begin{equation}\label{eq:jsigmadef}
j_\sigma(m)\equiv B_m\,J_{4,3}(m)
 %=B_m\,{\int_0^\infty}\!{d}\upsilon\, \upsilon^{m+3}\,
%\Phi_{m,d^*}^3(\upsilon)
\;, 
\end{equation}
where $J_{0,3}(m)$ and $J_{4,3}(m)$ are special cases of the integral
\begin{equation}\label{eq:Jps}
J_{p,s}(m)\equiv\int\limits_0^\infty \!\upsilon^{m-1+p}\,
\Phi_{m,d^*}^s(\upsilon)\,d\upsilon\;,
\end{equation}
previously considered (even for $d\ne d^*$) in
Ref.~\onlinecite{DS00a}. Further,
\begin{equation}\label{eq:jrhodef}
j_\rho(m)\equiv B_m\,{\int_0^\infty}\!{d}\upsilon\,\upsilon^{m+1}\,
\Phi_{m,d^*}^2(\upsilon)\,\Xi_{m,d^*}(\upsilon)\;,
\end{equation}
with
\begin{equation}\label{eq:Xidstar}
\Xi_{m,d^*}(\upsilon)=
\frac{{\upsilon}^{(4-m)/2}}{32\,(2\pi)^{\frac{4+m}{4}}}\,
{\left[
{I}_{\frac{m-4}{4}}{\left(\frac{\upsilon^2}{4}\right)}-
{{\mathop{\mathbf{L}}\nolimits}_{\frac{m-4}{4}}}{\left(\frac{\upsilon^2}{4}\right)}
\right]}\,, 
\end{equation}
where $I_\nu$ and $\mathop{\mathbf{L}}\nolimits_{\nu}$ denote modified
Bessel and Struve functions, respectively.\cite{rem:Xi}

Here we are interested in the surface critical behavior at the analog
of the \emph{ordinary} transition that occurs at the Lifshitz point
(LP).  Whenever it is necessary to distinguish this type of ordinary
transition from its counterpart taking place on the ferromagnetic
section of the critical line (CL), we shall refer to the former as LP
ordinary transition and to the latter as CL ordinary transition.  The
latter belongs, of course, to the surface universality class of the
usual ordinary transition at a critical point (CP), which is described
by the standard semi-infinite $n$-vector model.\cite{Die86a}

Let us note some characteristic features of the LP ordinary transition,
which serve to identify it and distinguish it from other types of
surface transitions such as the special and extraordinary ones:

(i) The surface is \emph{disordered} on the bulk disordered side of
the transition. Hence the order-parameter profile
$\langle\phi(\bm{x})\rangle$, and especially the local surface order
parameter $\langle\phi^{\mathfrak{B}}(\bm{r})\rangle=
\langle\phi(\bm{r},z{=}0)\rangle$, \emph{vanish} at the transition.

(ii) The surface does not become critical on its own but remains
\emph{noncritical} at the transition. That is, the ir singularities
which occur at the surface are induced by the bulk, and the surface
susceptibility
\begin{equation}
  \label{eq:Ggen11}
  \chi_{11}(\bm{p})=\hat{G}^{(0,2)}(\bm{p};z{=}0,z'{=}0)
\end{equation}
diverges neither at $\bm{p}=\bm{0}$ nor at any nonzero value of the
momentum $\bm{p}$, so that its inverse is strictly positive:
\begin{equation}\label{eq:chi11cond}
  \chi_{11}^{-1}(\bm{p})>0\;,\quad \forall\bm{p}\;.
\end{equation}

In the case of the LP special transition, condition (i) continues to
apply. However, Eq.~(\ref{eq:chi11cond}) holds only for nonzero
momenta $\bm{p}$ because the surface becomes critical as well; i.e.,
the surface susceptibility $\chi_{11}\equiv\chi_{11}(\bm{0})$
diverges, $\chi_{11}^{-1}(\bm{0})=0$. On the other hand, at the LP
extraordinary transition, $m_1\equiv\langle\phi^{\mathfrak{B}}\rangle$
does \emph{not} vanish.  Therefore (i) ceases to hold while (ii) still
applies for all $\bm{p}$ (barring eventual Goldstone singularities at
$\bm{p}=\bm{0}$ due to the spontaneous breaking of the continuous
$O(n)$ symmetry for $n\ge 2$ in sufficiently high
dimensions).\cite{com:GSSing}

From the form of the beta function $\beta_c$ in
Eq.~(\ref{eq:betauetc}) for $\rho=0$ we can read off the fixed-point
values $c^*=\pm\infty$ and $c^*=0$ of $c$; all others would require
that
\begin{equation}
  \label{eq:etaccond}
  \eta_c(u^*,\lambda^*)=-1\;.
\end{equation}
We ignore this possibility since we see no reason why the zeros
$\lambda=\lambda^*$ of the function $\beta_\lambda(u^*,\lambda)$
should satisfy Eq.~(\ref{eq:etaccond}). (By inspection of the
perturbation series for $\eta_c$ and $\beta_\lambda$ to low orders in
$u$ one can convince oneself that this condition cannot hold as an
identity in $\epsilon$.)

Recalling that the line of CP ordinary transitions is mapped onto a
fixed point with $c=c^*_{\text{ord}}\equiv\infty$, we expect the fixed
point describing the critical behavior at the LP ordinary transition
to be located at $c=\infty$ as well. In the special case $m=0$, in
which the variable $\mathring{\lambda}$ drops out, this is evident
because the results must reduce to those for the usual isotropic
semi-infinite $\phi^4$ model (cf.\ Refs.~\onlinecite{Die86a} and
\onlinecite{DD80,DD81a,DD81b,DD83a}).  However, for $m>0$ this remains
to be verified.

To this end we need some information about the flow in the $c\lambda$
hyperplane at $u=u^*$ and $\rho=\tau=0$.  Upon inserting the $O(u)$
result (\ref{eq:Plambdares}) for $P_\lambda$ into
Eq.~(\ref{eq:betalamres}), we arrive at a two-loop expression of the
form
\begin{equation}
  \label{eq:betalambdares}
  \beta_\lambda(u,\lambda)= -\frac{n+2}{6}\,i_1(\lambda;m)\,\lambda\,u
   +2u^2\,P^{(2,-1)}(\lambda)+ O(u^3)\;.
\end{equation}

\begin{figure}%[htbp]
  \centering
\includegraphics[width=\columnwidth]{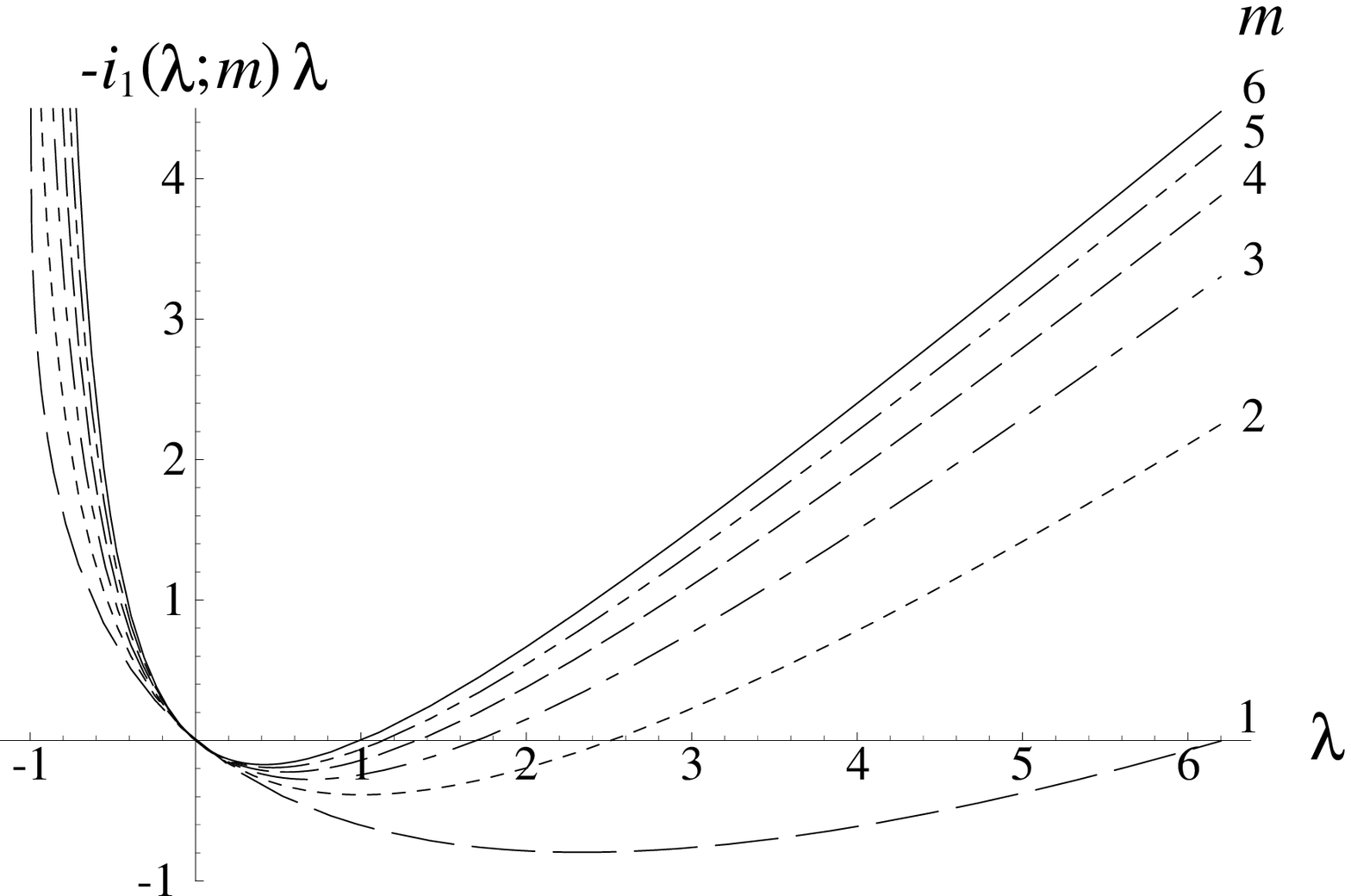}
  \caption{The functions $-i_1(\lambda;m)\,\lambda$ for
    $m=1,2,\ldots,6$.}
  \label{fig:betalamonel}
\end{figure}

Let us first consider $\beta_\lambda(u^*,\lambda)$ in the one-loop
approximation, i.e., to first order in $u$. Its fixed-point values are
given by the zeros of the functions $-\lambda\,i_1(\lambda;m)$, which
are plotted for $m=1,2\ldots, 6$ in Fig.~\ref{fig:betalamonel}.

From their form we see that the trivial zeros $\lambda=0$ are ir
\emph{unstable} in the $\lambda$~direction. There also exist
nontrivial positive roots $\lambda^*_+$ for $1\le m\le 6$, given by the
zeros $\lambda_0(m)$ of the functions $i_1(\lambda;m)$:
\begin{equation}
  \label{eq:lambda0}
  i_1(\lambda_0;m)=0\;\;\;\text{for}\;\;\lambda_0(m)=\left\{
    \begin{array}[c]{l@{\;\;\text{for}\;\;}l}
6.2092&m=1\;,\\ 2.5129&m=2\;,\\ 1.7018&m=3\;,\\ 1.3425&m=4\;,\\
1.1363&m=5\;,\\ 1&m=6\;. 
    \end{array}\right.
\end{equation}
Since $\beta_\lambda(u^*,\lambda)$ has positive slopes at $\lambda_0(m)$,
these fixed points are ir \emph{stable} in the $\lambda$~direction.

At two-loop order the trivial zeros get shifted to
\begin{equation}
  \label{eq:lambdastar}
  \lambda^*=\frac{72\,P^{(2,-1)}(0)}{(n+2)(n+8)}\,\epsilon
  +O(\epsilon^2)\;.
\end{equation}
In deriving this result we have substituted $u^*$ by it $\epsilon$
expansion (\ref{eq:ustar}).

The associated RG eigenvalues (which govern the behavior
$\bar{\lambda}(\ell)-\lambda^*\sim \ell^{-y_\lambda}$ of the running
variable $\bar{\lambda}$ near $\lambda_*$) are
\begin{equation}
\label{eq:ylambda}
  y_\lambda\equiv -(\partial_\lambda\beta_\lambda)(u^*,\lambda^*)
   =\frac{n+2}{n+8}\,\epsilon+O(\epsilon^2)\;.
\end{equation}
According to this $O(\epsilon)$ result, $y_\lambda>0$ for
$\epsilon>0$. Thus the fixed points with $\lambda=\lambda^*$ are
indeed ir unstable.

The two-loop function $P^{(2,-1)}(\lambda)$ is as yet unknown, but in
Appendix~\ref{app:Plam2} we calculate its value at $\lambda=0$,
showing that it is negative. In the special cases $m=2$ and $m=6$,
it can be computed analytically. One obtains
\begin{eqnarray}
\label{eq:P21zerom2}
  \left.P^{(2,-1)}(0)\right|_{m=2}&=&-\frac{n+2}{3}\,\frac{1}{192}\, 
{\big[}5\pi^2-16\ln 2\nonumber\\&&\quad\mbox{}
-6\ln^23-12\,{\text{Li}}_2(1/3)
{\big]}
\end{eqnarray}
and
\begin{equation}
  \label{eq:P21zerom6}
  \left.P^{(2,-1)}(0)\right|_{m=6}=-\frac{n+2}{3}\,\frac{1}{6}\;,
\end{equation}
where ${\text{Li}}_2(x)=\sum_{k=1}^\infty x^k/k^2$ is the dilogarithm,
giving ${\text{Li}}_2(1/3)=0.366213\ldots$. For other choices of $m$,
the integrals in terms of which $P^{(2,-1)}(0)$ is expressed in
Appendix~\ref{app:Plam2} [see Eqs.~(\ref{eq:Ione}), (\ref{eq:Itwo}),
and (\ref{eq:P11fin})] can be determined by numerical means (see
Appendix~\ref{app:numeval}). Our results
\begin{equation}
  \label{eq:P2min1numval}
  P^{(2,-1)}(0)=-\frac{n+2}{3}\,\left\{
    \begin{array}[c]{l@{\;\;\text{for}\;\;}l}
0.12473&m=1\;,\\
0.13865&m=2\;,\\
0.14885&m=3\;,\\
0.15652&m=4\;,\\
0.16231&m=5\;,\\
0.16667&m=6\;,
    \end{array}\right.
\end{equation}
demonstrate that $P^{(2,-1)}(0)$ does not vanish.

In order to understand the meaning of the fixed point at
$(u,c,\lambda)=(u^*,0,\lambda^*)$, one should note that the derivative
$\partial\chi^{-1}_{11}(\bm{p}) /\partial(p_\alpha
p_\alpha)\big|_{\bm{p}=\bm{0}}$ of the zero-loop inverse surface
susceptibility vanishes at $\lambda=0$. This behavior of
$\chi^{-1}_{11}(\bm{p})$ is reminiscent of the vanishing of the
derivative $\partial
\tilde{\Gamma}^{(2)}_{\text{b}}(\bm{q},\tau{=}0,\rho)/\partial(q_\alpha
q_\alpha)\big|_{\bm{q}=\bm{0}}$ of its bulk analog, the bulk vertex
function $\tilde{\Gamma}^{(2)}_{\text{b}}(\bm{q})={\int}d^dx\,
\Gamma_{\text{b}}^{(2)}(\bm{x})\, e^{i\bm{q}\cdot\bm{x}}$, at
$\rho=0$. Clearly, if $\lambda$ becomes negative, then stabilizing
surface contributions to the Hamiltonian of fourth order in
$\partial_\alpha$ and second order in $\bm{\phi}$ will be needed. In
sufficiently high space dimensions $d$, a phase with modulated surface
order and a surface Lifshitz point should exist for such a generalized
model, at least when the bulk parameters $\tau$, $\rho$, and $\sigma$
are in a regime for which the bulk is disordered. Beyond Landau
theory, fluctuations are expected to shift the location of the
``instability point'' of the surface susceptibility (i.e., where
$\partial\chi^{-1}_{11}(\bm{p}) /\partial(p_\alpha
p_\alpha)\big|_{\bm{p}=0}$ vanishes if $c=\rho=\tau=0$) to a nonzero value
of $\lambda$. Hence it is reasonable that the fixed point associated
with this surface wave-vector instability gets shifted to the
nontrivial value (\ref{eq:lambdastar}) of $\lambda$.

In $d=3$~(bulk) dimensions, no surface Lifshitz point is expected to
occur at absolute temperatures $T>0$ for models of the kind considered
(with short-range interactions) because the bulk phase in $d-1$
dimensions to which the required surface phase with \emph{long-range
  modulated order} would correspond to is believed to be
thermodynamically unstable.\cite{Die02,GP76} Thus the fixed point at
$(u,\lambda)=(u^*,\lambda^*)$ and $c=\rho=\tau=0$ is of little
interest. Since our main concern here is the ordinary LP transition,
we will refrain in the sequel from a detailed investigation of the
multicritical behavior it may describe in dimensions $d>3$.

Next, we consider the nontrivial zeros $\lambda_+^*$ of
$\beta_\lambda(\lambda,u^*)$. Using the two-loop expression
(\ref{eq:betalambdares}) in conjunction with the previously utilized
result (\ref{eq:ustar}) for $u^*$, one obtains
\begin{equation}
  \label{eq:lambdaplusstar}
  \lambda_+^*(m)=\lambda_0+  
  \frac{72\,P_\lambda^{(2,-1)}(\lambda_0)\,\epsilon}{(n+2)(n+8)\,%
    \lambda_0\, i_1'(\lambda_0;m)}  + O(\epsilon^2)\;,
\end{equation}
where $i_1'(\lambda;m)\equiv\partial i_1(\lambda;m)/\partial\lambda$.

For the associated correction-to-scaling exponent
\begin{equation}
  \label{eq:omegalam}
  \omega_\lambda \equiv (\partial_\lambda
  \beta_\lambda)(u^*,\lambda^*_+)\;, 
\end{equation}
which governs the behavior $\bar{\lambda}-\lambda^*_+\sim
\ell^{\omega_\lambda}$ of the running variable $\bar{\lambda}$ near
$\lambda^*_+$, we find the result
\begin{equation}
  \label{eq:omegalamres}
  \omega_\lambda= -\frac{n+2}{n+8} \,
  i_1'(\lambda_0;m)\,\lambda_0\,\epsilon+O(\epsilon^2) \;.
\end{equation}
By performing the differentiation $\partial/\partial\lambda$ inside of
the integral (\ref{eq:i1}), or from Fig.~\ref{fig:betalamonel}, one
can easily see that $i_1'(\lambda;m)<0$ for $\lambda>0$. Hence
$\omega_\lambda>0$ to linear order in $\epsilon>0$.

\begin{figure}%[htbp]
  \centering \includegraphics[width=\columnwidth]{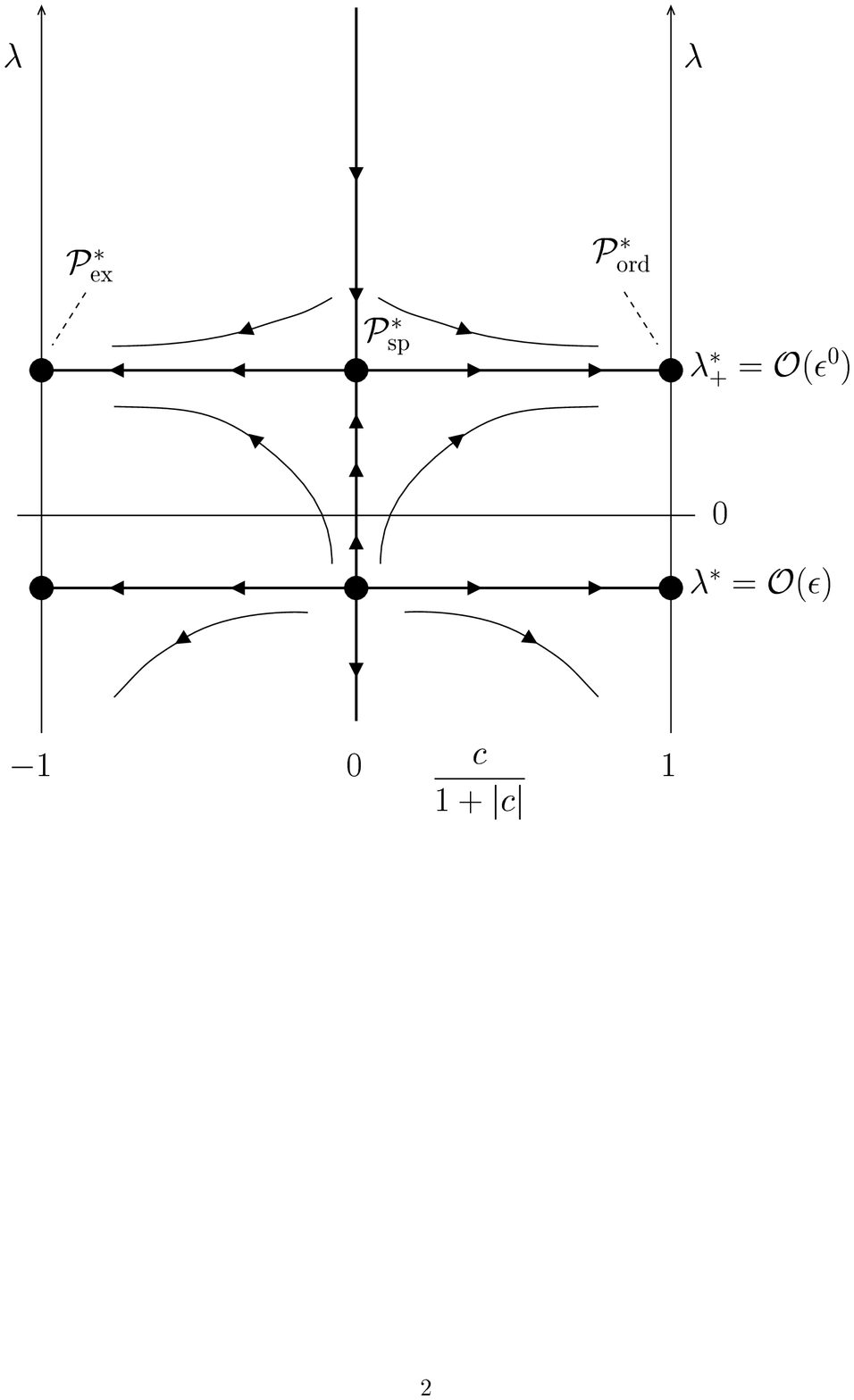}
  \caption{Schematic picture of the RG flow in the $c\lambda$~plane at
    $\rho=\tau=0$ and $u=u^*$, showing the fixed points
    ${\mathcal{P}}_{\text{ord}}^*$, ${\mathcal{P}}_{\text{sp}}^*$, and
    ${\mathcal{P}}_{\text{ex}}^*$ specified in Eq.~(\ref{eq:Pfix}).}
  \label{fig:flow}
\end{figure}

A schematic picture of the flow is depicted in Fig.~\ref{fig:flow}.
On the line $\lambda=\lambda^*_+$ in the hyperplane $u=u^*$ we can
identify the three fixed points
\begin{eqnarray}
  \label{eq:Pfix}
\begin{array}{llr}
  {\mathcal{P}}_{\text{ord}}^*:&
   \quad(c^*_{\text{ord}}=\infty,&\lambda=\lambda^*_+) \;,\\[0.5em]
  {\mathcal{P}}_{\text{sp}}^*:&
   \quad(c^*_{\text{sp}}=0,&\lambda=\lambda^*_+)\;,\\[0.5em]
  {\mathcal{P}}_{\text{ex}}^*:& 
   \quad(c^*_{\text{ex}}=-\infty,&\lambda=\lambda^*_+)\;.\end{array}
\end{eqnarray}
All three are ir stable along the $\lambda$~direction. As the notation
suggests, we therefore expect them to describe the LP ordinary,
special, and extraordinary transitions, respectively.

\section{RG analysis of the LP ordinary transition}
\label{sec:RGOT}

\subsection{Reduction to cumulants with $\mathring{c}=\infty$ and
  $\mathring{\lambda}=0$}
\label{sec:cbinf}

Assuming that the initial values $u$, $c$, and $\lambda$ belong to the
basin of attraction of the fixed point ${\mathcal{P}}_{\text{ord}}^*$,
we now specialize to the case of the LP ordinary transition.  From the
explicit form (\ref{eq:Ghatgen}) of the free propagator $G$ and the
boundary condition (\ref{eq:bcG}) it satisfies we see that in the
limit $\mathring{c}\to\infty$ at fixed $\mathring{\lambda}>0$, the
dependence on $\mathring{\lambda}$ drops out and $G$ turns into the
Dirichlet propagator
\begin{equation}
  \label{eq:GDir}
  {G}_{\text{D}}(\bm{x},\bm{x}')=G_{\text{b}}(\bm{x}-\bm{x}')-
  G_{\text{b}}(\bm{x}-\bm{x}'+2z'\bm{n})\;.
\end{equation}

Both properties, the Dirichlet boundary condition as well as the
$\mathring{\lambda}$ independence for $\mathring{c}\to\infty$, carry
over to each individual Feynman integral of the bare (dimensionally
regularized) correlation functions $G^{(N,M)}$. Thus they hold at
least to any finite order of perturbation theory.

Now, suppose we compute the renormalized functions
$G_{\text{ren}}^{(N,M)}$ using RG-improved perturbation theory. The RG
maps these functions with given initial values $u$, $c$, and $\lambda$
to their RG images at points on the RG trajectory with the values
$\bar{u}(\ell)$, $\bar{c}(\ell)$, and $\bar{\lambda}(\ell)$. To obtain
the asymptotic behavior of the $G_{\text{ren}}^{(N,M)}$ in the large
length-scale limit $\ell\to 0$, we may replace the running
interactions constants $\bar{u}$, $\bar{c}$, and $\bar{\lambda}$ by
their limiting values $u^*$, $c_{\text{ord}}^*=\infty$, and
$\lambda_+^*$ for $\ell\to 0$ provided the RG images of the
correlation functions remain finite and nonzero in this limit. Below
the upper critical dimension $d^*(m)$ (i.e., for $\epsilon>0$), there
is no reason to expect any problems in the limits $\bar{u}\to u^*$ and
$\bar{\lambda}\to \lambda_+^*$: According to RG-improved perturbation
theory, neither $\lambda-\lambda_+^*$ nor $u-u^*$ should be dangerous
irrelevant variables.\cite{com:dangirrel,Fis83} 

On the other hand, as $\bar{c}\to\infty$, all functions
$G_{\text{ren}}^{(N,M)}$ with $M>0$ should vanish. To see this, note
that the boundary condition (\ref{eq:bc}) yields
\begin{equation}
  \label{eq:bcphip}
  \hat{\bm{\phi}}_{\bm{p}}^{\mathfrak{B}}=
  \mathring{c}_{\bm{p}}^{-1}\,\partial_n\hat{\bm{\phi}}_{\bm{p}}
\end{equation}
for the Fourier transform of the bare boundary operator
$\bm{\phi}^{\mathfrak{B}}$, where $\mathring{c}_{\bm{p}}$ was defined
in Eq.~(\ref{eq:cbp}). Using this and expanding in powers of
$1/\mathring{c}$, one sees that the leading contribution of the bare
cumulants $G^{(N,M)}$ in the limit $\mathring{c}\to\infty$ (with all
other interaction constants kept fixed), is given by
\begin{eqnarray}
  \label{eq:GNMcexp}
  G^{(N,M)}(\bm{x};\bm{r})&=&\delta_{M,2}\,\delta_{N,0}
  \,{\mathring{c}}^{-1}\, 
  {\left[1-(\mathring{\lambda}/\mathring{c})\,
      \partial_\alpha\partial_\alpha \right]}
    \delta(\bm{r}_{12})\nonumber\\&&\mbox{}+
    \mathring{c}^{-M}\,G_\infty^{(N,M)}(\bm{x};\bm{r})\;,
\end{eqnarray}
where $\bm{r}_{12}$ denotes the displacement $\bm{r}_1-\bm{r}_2$ of
the two surface points in the case $(N,M)=(0,2)$. The functions on the
right-hand side are defined through
\begin{equation}
  \label{eq:GNMinfdef}
  G_\infty^{(N,M)}(\bm{x};\bm{r})
=\biggl\langle \prod_{j=1}^N\phi_{a_j}(\bm{x}_j)
\prod_{k=1}^M
\partial_n\phi_{b_k}(\bm{r}_k)
\biggr\rangle^{\text{cum}}_{\mathring{c}=\infty,\,
    \mathring{\lambda}=0}.
\end{equation}
The extra term for $(N,M)=(0,2)$ in Eq.~(\ref{eq:GNMcexp}) is produced
by the zero-loop contribution to $G^{(0,2)}$. Equation
(\ref{eq:GNMcexp}) generalizes the $m=0$ result (3.133) of
Ref.~\onlinecite{Die86a} to the $m>0$ case.

The large-$\mathring{c}$ behavior (\ref{eq:GNMcexp}) of the bare
functions $G^{(N,M)}$ implies that the RG images of their renormalized
counterparts $G^{(N,M)}_{\text{ren}}$ vary $\sim
{\left.\bar{c}\right.}^{\,M\,(-1+O(\epsilon))}$ as $\bar{c}\to
\infty$, aside from extra terms proportional to $ \delta(\bm{r}_{12})$
and $\partial_\alpha\partial_\alpha\,\delta(\bm{r}_{12})$ for
$(N,M)=(0,2)$. All in all, the following conclusions can be drawn: 

(i) The $\lambda$ dependence that remains when corrections to scaling
$\sim \lambda-\lambda_+^*$ are ignored (upon making the replacement
$\bar{\lambda}\to\lambda_+^*$) is restricted to nonuniversal
amplitudes (which are expressible via appropriate RG trajectory
integrals\cite{com:amplitudes}). 

(ii) The renormalized cumulants $G^{(N,M)}_{\text{ren}}$ approach zero
in the limit $\bar{c}\to\infty$ whenever $M>0$, and  satisfy
Dirichlet boundary conditions. In other words, taken at
$\bar{c}\to\infty$, they vanish if one or several positions $\bm{x}_j$
of the operators $\bm{\phi}_{\text{ren}}(\bm{x}_j)$ approach the
surface.\cite{com:smalleps} 

(iii) The critical behavior of the $G^{(N,M)}$ at the LP ordinary
transition must match that of the $G_\infty^{(N,M)}$, and hence should
be derivable in much the same way as in the $m=0$ case by a direct RG
analysis of the latter functions.

\subsection{Renormalization of the cumulants $G^{(N,M)}_\infty$}
\label{sec:cbinfren}

To set up such an analysis, let us consider the renormalization of the
cumulants $G^{(N,M)}_\infty$.  Adding source terms to the Hamiltonian,
we introduce the action
\begin{eqnarray}
  \label{eq:actioninf}
  {\mathcal
    A}[\bm{\phi};\bm{J},\bm{J}_{1,\infty}]&=&
  {\int_{\mathfrak{V}}}dV\,\bm{J}\cdot\bm{\phi} +
  {\int_{\mathfrak{B}}}dA\,\bm{J}_{1,\infty}\cdot\partial_n\phi
  \nonumber\\ && \mbox{} -{\left.{\mathcal
        H}[\bm{\phi}]\right|}_{\mathring{c}=\infty,
    \mathring{\lambda}=0}\;,
\end{eqnarray}
in which the Hamiltonian is given by
Eqs.~(\ref{eq:Hamf})--(\ref{eq:L1}) with $\mathring{c}=\infty$ and
$\mathring{\lambda}=0$.  As before, we need the bulk
reparametrizations (\ref{eq:bulkrep}). The surface operator
$\partial_n\bm{\phi}$ is multiplicatively renormalizable; we introduce
its renormalized counterpart $(\partial_n\bm{\phi})_{\text{ren}}$
via
\begin{equation}
  \label{eq:repinfren}
  \partial_n\bm{\phi}=
  \big[Z_\phi(u,\epsilon)\, Z_{1,\infty}(u,\epsilon)\big]^{1/2}
  \big(\partial_n\bm{\phi}\big)_{\text{ren}} \;.
\end{equation}
Since the Fourier transform $\hat{G}^{(0,2)}(\bm{p})$ has
$\mu$-dimension one, it may contain additional primitive uv
divergences localized on the surface of the form
$\mu\,(C_0+\rho\,C_1)+ \sigma^{1/2}\,p_\alpha p_\alpha\,C_2$, where
the renormalization functions $C_0$, $C_1$ and $C_2$ would diverge as
$C_0\sim\Lambda$, $C_1\sim\ln\Lambda$, and $C_2\sim\ln\Lambda$ if we
regularized the theory's uv singularities by a large-momentum cutoff
$\Lambda$. Thus the generating functional of the renormalized
functions $G^{(N,M)}_{\infty,\text{ren}}$ can be written as
\begin{eqnarray}
  \label{eq:Ginfgf}
  \lefteqn{{\mathcal
    G}_{\infty,\text{ren}}[\bm{J},\bm{J}_{1,\infty}]}&&
\nonumber\\&=&\ln\text{Tr}_{\bm{\phi}}\,\exp{\bigg\{} 
      {\mathcal A}{\Big[\bm{\phi};\bm{J}/{\sqrt{Z_\phi}},
          \bm{J}_{1,\infty}/\sqrt{Z_\phi Z_{1,\infty}}\Big]} 
      \nonumber\\ &&\mbox{}
      +\int_{\mathfrak{B}}dA\,\bigg[\mu\,(C_0+\rho\,C_1)\,
      {\bm{J}^2_{1,\infty}}
\nonumber\\&&\qquad\mbox{}+
      \frac{1}{2}\,C_2\,\sigma^{1/2}\,
      \sum_{\alpha=1}^m(\partial_\alpha\bm{J}_{1,\infty})^2\bigg]
      {\bigg\}}\;,
\end{eqnarray}
where $\text{Tr}_{\bm{\phi}}$ means functional integration over
$\bm{\phi}$. Hence we have
\begin{eqnarray}
  \label{eq:Ginfren}
  G^{(N,M)}_{\infty,\text{ren}}&=&{\big[\mu\,\big(C_0+\rho\,C_1\big)
    -\sigma^{1/2}\,C_2\,\partial_\alpha\partial_\alpha\big]}
   \,\delta_{M,2}^{N,0}\,\delta(\bm{r}_{12})\nonumber\\
&&\mbox{}+Z_\phi^{-(N+M)/2}Z_{1,\infty}^{-M/2}\,
  G^{(N,M)}_\infty
\end{eqnarray}
for the renormalized cumulants in position space.

In our dimensional regularization scheme, $C_0$ is finite and hence
can be chosen to vanish. The $\ln\Lambda$ singularities of $C_1$ and
$C_2$ translate into poles at $\epsilon=0$. For convenience, we fix
$Z_{1,\infty}$ and $C_2$ by requiring that the poles of
$\hat{G}^{(1,1)}(\bm{p},z)|_{\rho=0}$ and the additional ones of
$\hat{G}^{(0,2)}(\bm{p})|_{\rho=0}$ be minimally subtracted,
respectively. Likewise, $C_1$ could be determined by requiring that
the additional poles of
$[\partial_\rho\hat{G}^{(0,2)}(\bm{p})]_{\rho=0}$ be minimally
subtracted. However, the explicit two-loop expression for $C_1$ will
not be needed in the sequel, so we refrain from determining it.
%\newpage

\begin{figure}[htbp]
  \centering
  \begin{eqnarray*}
    \hat{G}_\infty^{(1,1)}(\bm{p},z)&=&\raisebox{-0.3em}{\includegraphics[width=4em]{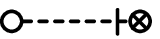}} + \raisebox{-0.3em}{\includegraphics[width=5em]{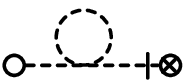}}+\raisebox{-0.3em}{\includegraphics[width=6em]{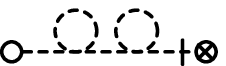}} \\
&&\mbox{}+\raisebox{-0.3em}{\includegraphics[width=5em]{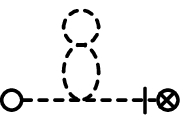}}+\raisebox{-0.3em}{\includegraphics[width=5em]{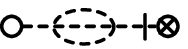}}+O(\mathring{u}^3)
  \end{eqnarray*}
  \caption{Graphs of $\hat{G}_\infty^{(1,1)}(\bm{p},z)$ to two-loop
    order. The dashed line denotes the free Dirichlet propagator
    $\hat{G}_D$. The vertical bars signify normal derivatives
    $\partial_n$. Open and crossed circles indicate external points
    off and on the surface, respectively.}
  \label{fig:Ginf11}
\end{figure}

The graphs of $\hat{G}_\infty^{(1,1)}(\bm{p},z)$ to two-loop order are
shown in Fig.~\ref{fig:Ginf11}.
To this order, the only graph contributing to $C_2$ is the analog of
the last one, namely {\raisebox{-0.4em}{\includegraphics[width=5em]{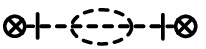}}\,}. In Appendixes~\ref{app:Ginfgraphs} and \ref{app:Ginf02} the
Laurent expansions of these graphs for $\rho=\tau=0$ are computed to
the required order in $\epsilon$. Using these one arrives at the
results
\begin{eqnarray}
  \label{eq:Z1inf}
  Z_{1,\infty}&=&1+\frac{n+2}{3}\,\frac{u}{2\epsilon}+
  \frac{(n+2)(n+5)}{36}\,\frac{u^2}{\epsilon^2} \nonumber\\&&
  \mbox{}+\frac{n+2}{3}\,{\left[j_1(m)-J_u(m)\right]}\,
  \frac{u^2}{4\epsilon} +O{\left(u^3\right)}\quad
\end{eqnarray}
and
\begin{eqnarray}
  \label{eq:C2}
  C_2(u,\epsilon)&=&-
  \frac{u^2}{\epsilon}\,\frac{n+2}{3}\,b_m
  \bigg[\frac{2}{3}\,\frac{J_{2,3}(m)}{6-m}
\nonumber\\&&\mbox{}
-I^{(1)}(m)+I^{(2)}(m)\bigg] +O(u^3)\;.
\end{eqnarray}

Here $b_m$ is the constant introduced by Eq.~(\ref{eq:bm}),
$I^{(1)}(m)$ and $I^{(2)}(m)$ are the integrals defined by
Eqs.~(\ref{eq:Ione}) and (\ref{eq:Itwo}), respectively, while
$J_{2,3}(m)$ denotes a particular one of the integrals (\ref{eq:Jps}).

The quantity
\begin{equation}
  \label{eq:j1m}
  j_1(m)\equiv B_m\, {\int_0^\infty}\!dy\,y^{m-5}\,
\Phi_{m,d^*}(y)
{\int_0^y}\!d\upsilon\,\upsilon^3\,\Phi_{m,d^*}^2(\upsilon)
\end{equation}
is a new integral, not encountered in the bulk case.  It is not
difficult to see that it can be reduced to a single integral.
Rewriting the double integral (\ref{eq:j1m}) as
${\int_0^\infty}\!d\upsilon{\int_\upsilon^\infty}\!dy\ldots$ gives
\begin{equation}
  \label{eq:j1Omegaform}
 j_1(m)= B_m\, {\int_0^\infty}\!d\upsilon\,\upsilon^3\,
 \Phi_{m,d^*}^2(\upsilon) \,\Omega_{m,d^*}(\upsilon)
\end{equation}
with
\begin{equation}
  \label{eq:Omega}
  \Omega_{m,d}(\upsilon)\equiv{\int_\upsilon^\infty}\! dy\, y^{m-5}\,
  \Phi_{m,d}(y)\;.
\end{equation}
The latter integration can be performed analytically; at $d=d^*(m)$
the result becomes
\begin{eqnarray}
  \label{eq:Omegacf} 
  \Omega_{m,d^*}(\upsilon)&=&
  \frac{\sqrt{\pi}\,\upsilon^{m-2}\,\mathop{_{1\!}{F}_{2}}\nolimits{\big(\frac{m-2}{4};
      \frac{m+2}{4},1+\frac{m}{4}; 
  \frac{\upsilon^4}{64}\big)}}{8\,(m-2)\,\Gamma{\big(1+\frac{m}{4}\big)}}
\nonumber\\&&\mbox{}
 -\frac{\upsilon^{m-4}\,\mathop{_{2}{F}_{3\!}}\nolimits{\big(1,\frac{m-4}{4};
     \frac{1}{2},\frac{m+2}{4},\frac{m}{4}; 
\frac{\upsilon^4}{64}\big)}}{(m-4)\,
  \Gamma{\big(\frac{m+2}{4}\big)}}
\nonumber\\
&&\mbox{}-\frac{2^{(3m/2)-7}\,
    \Gamma{\big(\frac{m-2}{4}\big)}}{\sin(m\pi/4)}\;.
\end{eqnarray}

If $m=2$ or $m=6$, then even the $\upsilon$-integration remaining in
Eq.~(\ref{eq:j1Omegaform}) can be performed to determine $j_1(m)$
analytically because the scaling function $\Phi_{m,d^*}$ reduces to an
elementary function.\cite{DS00a,SD01} Analogous statements apply to
the integrals $J_{2,3}(m)$, $I^{(1)}(m)$, and $I^{(2)}(m)$ (cf.\ 
Appendix~\ref{app:numeval}). One obtains
\begin{equation}
  \label{eq:j1val}
  j_1(2)=1-\frac{\ln 3}{2}\;,
\end{equation}
\begin{equation}
j_1(6)=-\frac{2}{3}+2\ln\frac{27}{16}\;,
\end{equation}
\begin{equation}
  \label{eq:J23m2}
 \left[ b_m\,\frac{2}{3}\,\frac{J_{2,3}(2)}{6-m}\right]_{m=2}
 =\frac{1}{54}\;,
\end{equation}
\begin{equation}
 \label{eq:J23m6}
 \left[ b_m\,\frac{2}{3}\,\frac{J_{2,3}(6)}{6-m}\right]_{m\to 6} =
 \frac{9\ln(4/3)-2}{27}\;, 
\end{equation}
\begin{eqnarray}
  \label{eq:I1m2res}
  b_2\,I^{(1)}(2)&=& \frac{1}{16}\big[{\text{Li}}_2(2/3)-
{\text{Li}}_2(1/4) - \ln(4/3)\ln 2\big]
       \nonumber\\&&\strut
 +\frac{\pi^2}{192}    - \frac{1}{12}\ln(3/2)
     \;,
\end{eqnarray}
\begin{equation}
  \label{eq:I1m6res}
 \left[b_m I^{(1)}(m)\right]_{m\to 6}=  \frac{\ln(4/3)}{6} \;,
\end{equation}
\begin{equation}
  \label{eq:I2m2res}
b_2\,  I^{(2)}(2)= \frac{{\pi }^2 + \ln(4/3)\,
     \left[3\,\ln(4/3)-8 \right]  + 
    6\,{\text{Li}}_2(1/4)}{96}\;,
\end{equation}
and
\begin{equation}
  \label{eq:I2m6res}
   \left[b_m I^{(2)}(m)\right]_{m\to 6}=\frac{1-\ln(4/3)}{6}\;,
\end{equation}
where it should be noted that $b_m/(6-m)$ approaches the finite value
$2^{14}\,\pi^9/3$ as $m\to 6$.

The limit
\begin{equation}
  \label{eq:j1m0}
  j_1(0+)=\lim_{m\to 0+}j_1(m)=\frac{1}{2}
\end{equation}
can be determined in a similar manner as the $m\to 0$ limits of the
integrals $j_\phi$, \ldots, $j_u$ were in Ref.~\onlinecite{SD01}. For
other values of $m$, no analytical results for $j_1(m)$ and the other
integrals are available.  We therefore resorted to numerical means,
proceeding in a manner analogous to the one described in appendix E of
Ref.~\onlinecite{SD01} (see Appendix~\ref{app:numeval} below). The
resulting numerical values of $j_1(m)$, $J_{2,3}(m)$, $I^{(1)}(m)$,
and $I^{(2)}(m)$ are gathered in Table~\ref{tab:jis}, along with those
for the (previously computed\cite{SD01}) integrals $j_\phi(m)$,
$j_\sigma(m)$, $j_\rho(m)$, $j_u(m)$, and $J_u(m)$.

\begin{squeezetable}
\begin{table*}[htb]
\caption{Numerical values of the integrals $j_\phi(m)$, $j_\sigma(m)$,
  $j_\rho(m)$, $j_u(m)$, $J_u(m)$, $j_1(m)$, 
  $100\times \frac{2}{3}\,\frac{b_m}{6-m}\,J_{2,3}(m)$, $10\times
  b_m\,I^{(1)}(m)$, and  $10\times b_m\,I^{(2)}(m)$ \label{tab:jis}}
\begin{ruledtabular}
\begin{tabular}{cdddddd}
$m$&1&2&3&4&5&6\\ \hline
$j_\phi(m)$&1.642(9)&1.33333&1.055(6)&0.803(7)&0.57(4)&0.36521\\
$j_\sigma(m)$&1.339(4)&4.74074&10.804(3)&20.067(7)&32.95(4)&49.77778\\
$j_\rho(m)$&0.190(6)&0.88889&1.999(9)&3.464(1)&5.23(4)&7.26958\\
$j_u(m)$&-0.203(7)&-0.40547&-0.624(2)
&-0.880(1)&-1.21(1)&-1.72286\\
$J_u(m)$&0.383(8)&0.28768&0.200(8)&0.119(8)&0.04(3)&-0.02971\\
$j_1(m)$&0.47289&0.45069&0.43092&0.41273&0.39577&0.37983\\
$\frac{2}{3}\,\frac{100\,b_m}{6-m}\,J_{2,3}(m)$ &1.66642&1.85185
&1.98142& 2.07349&2.13819&2.18198\\
$10\,b_m\,I^{(1)}(m)$&0.3644(5)& 0.40504& 0.4337(1)& 0.45438&
0.4690(0)& 0.47947\\
$10\,b_m\,I^{(2)}(m)$&0.882(2)& 0.98150& 1.05(4)& 1.110(7)& 1.15(3)&
1.18720\\
\end{tabular}
\end{ruledtabular}
\end{table*}
\end{squeezetable}

\subsection{RG equations and scaling}
\label{sec:RGEGinf}

Exploiting the relation (\ref{eq:Ginfren}) between the bare and
renormalized cumulants in a standard fashion yields the RG equations
\begin{eqnarray}
  \label{eq:RGinf}
 {\left[{\mathcal D}^{(\infty)}_\mu +\frac{(N+M)\,\eta_\phi}{2}+
      \frac{M\,\eta_{1,\infty}}{2}\right]}
 G ^{(N,M)}_{\infty,\text{ren}}
&&\nonumber\\
=\delta^{N,0}_{M,2}\,{\big(\mu\,{\mathcal R}_1 -\sigma^{1/2}\,
  {\mathcal R}_2\,
  \partial_\alpha\partial_\alpha\big)}\delta(\bm{r}_{12})
\end{eqnarray}
with
\begin{equation}
 \label{eq:Dmuinf}
  {\mathcal D}^{(\infty)}_\mu\equiv\mu\partial_\mu+\sum_{\wp
    =u,\sigma,\tau,\rho}\beta_\wp\,\partial_\wp\;,
\end{equation}
\begin{equation}
  \label{eq:inhR1}
  {\mathcal R}_1(\mu,\rho, u)\equiv Z_\phi^{-1}Z_{1,\infty}^{-1}\,
  {\left.\partial_\mu\right|_0}
  {\big[Z_\phi Z_{1,\infty}\,\mu\,(C_0+\rho\,C_1)\big]}\;, 
\end{equation}
and
\begin{equation}
  \label{eq:inhR2}
  {\mathcal R}_2(\sigma, u)\equiv \sigma^{-1/2}\,
  Z_\phi^{-1}Z_{1,\infty}^{-1}\,
  {\left.\mu\partial_\mu\right|_0}
  {\big[Z_\phi Z_{1,\infty}\,
    \sigma^{1/2}\,C_2\big]}\;, 
\end{equation}
where $\eta_{1,\infty}(u)$ is defined by Eq.~(\ref{eq:etasdef}) with
$\wp=(1,\infty)$.

The RG equation for $G^{(0,2)}_{\infty,\text{ren}}$ is inhomogeneous
because of the additive counterterm in Eq.~(\ref{eq:Ginfgf}) and the
implied additive term in the reparametrization relation
(\ref{eq:Ginfren}). If $C_0$, $C_1$ and $C_2$ are determined by
minimal subtraction of poles (so that $C_1$ and $C_2$ are Laurent
series in $\epsilon$ with a vanishing regular part while $C_0=0$),
then the uv finiteness of ${\mathcal R}_1$ and ${\mathcal R}_2$
together with the familiar expressions (\ref{eq:betauetc}) and
(\ref{eq:minsubformeta}) for $\beta_u$ and the RG functions
$\eta_\iota$ imply that Eqs.~(\ref{eq:inhR1}) and (\ref{eq:inhR2})
simplify to
\begin{equation}
  \label{eq:minRform}
  {\mathcal R}_1=-\rho\,u\partial_u
  \mathop{\text{Res}}_{\epsilon=0}C_1(u,\epsilon)\,,\quad
  {\mathcal R}_2=-u\partial_u
  \mathop{\text{Res}}_{\epsilon=0}C_2(u,\epsilon)\,.
\end{equation}

Thus, to order $u^2$, ${\mathcal R}_2 $ is simply given by $-2\epsilon$
times the term $\propto u^2/\epsilon$ on the right-hand side of
Eq.~(\ref{eq:C2}). For the exponent function $\eta_{1,\infty}(u)$ we
find from Eq.~(\ref{eq:Z1inf}) the result
\begin{equation}
  \label{eq:etainf1}
  \eta_{1,\infty}(u)=
  -\frac{n+2}{6}\,u\,{\big\{1+
    u\big[j_1(m)- J_u(m)\big]\big\}}+O{\left(u^3\right)}\;.
\end{equation}

We are now ready to exploit the RG equations (\ref{eq:RGinf}) by an
appropriate generalization of the analysis given in
Ref.~\onlinecite{DS00a}. Owing to the $\rho$~dependent term of the
function $\beta_\tau$ [cf.\ the set of equations (\ref{eq:betauetc})],
the flow equations (\ref{eq:floweq}) for $\bar{\tau}$ and $\bar{\rho}$
are coupled. We therefore introduce a nonlinear scaling field
\begin{equation}
  \label{eq:gtaudef}
  g_\tau(\tau,\rho,u)=\tau+c^\tau_{\rho^2}(u)\,\rho^2
\end{equation}
such that the associated running variable is a solution to
\begin{equation}
  \label{eq:gtauflow}
 \ell\frac{d}{d\ell}\,
 \bar{g}_\tau(\ell)=-[2+\eta_\tau(\bar{u})]\,\bar{g}_\tau(\ell)\;,\quad
 \bar{g}_\tau(1)=g_\tau\;.
\end{equation}
Substituting the ansatz (\ref{eq:gtaudef}) into the flow equations
(\ref{eq:floweq}) leads to the condition
\begin{equation}
  \label{eq:ctaurhocond}
  \beta_u(\bar{u},\epsilon)\, \frac{\partial
    c^\tau_{\rho^2}(\bar{u})}{\partial \bar{u}}
  +b_\tau(\bar{u}) 
  =[\eta_\tau(\bar{u})-2\,\eta_\rho(\bar{u})]\,c^\tau_{\rho^2}(\bar{u})\;. 
\end{equation}
which is easily solved for $\bar{u}$ near $u^*$ to obtain
\begin{equation}
  \label{eq:ctaurhosol}
  c^\tau_{\rho^2}(\bar{u})=\frac{b_\tau^*}{\eta_\tau^*-2\eta_\rho^*}
  +O(\bar{u}-u^*)\;. 
\end{equation}
Since we shall ignore corrections to scaling $\bar{u}-u^*$ below, it
will be sufficient to keep the first term on the right-hand side. Our
one-loop result (\ref{eq:btaures}) for $b_\tau(u)$ yields for it:
\begin{equation}
  \label{eq:gtaupert}
   c^{\tau}_{\rho^2}(u^*)=\frac{b_\tau^*}{\eta_\tau^*
    -2\eta_\rho^*} 
  =-\frac{m}{8}+O(u^*)\;.
\end{equation}

Let us define the RG trajectory integrals
\begin{eqnarray}
\label{eq:Eiota}
E_{\iota}[\bar{u},u]&=&\exp\left\{
\int_u^{\bar{u}(\ell)}\!{dx}\,
\frac{\eta_\iota^*-\eta_\iota(x)}{\beta_u(x)}
\right\}\;,
\nonumber\\\iota&=&\phi,\,\sigma,\,\rho,\,\tau,\,(1,\infty),
\end{eqnarray}
and the nonuniversal constants
\begin{equation}
 \label{eq:Eiotastar}
E_\iota^*(u)\equiv E_{\iota}(u^*,u)
\;,\;\;\iota=\phi,\,\sigma,\,\rho,\,\tau,\,(1,\infty),
\end{equation}
they approach in the infrared limit $\ell\to 0$. Then the solutions to
the flow equations (\ref{eq:floweq}) and (\ref{eq:gtauflow}) for
$\bar{\sigma}$, $\bar{\rho}$, and $\bar{g}_\tau$ can be written as
\begin{equation}
  \label{eq:sigmaflow}
\bar{\sigma}(\ell)=\ell^{-\eta_\sigma^*}\,
E_\sigma[\bar{u}(\ell),u]\,\sigma
\mathop{\approx}\limits_{\ell\to 0}
\ell^{-\eta_\sigma^*}\,E_\sigma^*(u)\,\sigma\;,
\end{equation}
\begin{equation}
  \label{eq:rhoflow}
\bar{\rho}(\ell)=\ell^{-(1+\eta_\rho^*)}\,
E_\rho[\bar{u}(\ell),u]\,\rho
\mathop{\approx}\limits_{\ell\to 0}
\ell^{-\varphi/\nu_{l2}}\,E_\rho^*(u)\,\rho\;,
\end{equation}
and
\begin{equation}
\label{eq:tauflow}
\bar{g}_\tau(\ell)=\ell^{-(2+\eta_\tau^*)}\,E_\tau[\bar{u}(\ell),u]\,
g_\tau \mathop{\approx}\limits_{\ell\to 0}
\ell^{-1/\nu_{l2}}\,E_\tau^*(u)\,g_\tau\;,
\end{equation}
where we have introduced the familiar correlation-length exponent
\begin{equation}
  \label{eq:nul2}
  \nu_{l2}\equiv(2+\eta_\tau^*)^{-1}
\end{equation}
and the crossover exponent
\begin{equation}
\label{eq:coexp}
\varphi=\nu_{l2}\,(1+\eta_\rho^*)\;.
\end{equation}

We choose $\ell\equiv\ell_\tau$ such that $\bar{g}_\tau(\ell_\tau)=\pm
1$ for $\pm g_\tau>0$, consider the limit $g_\tau \to 0\pm$, and
introduce the scaling lengths\cite{rem:corrl}
\begin{equation}
  \label{eq:xil2}
\xi_{l2}\equiv \mu^{-1}/\ell_\tau\approx
\mu^{-1}\,\left[E_\tau^*(u)\,|g_\tau|\right]^{-\nu_{l2}}
\end{equation}
and
\begin{equation}
  \label{eq:xil4}
\xi_{l4}\equiv \left[\frac{\bar{\sigma}(\ell_\tau)}
  {\mu^2\,{\ell_\tau}^2}\right]^{1/4}\approx
\mu^{-1/2}\,\left[E_\tau^*(u)\,|g_\tau|\right]^{-\nu_{l4}}\;,
\end{equation}
where $\nu_{l4}$ denotes the correlation-length exponent
\begin{equation}
  \label{eq:nul4}
\nu_{l4}\equiv\theta\,\nu_{l2}=\frac{2+\eta_\sigma^*}{4(2+\eta_\tau^*)}\,.
\end{equation}

The RG equations (\ref{eq:RGinf}) can now be solved in a
straightforward manner via characteristics. It is convenient to
utilize instead of $\tau$ the nonlinear scaling field $g_\tau$,
considering the cumulants $G^{(N,M)}_{\infty,\text{ren}}$ as functions
of $\rho$, $g_\tau$, and $u$. Upon substituting the trajectory
integrals (\ref{eq:Eiota}) by their fixed-point values
(\ref{eq:Eiotastar}) and the running interaction constants
$\bar{\sigma}$, $\bar{\rho}$, and $\bar{g}_\tau$ by their asymptotic
expressions displayed on the far right of
Eqs.~(\ref{eq:sigmaflow})--(\ref{eq:tauflow}), one finds that the
$G^{(N,M)}_{\text{ren}}$ behave near the Lifshitz point as
\begin{eqnarray}
\label{eq:RGGscalingform}
  \lefteqn{G^{(N,M)}_{\infty,\text{ren}} (r_\alpha,r_\beta, z; 
                           \rho, g_\tau, u, \sigma, \mu )}&&
\nonumber\\ &\approx&
 {\bigg[\frac{\mu^{-\eta_\phi^*}}{E_\phi^*}\,
 \xi_{l2}^{-(d-m-2+\eta_\phi^*)}\, \xi_{l4}^{-m} \bigg]}^{(N+M)/2}
\nonumber\\&&\times
 {\bigg[\frac{\mu^{-\eta_{1,\infty}^*}}{E_{1,\infty}^*}\,
 \xi_{l2}^{-2-\eta_{1,\infty}^*} \bigg]}^{M/2}\nonumber\\
 &&\times \;
    {\mathcal G}_{\infty,\pm}^{(N,M)}{\bigg[ \frac{r_\alpha}{\xi_{l4}}, 
          \frac{r_\beta}{\xi_{l2}},
          \frac{z}{\xi_{l2}};
          E_\rho^* \,\rho \,(\mu\,\xi_{l2})^{\varphi/\nu_{l2}}
          \bigg]}\,.\;\; 
\end{eqnarray}
Here $r_\alpha$ and $r_\beta$ stand for the sets of all $\alpha$ and
$\beta$ coordinates parallel to the surface, respectively, while $z$
represents the set of all coordinates perpendicular to it.  Further,
${\mathcal G}_{\infty,\pm}^{(N,M)}$ is a universal scaling function,
given by
\begin{eqnarray}
  \label{eq:universalG}
\lefteqn{{\mathcal G}_{\infty,\pm}^{(N,M)}(r_\alpha,
  r_\beta,z;\rho)}&& \nonumber\\&\equiv&
    G^{(N,M)}_{\infty,\text{ren}}[r_\alpha, r_\beta, z;
    \rho,g_\tau{=}1,u^*,1,1]\;.
\end{eqnarray}

In the special case of $G^{(0,2)}_{\infty,\text{ren}}(\bm{r}_{12})$,
the inhomogeneity implies a contribution $\propto \delta(\bm{r}_{12})$
linear in $\rho$ we suppressed in Eq.~(\ref{eq:universalG}),
regarding $G^{(0,2)}_{\infty,\text{ren}}(\bm{r}_{12})$ as a
conventional function (rather than a distribution) with
$\bm{r}_{12}\ne\bm{0}$. In momentum space this term yields a regular
contribution. It reflects the fact that the momentum-dependent surface
susceptibility $\chi_{11}(\bm{p})$ [cf.\ Eq.~(\ref{eq:Ggen11})] does
not diverge at the ordinary transition and may be viewed as analog of
the term $\propto \delta_{M,2}^{N,0}$ in Eq.~(\ref{eq:GNMcexp}).

Owing to the relations (\ref{eq:GNMcexp}) between the bare cumulants
$G^{(N,M)}$ in the limit $\mathring{c}\to\infty$ and
$G_\infty^{(N,M)}$ the scaling forms (\ref{eq:universalG}) carry over
to the behavior of the former at the LP ordinary transition. An
immediate consequence is that the critical exponents at the LP
ordinary transition can be expressed in terms of four independent bulk
critical indices, e.g., the correlation exponent
\begin{equation}
  \label{eq:bulkexp}
  \eta_{l2}\equiv\eta_\phi^*
\end{equation}
and the previously introduced anisotropy, correlation-length, and
crossover exponents $\theta$, $\nu_{l2}$, and $\varphi$ [cf.\ 
Eqs.~(\ref{eq:nul2}) and (\ref{eq:coexp})], and a single surface
critical index. As the latter one can choose the critical exponent
$\beta_1^{\text{ord}}$ governing the asymptotic dependence
\begin{equation}
  \label{eq:m1ordas}
  m_1(\tau,\rho=0) \mathop{\sim}_{\tau\to 0-}
  |\tau|^{\beta_1} 
\end{equation}
of the surface magnetization $m_1$ at $\rho=0$ (and given values of
the remaining interaction constants $c$, $u$, and $\lambda$). From the
behavior of $G^{(0,1)}_{\infty,\text{ren}}$ predicted by
Eq.~(\ref{eq:universalG}) one reads off that
\begin{equation}
  \label{eq:beta1ord}
  \beta_1^{\text{ord}}={\nu_{l2}}\,{\big[
d+\eta_{l2}+\eta_{1,\infty}^*+m\,(\theta-1)\big]}/2\;.
\end{equation}

\subsection{Boundary operator expansion}
\label{sec:boe}
In the above derivation of the scaling behavior of the cumulants
$G^{(N,M)}$ at the ordinary transition we relied on the
large-$\mathring{c}$ form (\ref{eq:GNMcexp}) of these functions. An
alternative way to arrive at the same conclusions, which gives further
insight, is the use of the boundary operator expansion
(BOE).\cite{DD81a,Die86a} Since both the bare and renormalized
theories satisfy Dirichlet boundary conditions for
$\mathring{c}=\infty$ and $c=\infty$, there is no term $\propto
\bm{\phi}^{\mathfrak{B}}$ contributing to the BOE of $\bm{\phi}(\bm{r},z)$.
Just as in the CP case, the leading operator should be $\partial_n
\bm{\phi}$. Hence we write
\begin{equation}
  \label{eq:boe}
  \bm{\phi}_{\text{ren}}(\bm{r},z)\mathop{\approx}_{z\to 0}C_\infty(z)\,
  (\partial_n\bm{\phi})_{\text{ren}}(\bm{r})\;. 
\end{equation}
Consistency with the RG equations (\ref{eq:RGinf}) requires that the
function $C_\infty(z)$ must satisfy the RG equation
\begin{equation}
  \label{eq:RGCinf}
 {\big({\mathcal D}^{(\infty)}_\mu-{\eta_{1,\infty}}/{2}\big)}\,
 C_\infty(z)=0\;.
\end{equation}
Solving this (with $\rho$ set to zero) yields a short-distance
singularity for $z\ll \xi_{l2}$ of the form
\begin{equation}
  \label{eq:Cinfas}
  C_\infty(z)\approx \text{const}\,\mu^{-1}\, (\mu
  z)^{1+\eta_{1,\infty}^*/2}\;.
\end{equation}
By analogy to the CP case, the exponent is governed by the difference
between
\begin{equation}
  \label{eq:betal}
\beta_l=\nu_{l2}\,[d-2+\eta_{l2}+m\,(\theta-1)]/2\;,
\end{equation}
the critical index of the bulk order parameter
$m_{\text{b}}=\langle\phi|_{z=\infty}\rangle$, and
$\beta_1^{\text{ord}}$; we have
\begin{equation}
  \label{eq:betadiff}
  (\beta_1^{\text{ord}}-\beta_l)/\nu_{l2}=1+\eta_{1,\infty}^*/2\;.
\end{equation}

\subsection{Exact critical exponent of the surface energy density}
\label{sec:sed}

In the foregoing subsections~\ref{sec:cbinf}--\ref{sec:boe} we have
shown that the critical exponents of the cumulants $G^{(N,M)}$ at the
ordinary transitions can all be expressed in terms of four bulk
critical indices (e.g., $\eta_{l2}$, $\theta$, $\nu_{l2}$, $\varphi$)
and a single surface critical index such as $\beta_1^{\text{ord}}$. We
now turn to a discussion of the behavior of the surface energy density
\begin{equation}
  \label{eq:seddef}
{\mathcal E}_1\equiv
\big\langle{\big[\bm{\phi}^{\mathfrak{B}}(\bm{r})\big]}^2/2\big\rangle
\end{equation}
at the LP ordinary transition. 

The surface energy density at the CP ordinary transition is a
well-known example of a local boundary density whose critical
exponents are expressible in terms of bulk exponents: Its leading
thermal singularity is of the same form $|\tau|^{2-\alpha}$ as the
bulk free energy.\cite{DD81c,DDE83,BC87,BD94} Our aim here is to show
that the same holds true for the surface energy density at the LP
ordinary transition.

A particularly easy way to obtain this result is to argue along the
lines of Ref.~\onlinecite{BD94} and consider a finite system
that extends in the $z$-direction from ${z=-L_2}$ to $z=L_1$, has
cross-sectional area $A$, and different values $\mathring{\tau}_1$,
$\mathring{\rho}_1$ and $\mathring{\tau}_2$, $\mathring{\rho}_2$ in
the regions ${\mathfrak{V}}_1$ with $z>0$ and ${\mathfrak{V}}_2$ with
$z<0$, respectively. Its Hamiltonian reads
\begin{equation}
  \label{eq:HamBD}
  {\mathcal H}_2=\int_{{\mathfrak{V}}_1}d^dx\, {\mathcal
    L}_{\text{b}}(\bm{x};\mathring{\tau}_1,\mathring{\rho}_1)+
  \int_{{\mathfrak{V}}_2}d^dx\, {\mathcal
    L}_{\text{b}}(\bm{x};\mathring{\tau}_2,\mathring{\rho}_2)\;,
\end{equation}
where ${\mathcal
  L}_{\text{b}}(\bm{x};\mathring{\tau},\mathring{\rho})$ is the bulk
density (\ref{eq:Lb}). We now choose values of $\mathring{\tau}_2$ and
$\mathring{\rho}_2$ such that the bulk at large $-z>0$ is not critical
and in the disordered phase when the thermodynamic limit is
taken.\cite{com:bulknc} Since the half-space $z<0$ for the chosen
values of $\mathring{\tau}_2$ and $\mathring{\rho}_2$ is neither
critical nor has no long-range order, its net effect is to generate
modified effective short-ranged $O(n)$ invariant interactions at the
interface $z=0$.  The boundary critical behavior which occurs at this
interface as $\mathring{\tau}_1\to\mathring{\tau}_{\text{LP}}$ and
$\mathring{\rho}_1\to\mathring{\rho}_{\text{LP}}$ should be in the
universality class of the LP ordinary transition.

Suppose the interface is displaced a small distance $\Delta L$
downwards, so that the heights $L_1$ and $L_2$ of the regions
${\mathfrak{V}}_1$ and ${\mathfrak{V}}_2$ increase and decrease by
$\Delta L$, respectively. The change of total free energy per volume
$A\,\Delta L$ caused by this displacement in the limit $A\to \infty$
and $L_1,\,{L_2\to\infty}$ becomes
\begin{eqnarray}
  \label{eq:WTid}
  \lefteqn{f_{\text{b}}(\mathring{\tau}_1,\mathring{\rho}_1)-
    f_{\text{b}}(\mathring{\tau}_2,\mathring{\rho}_2)} 
  &&\nonumber\\ &=& 
  (\mathring{\tau}_1-\mathring{\tau}_2)\,{\mathcal E}_1+
  \frac{\mathring{\rho}_1-\mathring{\rho}_2}{2}\,
  \sum_{\alpha=1}^m{\big\langle(\partial_\alpha
  \bm{\phi}^{\mathfrak{B}})^2\big\rangle} \;.  
\end{eqnarray}
Here the left-hand side is simply the change of bulk free energy per
unit volume caused by the alteration of values
$(\mathring{\tau}_2,\mathring{\rho}_2)\to
(\mathring{\tau}_1,\mathring{\rho}_1)$ within the region $-\Delta L\le
z\le 0$. The right-hand side follows by expressing this change in
terms of
\begin{equation}
  \label{eq:DeltaH}
  \Delta{\mathcal H}={\int_{0}^{\Delta L}}\!dz{\int} dA
  {\bigg[\frac{\mathring{\tau}_1-\mathring{\tau}_2}{2}\,\bm{\phi}^2+
  \frac{\mathring{\rho}_1-\mathring{\rho}_2}{2}\sum_{\alpha=1}^m
  (\partial_\alpha\bm{\phi})^2\bigg]}\,, 
\end{equation}
the corresponding change of the Hamiltonian, as $-(A\,\Delta L)^{-1}
\ln\,\langle e^{-\Delta{\mathcal H}}\rangle\approx \langle\Delta
{\mathcal H}\rangle /(A\,\Delta L)$, where $\langle.\rangle$ indicates
a thermal average with the Boltzmann factor $e^{-{\mathcal H}_2}$ of
the original unperturbed system. For small $\Delta L$ we have
${\int_0^{\Delta L}}dz\,f(z)\approx f(0)\,\Delta L$. Assuming
translational invariance parallel to the interface then yields the
right-hand side of Eq.~(\ref{eq:WTid}). (This assumption excludes
situations in which the order parameter varies along the interface. In
such cases of broken translational invariance along the interface the
right-hand side of the identity (\ref{eq:WTid}) must be replaced by
its spatial average $A^{-1}\int dA\ldots$ over the interface.)

For small deviations $\mathring{\tau}_1-\mathring{\tau}_{\text{LP}}$
and $\mathring{\rho}_1-\mathring{\rho}_{\text{LP}}$ from the LP, i.e.,
small values of the scaling fields $g_\tau$ [cf.\ 
Eq.~(\ref{eq:gtaudef})] and $\rho$, the first term on the right-hand
side of Eq.~(\ref{eq:WTid}) behaves as
\begin{equation}
  \label{eq:scffb}
  f_{\text{b}}(\mathring{\tau}_1,\mathring{\rho}_1)\approx 
  |g_\tau|^{2-\alpha_l}\,{\mathcal
    F}_\pm{\big(\rho\,|g_\tau|^{-\varphi}\big)}+\text{reg}\;.
\end{equation}
Here ``reg'' stands for regular contributions, and the plus and minus
signs of the subscript $\pm$ refer to the cases $g_\tau>0$ and $g_\tau<0$
respectively.

Noting that ${\mathcal F}(0)\ne 0$, we can equate the leading singular
parts of the identity (\ref{eq:WTid}) to obtain the asymptotic
behavior
\begin{eqnarray}
\label{eq:sedasbeh}
 (\mathring{\tau}_{\text{LP}}-\mathring{\tau}_2)\, {\mathcal
   E}^{\text{sing}}_1+ 
\frac{\mathring{\rho}_{\text{LP}}-\mathring{\rho}_2}{2}\,
  \sum_{\alpha=1}^m{\big\langle(\partial_\alpha
  \bm{\phi}^{\mathfrak{B}})^2\big\rangle}^{\text{sing}}\nonumber\\ 
 \approx {\mathcal F}_\pm(0)\,|\tau|^{2-\alpha_l}  
\end{eqnarray}
for $\tau\to 0\pm$ at $\rho=0$. Hence one of the two surface
quantities on the left-hand side must contain the singularity on the
right-hand side.

The critical line $\mathring{\tau}_c=f_{\text{CL}}(\mathring{\rho})$
is not a straight line in the vicinity of the {LP}. On the section
separating the disordered phase from the modulated ordered one, its
asymptotic behavior near the LP is governed by the crossover exponent
$\varphi$, i.e., $\mathring{\tau}_c-\mathring{\tau}_{\text{LP}}\sim
(\mathring{\rho}-\mathring{\rho}_{\text{LP}})^{1/\varphi}$.
Nevertheless, choices of the point
$(\mathring{\tau}_2,\mathring{\rho}_2)$ in the disordered phase with
$\mathring{\rho}_2=\mathring{\rho}_{\text{LP}}$ and
$\mathring{\tau}_2\ne\mathring{\tau}_{\text{LP}}$ should be possible.
Since for such choices only the term proportional to the surface
energy density ${\mathcal E}_1$ contributes on the left-hand side of
Eq.~(\ref{eq:sedasbeh}), it is clear that ${\mathcal E}_1$ must have
the bulk free-energy singularity on the right-hand side.

This result means that the scaling dimension
$\Delta[(\bm{\phi}^{\mathfrak{B}})^2]$ of the surface energy density
operator $(\bm{\phi}^{\mathfrak{B}})^2$ coincides with that of the
volume $\int dV$, i.e.,
\begin{equation}
  \label{eq:Delsed}
  \Delta[(\bm{\phi}^{\mathfrak{B}})^2] = d+\theta\,(m-1)\;. 
\end{equation}

As usual, the amplitudes ${\mathcal F}_\pm(0)$ are nonuniversal, but
the ratio ${\mathcal F}_+(0)/{\mathcal F}_-(0)$ is a universal bulk
amplitude ratio. The corresponding amplitude ratio of ${\mathcal E}_1$
must have the same value.

There is an instructive alternative way to confirm this result.
Knowing that the ordinary fixed point ${\mathcal{P}}_{\text{ord}}^*$
is located at $c=\infty$, one can study the effect of the
Hamiltonian's boundary term
$(\mathring{c}/2)\int_{\mathfrak{B}}\phi^2$ for large but finite
$\mathring{c}$ to leading order in $1/\mathring{c}$.

Expanding the free propagator (\ref{eq:Ghatgen}) in powers of
$1/\mathring{c}_{\bm{p}}$ gives
\begin{eqnarray}
  \label{eq:oneovcfreeprop}
  \lefteqn{\hat{G}(\bm{p};z,z')-\hat{G}_{\text{D}}(\bm{p};z,z')} &&
  \nonumber\\ 
  &=&\mathring{c}_{\bm{p}}^{-1}
  e^{-\mathring{\kappa}_{\bm{p}}\,(z-z')}+O(\mathring{c}_{\bm{p}}^{-2}) \\
&=&\mathring{c}_{\bm{p}}^{-1}\,(\hat{G}_{\text{D}}
\loarrow{\partial}_n)(\bm{p};z,0)
\,(\partial_n\hat{G}_{\text{D}})(\bm{p};0,z')+
O(\mathring{c}_{\bm{p}}^{-2})\;, 
\nonumber 
\end{eqnarray}
where $\loarrow{\partial}_n$ acts to the left. More generally, one
notices that the expansion of the free propagator in powers of
$1/\mathring{c}_{\bm{p}}$ in the present $m\ne 0$ case is completely
analogous to the expansion of its $m=0$ counterpart in powers of
$1/\mathring{c}$, with \emph{identical} expansion coefficients [cf.\ 
Eq.~(11) of Ref.~\onlinecite{DDE83}].

If we restrict ourselves to first order in $\mathring{c}^{-1}$, we can
make the replacement $\mathring{c}^{-1}_{\bm{p}}\to\mathring{c}^{-1}$.
In analogy with the $m=0$ case the expansion of the $N$-point cumulant
to first order in $\mathring{c}^{-1}$ therefore becomes
\begin{equation}
  \label{eq:GGinfins}
  G^{(N,0)}=G^{(N,0)}_\infty+\mathring{c}^{-1}\,G^{(N,0;1)}_\infty
  +O(\mathring{c}^{-2})\;,
\end{equation}
where $G^{(N,0;1)}_\infty$ means a cumulant involving $N$ fields
$\phi_a$ and a single insertion of the boundary operator
${\int_{\mathfrak{B}}}dA\,(\partial_n\bm{\phi})^2/2$, evaluated at
$\mathring{c}=\infty$ and $\mathring{\lambda}=0$. Proceeding as in
Appendix C of Ref.~\onlinecite{DDE83}, one can easily derive the
relation
\begin{equation}
  \label{eq:WTid2}
  \sum_{i=1}^N\partial_{z_i}G^{(N,0)}_\infty(\bm{x}_1,\ldots,\bm{x}_N)
  =G^{(N,0;1)}_\infty(\bm{x}_1,\ldots,\bm{x}_N)\;,
\end{equation}
where $z_i$ is the $z$-component of $\bm{x}_i$. Since the
renormalization factor $Z_\phi^{-N/2}$ of the cumulant on the
left-hand side must also renormalize the one on the right-hand side,
the inserted boundary operator needs no renormalization. Thus the
associated interaction constant $\sim \mathring{c}^{-1}$ must scale
naively as $1/\mu$ at the ordinary fixed point. Noting that the RG
eigenexponent $y_1$ of a boundary scaling field $g_1$ and the scaling
dimension $\Delta[{\mathcal O}_1]$ of a local boundary operator
${\mathcal O}_1$ (in the case of the parallel surface orientation
considered in this paper) are related via
\begin{equation}
  \label{eq:y1Del1}
  \Delta[{\mathcal O}_1]=d-1+m\,(\theta-1)-y_1\;,
\end{equation}
one immediately concludes that the scaling dimension of
$\Delta[(\partial_n\bm{\phi})^2]$ of the inserted operator
$(\partial_n\bm{\phi})^2/2$, which represents the surface energy density
at the ordinary transition, is indeed given by (\ref{eq:Delsed}).

The above findings indicate that the boundary operator with smallest
scaling dimension contributing to the BOE of the local energy density
operator $\phi^2(\bm{x})$ (besides the one-operator $\openone$) is
$T_{zz}^{\mathfrak{B}}$, the $zz$-component of the stress-energy
tensor $T_{\mu\nu}$, taken at the boundary (cf., e.g.,
Ref.~\onlinecite{ES94}).

\section{Critical exponents of the ordinary transition}
\label{sec:critexp}

\subsection{Definition of surface correlation exponents}
\label{sec:scorrexp}

In order to characterize the power-law decay of the pair correlation
function $G^{(2,0)}(\bm{x}_1,\bm{x}_2)$ of semi-infinite systems at a
critical point in the limits where the separation $\bm{x}_1-\bm{x}_2$
increases parallel or perpendicular to the surface, one conventionally
introduces analogs of the usual correlation exponent $\eta$ termed
$\eta_\|$ and $\eta_\perp$, respectively. In the case of anisotropic
scale invariant systems with boundaries we must be more careful
because distances along different axes scale differently and the
orientation of the surface matters.

Clearly, distinct sets of surface correlation exponents
$\eta_{l2,\|}$, $\eta_{l2,\perp}$, and $\eta_{l4,\|}$,
$\eta_{l4,\perp}$, could be introduced in analogy with the bulk
correlation exponents $\eta_{l2}$ and $\eta_{l4}$. However, this is
unnecessary since the ``$l4$ exponents'' are simply related to their
``$l2$ counterparts'' via the anisotropy exponent $\theta$, just as
the introduction of $\eta_{l4}$ could be avoided by writing the
exponent $4-\eta_{l4}$ as $\theta\,(2-\eta_{l2})$. In order to define
the exponents $\eta_\|\equiv\eta_{l2,\|}$ and
$\eta_\perp\equiv\eta_{l2,\perp}$, consider the pair correlation
function $G^{(2,0)}(\bm{r};z,z')$ at the Lifshitz point between two
points $\bm{x}=(\bm{r},z)$ and $\bm{x}'=(\bm{0},z')$, and let us
introduce 
\begin{equation}
  \label{eq:Rdef}
  \check{r}\equiv\sqrt{{r_\alpha}{r_\alpha}}\quad\;\;\text{and}\;\;\quad
   R\equiv\sqrt{{r_\beta}{r_\beta}}\;,
\end{equation}
the lengths of the components $(r_\alpha)$ and $(r_\beta)$ of their
parallel separation $\bm{r}$.  In the limits $z\to\infty$,
$R\to\infty$ or $\check{r}\to\infty$, with $z'$ and the respective
other distances kept fixed, this function decays as
\begin{equation}
  \label{eq:paircorrdec}
  G^{(2,0)}(\bm{r};z,z')\sim\left\{
  \begin{array}[c]{l@{\;,\;\;}l}
z^{-[d-2+m\,(\theta-1)+\eta_\perp]}&z\to\infty\;,\\
R^{-[d-2+m\,(\theta-1)+\eta_\|]}&R\to\infty\;,\\
{\check{r}}^{-[d-2+m\,(\theta-1)+\eta_\|]/\theta}&\check{r}\to\infty\;.
  \end{array}\right.
\end{equation}
These relations translate into the small-momentum behavior
\begin{equation}
  \label{eq:chi11psing}
  \chi_{11}(\bm{p})^{\text{sing}}\sim\left\{
\begin{array}[c]{l@{\;\;\text{for}\;\;}l}
p^{\eta_\|-1}&(p_\alpha)=(0)\;,\\
p^{(\eta_\|-1)/\theta}&(p_\beta)=(0)\;,
\end{array}\right.
\end{equation}
and\cite{com:etaperp}
\begin{equation}
  \label{eq:chi1psing}
  \chi_1(\bm{p})\sim  \left\{
\begin{array}[c]{l@{\;\;\text{for}\;\;}l}
p^{\eta_\perp-2}&(p_\alpha)=(0)\;,\\
p^{(\eta_\perp-2)/\theta}&(p_\beta)=(0)\;,
\end{array}\right.
\end{equation}
of the local surface susceptibility (\ref{eq:Ggen11}) and the layer
surface susceptibility
\begin{equation}
  \label{eq:chi1def}
  \chi_1(\bm{p})={\int_0^\infty}\hat{G}^{(1,1)}(\bm{p};z)\,dz\;,
\end{equation}
respectively.

The large-$\mathring{c}$ behavior (\ref{eq:GNMcexp}) in conjunction
with the scaling forms (\ref{eq:universalG}) implies that
\begin{equation}
  \label{eq:etaord}
  \eta_\|^{\text{ord}}=2+\eta_{l2}+\eta_{1,\infty}^*\;.
\end{equation}
The corresponding result for $\eta_\perp^{\text{ord}}$ is equivalent
to the scaling law
\begin{equation}
  \label{eq:etaperpsl}
  \eta_\perp=\big(\eta_{l2}+\eta_\|\big)/2\;,
\end{equation}
which is well known from the CP case.\cite{Bin83,DD80,DD81a,Die86a}

Let us also note that the decay law of $G^{(2,0)}$ (involving
$\eta^{\text{ord}}_\perp$) displayed in the first line on the
right-hand side of Eq.~(\ref{eq:paircorrdec}) carries over to the case
where $\bm{x}$ moves away from the surface along an arbitrary
direction not parallel to the surface. This can be shown as in the CP
case,\cite{DD81a,Die86a} utilizing the BOE (\ref{eq:boe}) and
Eq.~(\ref{eq:Cinfas}).

\subsection{$\epsilon$ expansion of surface critical exponents and
  estimates for three dimensions}
\label{sec:3dest}

We now turn to the $\epsilon$ expansions of the critical exponents at
the ordinary transition.  Substituting the $\epsilon$~expansion
(\ref{eq:ustar}) of the nontrivial root $u^*$ of $\beta_u$ into
Eq.~(\ref{eq:etainf1}) yields
\begin{eqnarray}
\label{eq:eta1infeps}
\eta_{1,\infty}^*&=&\eta_\|^{\text{ord}}-\eta_{l2}-2\nonumber\\
&=&-\frac{n+2}{n+8}\,\epsilon-\frac{(n+2)^2}{(n+8)^3}
\bigg[ 
\frac{j_\sigma(m)}{16\,(m+2)}-\frac{j_\phi(m)}{2}\nonumber\\
&&\mbox{}+2\,\frac{20+7\,n}{n+2}\,J_u(m)
+6\,\frac{n+8}{n+2}\,j_1(m)\bigg]\epsilon^2
\nonumber\\&&\qquad\mbox{}+O(\epsilon^3)\;.
\end{eqnarray}

For $m\to 0$, $m=2$, and $m=6$, the exactly known values\cite{SD01}
\begin{equation}
  \label{eq:jsexm0}
  j_\phi(0)=2\;,\quad j_\sigma(0)=0\;,\quad J_u(0)=1/2\;,
\end{equation}
\begin{equation}
  \label{eq:jsexm2}
 j_\phi(2)=\frac{4}{3}\;,\quad j_\sigma(2)=\frac{128}{27}\;, \quad
 J_u(2)=\ln\frac{4}{3}\;,
\end{equation}
and
\begin{eqnarray}
  \label{eq:jsexm6}
  j_\phi(6)&=&\frac{8}{3}\,{\Big[1-3\ln\frac{4}{3}\Big]}\;, \quad
  j_\sigma(6)=\frac{448}{9}\;,\nonumber\\
 J_u(6)&=&\frac{5}{6}-3\ln\frac{4}{3}\;,
\end{eqnarray}
can be inserted into Eq.~(\ref{eq:eta1infeps}), along with those of
$j_1$ given in Eqs.~(\ref{eq:j1m0}) and (\ref{eq:j1val}), to convert
it into an analytical expression. Doing so one finds that the result
(\ref{eq:eta1infeps}) reduces in the limit $m\to 0$ to its established
analog\cite{DD80,RG80,DD81a} for the CP ordinary transition,
Eq.~(IV.35) of Ref.~\onlinecite{DD81a}, as it should.

We can now replace $\eta_{l2}$ in Eq.~(\ref{eq:eta1infeps}) by its
known $O(\epsilon^2)$ expression [given in Eq.~(85) of
Ref.~\onlinecite{DS00a} and Eq.~(61) of Ref.~\onlinecite{SD01}] to
obtain the $\epsilon$ expansion of $\eta_\|^{\text{ord}}$ to order
$\epsilon^2$. The analogous ``direct'' series expansions to order
$\epsilon^2$ of the surface exponents $\beta_1^{\text{ord}}$,
$\eta_\perp^{\text{ord}}$, and the surface susceptibility exponents
$\gamma_1^{\text{ord}}$ and $\gamma_{11}^{\text{ord}}$ follow in a
similar fashion by combining the $\epsilon$-expansion results for the
bulk exponents given in Eqs.~(61)--(66) of Ref.~\onlinecite{SD01} with
ours for $\eta_{1,\infty}^*$ and the scaling relations
(\ref{eq:etaperpsl}),
\begin{equation}
  \label{eq:beta1scal}
  \beta_1=\nu_{l2}\big[d-2+\theta\,(m-1)+\eta_\|\big]/2\;,
\end{equation}
\begin{equation}
  \label{eq:gamma11}
  \gamma_{11}=\nu_{l2}\,(1-\eta_\|)\;,
\end{equation}
and
\begin{equation}
  \label{eq:gamma1}
  \gamma_{1}=\nu_{l2}\,(2-\eta_\perp)\;,
\end{equation}
all of which are straightforward consequences of the scaling forms
(\ref{eq:universalG}), the large-$\mathring{c}$ form of the cumulants
$G^{(N,M)}$, and the short-distance behavior (\ref{eq:Cinfas}). They
can be derived in much the same way as in the $m=0$
case,\cite{DD81a,Die86a} for the parallel orientation of the surface
considered in this paper.

In Table~\ref{tab:exponentsm1n1} we have gathered the numerical
estimates of various critical exponents of the LP ordinary transition
one obtains for the uniaxial one-component case in three dimensions by
evaluating their $\epsilon$ expansions to zeroth, first, and second
order at $\epsilon=3/2$. The critical indices $\beta_{\text{s}}$ and
$\gamma_{\text{s}}$, defined through the behavior
$m_{\text{s}}\sim |\tau|^{\beta_{\text{s}}}$ and
$\chi_{\text{s}}\sim |\tau|^{-\gamma_{\text{s}}}$ for $\tau\to 0$,
$\rho=0$, of the excess magnetization
\begin{equation}
  \label{eq:msdef}
  m_{\text{s}}={\int_0^\infty} dz\,[G^{(1,0)}(z)-G^{(1,0)}(\infty)]
\end{equation}
and the excess susceptibility
\begin{equation}
  \label{eq:chisdef}
  \chi_{\text{s}}={\int_0^\infty}\!\!dz{\int_0^\infty}\!\!dz'
 {\big[\hat{G}^{(2,0)}(\bm{p};z,z')
  -\hat{G}^{(2,0)}(\bm{p};\infty,\infty)\big]}_{\bm{p}=\bm{0}}
\end{equation}
do not involve an independent surface exponent because they can be
expressed in terms of bulk exponents as
\begin{equation}
  \label{eq:betas}
  \beta_{\text{s}}=\beta_l-\nu_{l2}
\end{equation}
and
\begin{equation}
  \label{eq:gammas}
  \gamma_{\text{s}}=\gamma_l+\nu_{l2}\;.
\end{equation}
We have included them in Table~\ref{tab:exponentsm1n1} to compare also
them with the Monte Carlo results of Ref.~\onlinecite{Ple02}.

\begin{table}[htb]
 \caption{\label{tab:exponentsm1n1} Estimated values of critical
   exponents at the LP  ordinary transition in $d=3$ dimensions for
   $m=1$ and $n=1$. The values in the columns labeled ``MFT'',
   ``$O(\epsilon)$'', ``$O(\epsilon^2)$'', and ``MC'' correspond to
   mean field theory, to the estimates obtained by setting
   $\epsilon=3/2$ in the $\epsilon$ expansions to first and second
   order, and to Monte Carlo results, respectively.}
\begin{ruledtabular}
  \begin{tabular}{cdddd}
    & \text{MFT} 
        & O(\epsilon)
        & O(\epsilon^2)
        & \text{MC}\footnote{Taken from Ref.~\onlinecite{Ple02}.}\\\hline 
   $\eta^{\text{ord}}_\parallel$ & 2 & 1.5 & 1.133 
             & -\\
   $\eta^{\text{ord}}_{\perp}$ & 1 & 0.75 & 0.586 
             &-\\
   $\beta^{\text{ord}}_1$ & 1 & 0.75 & 0.697 
             & 0.687(5)\\
   $\gamma^{\text{ord}}_1$ & 1/2 & 0.75 & 0.947 
             & 0.82(4)\\
   $\gamma^{\text{ord}}_{1,1}$ & -1/2 & -0.375 & -0.212 
             & -0.29(6)\\
   $\beta_{\text{s}}$ & 0 & -0.375 & -0.462 & -0.46(3)\\
   $\gamma_{\text{s}}$ & 3/2 & 1.875 & 2.106 & 1.98(8)\\
  \end{tabular}
\end{ruledtabular}
\end{table}

The direct series expansions to $O(\epsilon^2)$ of the exponents
listed in Table~\ref{tab:exponentsm1n1} are reasonably well behaved.
We therefore expect the corresponding estimates to be fairly reliable.
Yet for some of the series, Pad\'e [1,1] approximants give
significantly different values. In other, more favorable cases the
differences are small. One example of this kind is
$\beta_1^{\text{ord}}$. The Pad\'e [1,1] approximant and the solution
to the scaling relation (\ref{eq:betadiff}) yield
\begin{equation}
  \label{eq:beta1Pade}
  \beta_1^{\text{ord}}(3,1,1)\simeq \left\{
    \begin{array}[c]{ll}
    0.683\;,&\text{ Pad\'e }[1,1]\;,\\
    0.628\;,&\text{ with bulk exponents}\;,
    \end{array}\right.
\end{equation}
respectively. To obtain the second number, the ``best estimates''
$\beta_l\simeq 0.220$ and $\nu_{l2}\simeq 0.746$ of
Ref.~\onlinecite{SD01} were inserted into Eq.~(\ref{eq:betadiff})
together with the direct series estimate $\eta_{1,\infty}^*\simeq
-0.9060$. Both so-obtained numbers are approximately within $10\%$ of
the estimate given in Table~\ref{tab:exponentsm1n1}.
 
The exponent $\gamma_{1}^{\text{ord}}$ provides an example of a case
in which the $d=3$ value predicted by the Pad\'e approximant, namely
$\gamma_{1}^{\text{ord}}\simeq 1.670$, deviates considerably from the
direct series estimate. However, the relation
$\gamma_1^{\text{ord}}=\nu_{l2}(1-\eta_{l2}-\eta_{1,\infty}^*/2)$,
applied with the just mentioned values of $\nu_{l2}$ and
$\eta_{1,\infty}^*$ and the ``best estimate'' $\eta_{l2}\simeq 0.124$
of Ref.~\onlinecite{SD01}, yields $\gamma_{1}^{\text{ord}}\simeq
0.99$, which is reasonably close to the direct series estimate
(Table~\ref{tab:exponentsm1n1}). We therefore discard the
Pad\'e [1,1] estimate in this case.

As can be seen from Table~\ref{tab:exponentsm1n1}, our estimates are
in good agreement with Pleimling's\cite{Ple02} Monte Carlo estimates.
Normally the Monte Carlo results for $\beta_1$ are more accurate than
for susceptibilities such as $\gamma_1$ and $\gamma_{11}$. It is
therefore gratifying that the agreement is particularly good in the
case of $\beta_1^{\text{ord}}$.

Of interest is also the case of the ordinary transition at the
uniaxial Lifshitz point of the axial next-nearest neighbor XY model
(ANNNXY model), with $n=2$ and $m=1$. This model was recently argued
to describe the bulk critical behavior of Tb.\cite{Bar95} To our
knowledge, neither experimental nor theoretical results for the
surface critical exponents at the corresponding ordinary transition
have been published. On the other hand, Monte Carlo results for the
bulk critical exponents of the three-dimensional ANNNXY model have
been obtained a long time ago.\cite{Sel78b,Sel80} As discussed in
Ref.~\onlinecite{SD01} (see its tables 4 and 5), Selke's
values\cite{Sel80} $\beta_l=0.2\pm 0.02$ and $\gamma_l=1.5\pm 0.1$ are
in reasonable agreement with the $O(\epsilon^2)$ estimates
$\beta_l\simeq 0.276$, and $\gamma_l\simeq 1.495$ obtained there.
Inserting the latter together with the $O(\epsilon^2)$ result
$\nu_{l2}\simeq 0.757$ into Eqs.~(\ref{eq:betas}) and
(\ref{eq:gammas}) yields $\beta_s\simeq -0.48$ and $\gamma_s\simeq
2.25$.

To obtain estimates of $\beta_1^{\text{ord}}$ for $d=3$, $m=1$, and
$n=2$, we exploited our above results for its series expansion to
order $\epsilon^2$. Direct extrapolation of this (truncated) series
and Pad{\'e} [1,1] and [0,2] approximants yielded the values
$\beta_1^{\text{ord}}\simeq 0.72$, $0.71$, and $0.75$, respectively.
Aside from experimental tests of these predictions, checks via
high-precision Monte Carlo simulations would be very welcome.

\section{Summary and conclusions}
\label{sec:concl}

We have investigated the critical behavior of semi-infinite
$d$-dimensional systems at an $m$-axial bulk Lifshitz point. To this
end we have considered systems whose bulk universality class is
described by an $n$-component $\phi^4$ Hamiltonian with the bulk
density (\ref{eq:Lb}). Assuming that the surface plane of the system
is oriented parallel to all potential modulation axes in the modulated
ordered phase, we have constructed appropriate ``minimal''
semi-infinite extensions of these continuum models with the following
properties:

(a) They are compatible with the bulk universality class of Lifshitz
critical behavior described by an infinite-space model whose
Hamiltonian has the density (\ref{eq:Lb}).

(b) They are minimal in the sense that redundant and irrelevant
surface terms compatible with the presumed $O(n)$ symmetry and
short-range nature of the interactions and their surface-induced
perturbations have been dropped.

(c) Taken at their respective RG fixed points, these models represent
the (surface) universality classes of the associated ordinary,
special, and extraordinary transitions.
 
The resulting contributions to the Hamiltonian (\ref{eq:Hamf})
localized on the surface correspond to the surface density
(\ref{eq:L1}). They differ from the usual one known from the much
studied ($m=0$) case of a critical point,\cite{Die86a} namely
$(\mathring{c}/2)\,\bm{\phi}^2$, by a similar derivative term. We have
shown that the latter is required for renormalizability of the model.
Physically, this means that such a term, if initially absent, gets
generated as short wave-length degrees of freedom are integrated out.
The fluctuating boundary conditions implied by the Hamiltonian
(\ref{eq:Hamf})--(\ref{eq:L1}) are given in Eq.~(\ref{eq:bc}); in the
$\bm{p}z$ representation they correspond to Robin boundary conditions
with a momentum-dependent $\mathring{c}_{\bm{p}}$ [see
Eqs.~(\ref{eq:bc}), (\ref{eq:cbp}) and (\ref{eq:bcphip})].

The boundary term $\propto\mathring{\lambda}$ gives rise to a
dimensionless interaction constant $\lambda$, on which the surface
counterterms and renormalization functions depend for general values
of $\mathring{c}$.  We have clarified the fixed-point structure of the
model, identifying the fixed points ${\mathcal{P}}_{\text{ord}}^*$,
${\mathcal{P}}_{\text{sp}}^*$, and ${\mathcal{P}}_{\text{ex}}^*$
describing the ordinary, special, and extraordinary transitions (cf.\ 
Fig~\ref{fig:flow}), which are located at a nontrivial value
(\ref{eq:lambdaplusstar}) of $\lambda$.

In order to investigate the LP special transition, dealing with the
$\lambda$ dependence cannot be avoided. However, as we have shown,
this is possible in the case of the ordinary transition by taking the
limit $\mathring{c}\to\infty$ from the outset. Just as in the CP
($m=0$) case, both the regularized bare and the renormalized theories
obey Dirichlet boundary conditions if $\mathring{c}=\infty$, and the
single required surface critical exponent $\beta_1^{\text{ord}}$
follows from the scaling dimension of $\partial_n\bm{\phi}$. Likewise,
the proofs\cite{DDE83,BC87,BD94} for the CP case showing that the
surface energy density at the ordinary transition has a leading
singularity of the bulk free-energy form $\sim |\tau|^{2-\alpha}$
carry over to the present $m\ne 0$ one.

Performing a two-loop RG analysis of the LP ordinary transition for
general values of $m$, we have been able to determine the $\epsilon$
expansions of the surface critical indices of this transition to
second order in $\epsilon$. Extrapolations of these series expansions
to $d=3$ dimensions yield values of the surface critical exponents for
the uniaxial one-component case ${m=n=1}$ in good agreement with
recent Monte Carlo results\cite{Ple02} for the ANNNI model.

There are several obvious directions for extensions of the present
work.  First of all, the special transition should be analyzed,
preferably by means of a two-loop RG analysis giving the
$\epsilon$~expansions of its surface critical exponents to second
order in $\epsilon$. From the results presented above, the $\epsilon$
expansions of the two independent surface critical exponents of this
transition, the surface correlation exponent $\eta_\|^{\text{sp}}$ and
the surface crossover exponent $\Phi$, can be derived to first order
in $\epsilon$ with moderate effort.\cite{DGRunpub02} However, a
detailed investigation of the special transition requires more work,
is beyond the scope of the present paper and left to a planned
subsequent one.  Suffice it here to say that for our $O(n)$ symmetric
Hamiltonian with $n>1$, the existence of a surface-ordered phase at
temperatures $T>0$ is ruled out for bulk dimensions $d\le
d_*^{O(n)}+1=3+m/2$. Extrapolations of the $\epsilon$ expansions of
the special transition's critical exponents to $d=3$ thus make sense
only in the scalar case, $n=1$. For the interesting case $m=n=1$,
Monte Carlo results for the exponent $\beta_1^{\text{sp}}$ of the
ANNNI model are available\cite{Ple02} which may be used to check the
resulting $d=3$ extrapolation values.

A second important line of extension is to consider a surface
orientation perpendicular to one of the potential modulation axes (the
``perpendicular case'' of the Introduction). This case is somewhat
more difficult to handle than the parallel surface orientation treated
in this paper. Since the $z$-direction then scales naively as
$\mu^{-1/2}$ rather than as $1/\mu$, more potentially dangerous
boundary terms (whose coupling constants have nonnegative RG
eigenexponents at the Gaussian fixed point) exist. Since the
Landau-theory equations for the order parameter profile and the free
propagator involve the fourth derivative $\partial^4/\partial z^4$,
two boundary conditions rather than the single one, Eq.~(\ref{eq:bc})
or Eq.~(\ref{eq:bcG}), are required and found.\cite{DGunpub} A proper
study of this case beyond Landau theory\cite{BF99} also requires a
careful identification of redundant surface operators.

Another obvious limitation of the present work is its lack of detailed
investigations of the homogeneous and modulated ordered phases, the
extraordinary transitions, as well as of profiles of the order
parameter and other densities at the {LP}. The progress made here
indicates that such studies, though technically demanding, should be
possible, involving no fundamental new difficulties.

Unfortunately, there are, to our knowledge, yet no experimental
results on surface critical behavior at bulk Lifshitz points available
with which we could compare. Owing to the richness of the physics
involved, detailed experimental studies would be very interesting. We
hope that the progress made recently on the theoretical side will
foster such experimental activities.

\begin{acknowledgments}
  It is our pleasure to thank M.\ Shpot for helpful discussions. We
  gratefully acknowledge the support of this work provided by the
  Deutsche Forschungsgemeinschaft (DFG) --- in its initial phase via
  Sonderforschungsbereich 237, in its final phase via DFG grant \#
  Di-378/3.
\end{acknowledgments}

\appendix
\begin{widetext}
\section{Laurent expansion of the Feynman integrals of
  $\hat{G}_\infty^{(1,1)}$}
\label{app:Ginfgraphs}

In this appendix, we briefly describe the calculation of the Laurent
expansion of the one and two-loop graphs of the function
$G_\infty^{(1,1)}(\bm{x},\bm{x}')$ to the required orders in
$\epsilon^{-1}$.

The Dirichlet propagator (\ref{eq:GDir}) to which the free propagator
(\ref{eq:Ghatgen}) reduces for $\mathring{c} = \infty$ can be written as
\begin{equation}
  \label{eq:frD}
\hat{G}_{\text{D}}(\bm{p};z_1,z_2)=\frac{1}{2\mathring{\kappa}_{\bm{p}}}
  {\Big[}e^{-\mathring{\kappa}_{\bm{p}}|z_1-z_2|} 
-e^{-\mathring{\kappa}_{\bm{p}}(z_1+z_2)}{\Big]}=
\hat{G}_{\text{b}}(\bm{p};z_1-z_2)-\hat{G}_{\text{s}}(\bm{p};z_1,z_2) 
\end{equation}
in the $\bm{p}z$ representation. The part depending on $|z_1-z_2|$ is
the bulk propagator $\hat{G}_{\text{b}}$. The subtracted piece
$\hat{G}_{\text{s}}(\bm{p};z_1,z_2)=\hat{G}_{\text{b}}(\bm{p};z_1+z_2)$
is the image term which ensures that the Dirichlet condition holds.

As mentioned in the caption of Fig.~\ref{fig:Ginf11}, we use open and
crossed circles to indicate external points off and on the surface
${\mathfrak{B}}$, respectively, a broken line to represent
$\hat{G}_{\text{D}}$, and perpendicular strokes to mark
$z$-derivatives $\partial_z$. Accordingly, Eq.~(\ref{eq:frD}) can be
depicted as
\begin{equation}
  \label{eq:GDgraph}
  \raisebox{-0.8em}{\begin{texdraw} \drawdim pt \setunitscale 2.5 \linewd
    0.4 \move(-6 0) \move(-3 0) \lcir r:1   \move(9 0) \lcir r:1
\lpatt(0.75 0.75) \move(-2 0) \rlvec(10 0)
\htext(-4 -4.5){$z_1$} \htext(7.75 -4.5){$z_2$}\move(12 0)
\end{texdraw}}
=\raisebox{-0.8em}{\begin{texdraw} \drawdim pt \setunitscale 2.5 \linewd
    0.4 \move(-6 0)\move(-3 0) \lcir r:1   \move(9 0) \lcir r:1
 \move(-2 0) \rlvec(10 0)
\htext(-4 -4.5){$z_1$} \htext(7.75 -4.5){$z_2$}
\htext(1.8 0.75){\small b} \move(12 0)
\end{texdraw}} -
\raisebox{-0.8em}{\begin{texdraw} \drawdim pt \setunitscale 2.5 \linewd
    0.4 \move(-6 0) \move(-3 0) \lcir r:1   \move(9 0) \lcir r:1
 \move(-2 0) \rlvec(10 0)
\htext(-4 -4.5){$z_1$} \htext(7.75 -4.5){$z_2$}
\htext(1.8 0.75){\small s} \move(12 0)
\end{texdraw}}.
\end{equation}
The Feynman graph expansion of the two-point function
${\hat{G}}_\infty^{(1,1)} (\bm{p};z_1)$ to two-loop order is shown in
Fig.~\ref{fig:Ginf11}. The zero-loop term
\begin{equation}
  \label{eq:Ginf110}
  \raisebox{-0.75em}{$z_1$}\hspace{-0.75em}\raisebox{-0.3em}{\includegraphics[width=4em]{./ginf11.0.eps}} =
  \big({\hat{G}_{\text{D}}}\loarrow{\partial}_n\big)(\bm{p};z_1) 
  \equiv  \partial_{z_2}\hat{G}_{\text{D}}(\bm{p};z_1,z_2=0)=
  e^{-\mathring{\kappa}_{\bm{p}}z_1} 
\end{equation}
follows directly from Eq.~(\ref{eq:frD}).

In order to determine the one and two-loop terms, we amputate the
external free legs of the graphs and compute the Laurent expansion of
the resulting amputated graphs, considered as generalized functions
(distributions) of the dimensionless variable $\mu z$ and $\mu z'$ in
the $\bm{p}z$ representation.\cite{Die86a} The amputated legs play the
role of test functions\cite{com:testfct} on which the latter
distributions act.  Once the Laurent expansions of the amputated
graphs are known, those of the full graphs can be computed in a
straightforward fashion by performing the remaining integrations over
$z$ and $z'$. For convenience, we will set $\mu=\sigma=1$ in most of
the sequel. Whenever necessary, the dependences on $\mu$ and $\sigma$
can easily be reintroduced by dimensional arguments.

The one-loop term of ${\hat{G}}_\infty^{(1,1)} (\bm{p};z_1)$ reads
\begin{equation}
\label{eq:exp2loop}
\raisebox{-0.75em}{$z_1$}\hspace{-0.85em}\raisebox{-0.3em}{\includegraphics[width=5em]{./ginf11.1.eps}} =
-\frac{n+2}{3}\,\frac{\mathring{u}}{2}\, {\int_0^\infty}dz \, 
\hat{G}_{\text{D}} (\bm{p};z_1,z)\,  
\big({\hat{G}_{\text{D}}}\loarrow{\partial}_n\big)(\bm{p};z)\,
G_{\text{D}}(\bm{x},\bm{x})\;, 
\end{equation}
where the internal loop, $G_{\text{D}}(\bm{x},\bm{x})$, according to
Eq.~(\ref{eq:Gbxxs}) is given by
\begin{equation}
\label{eq:GD}
G_{\text{D}}(\bm{x},\bm{x})=-G_b(2z\,\bm{n})=
-F_{m,\epsilon}\,\frac{\Gamma(2 - \epsilon)\,  
      \sin ({\epsilon\,\pi }/{2})}
    {\epsilon\,\pi\,\mathring{\sigma}^{m/4}}\,z^{\epsilon-2}\;.
\end{equation}

Considering $z^{\epsilon-2}$ as a generalized function, we can
calculate its action on a test function $\varphi(z)$ and expand the
result about $\epsilon=0$ to obtain\cite{Die86a,GS64}
\begin{equation}
\label{eq:zmtwoplusE}
{\int_0^\infty}dz \,z^{\epsilon-2}\,\varphi(z)  =
\frac{1}{\epsilon}\frac{\partial\varphi(z)}{\partial z} {\bigg|}_{z=0}
+\big(z_+^{-2}, \varphi(z)\big)+O(\epsilon)\;,
\end{equation}
where $z_+^{-2}$ is a standard generalized function, defined as
\begin{equation}
\label{eq:zpm2}
\big(z_+^{-2}, \varphi(z)\big)\equiv {\int_0^\infty}
\frac{dz}{z^2}\,\big[\varphi(z)-\varphi(0)- 
\theta(1-z)\, z \, \varphi'(0) \big]\;.
\end{equation}
Equations~(\ref{eq:exp2loop})--(\ref{eq:GD}) yield
\begin{equation}
  \label{eq:Ginf111}
   \raisebox{-0.75em}{$z_1$}\hspace{-0.85em}\raisebox{-0.3em}{\includegraphics[width=5em]{./ginf11.1.eps}}
   =\mathring{U}\,\frac{n+2}{3}\, \frac{1}{4\epsilon}{\left[ 
1+R_{\text{D}}(\bm{p},z_1)\,\epsilon+O(\epsilon^2)\right]}\,
\raisebox{-0.75em}{$z_1$}\hspace{-0.8em}\raisebox{-0.3em}{\includegraphics[width=4em]{./ginf11.0.eps}} 
\end{equation}
with
\begin{equation}
  \label{eq:U0}
  \mathring{U}=\mu^{-\epsilon}\,\mathring{u} \,\mathring{\sigma}^{-m/4}\,F_{m,\epsilon}
\end{equation}
and 
\begin{eqnarray}
  \label{eq:Rpdef}
  R_{\text{D}}(\bm{p},z_1)&\equiv& C_E-1 +
  {\left(z_+^{-2},\hat{G}_{\text{D}}(\bm{p};z_1,z)\,
      \big({\hat{G}_{\text{D}}}
      \loarrow{\partial}_n\big)(\bm{p};z)\right)}/
  \raisebox{-0.75em}{$z_1$}\hspace{-0.8em} \raisebox{-0.3em}{\includegraphics[width=4em]{./ginf11.0.eps}}\\ 
  \label{eq:zpexpGD}
  &=&
  \text{Ei}(-2\,\mathring{\kappa}_{\bm{p}}z_1)\,e^{2\,\mathring{\kappa}_{\bm{p}}z_1}
  -\ln(2\,\mathring{\kappa}_{\bm{p}})\;,
\end{eqnarray}
where $\text{Ei}(x)$ is the exponential-integral function.

We next consider the two-loop diagram
\begin{eqnarray}
\label{eq:foota}
\raisebox{-0.75em}{$z_1$}\hspace{-0.8em}\raisebox{-0.3em}{\includegraphics[width=5em]{./ginf11.2b.eps}} 
&=&\bigg[\frac{\mathring{u}\,(n+2)}{6}\bigg]^2\, {\int_0^\infty}dz \, 
\hat{G}_{\text{D}}(\bm{p};z_1,z)\,
\big({\hat{G}_{\text{D}}} \loarrow{\partial}_n\big)(\bm{p};z) \nonumber\\
&&\times {\int_0^\infty}\!dz' \,G_{\text{D}}(\bm{x}',\bm{x}') {\int}
\frac{d^{d-1}\bm{p}'}{(2\pi)^{d-1}}\,
\hat{G}_{\text{D}}(\bm{p}';z,z')\,\hat{G}_{\text{D}}(-\bm{p}';z,z') 
\;. 
\end{eqnarray}
Performing the $\bm{p}'$ integration associated with the lower loop
yields
\begin{equation}
2\times
\raisebox{-1.4em}{%
\begin{texdraw}  \drawdim pt \setunitscale 2 \linewd
    0.4 \lpatt(1 1)
\htext(-1 -4){$z$} \htext(-1.5 12){$z'$} \move(0 0)  \fcir f:0.0 r:0.7
\move(0 10) 
 \fcir f:0.0 r:0.7 \move(0 5) \lellip rx:3 ry:5 \move(5 0)
\end{texdraw}}
=\frac{1}{2}\,F_{m,\epsilon} \frac{\Gamma(1 - \epsilon)\,
      \sin ({\epsilon\,\pi }/{2})}
    {\epsilon\,\pi\,\mathring{\sigma}^{m/4}}\, 
{\Big\{|z-z'|^{\epsilon-1}+(z+z')^{\epsilon-1}-2\,
[\max(z,z')]^{\epsilon-1}\Big\}}\;,  
\end{equation}
where the multiplication by $2$ on the left-hand side compensates the
graph's line-symmetry factor of $1/2$, so that the result is precisely
the integral over $\bm{p}'$ (last integral) in Eq.~(\ref{eq:foota}).

The subsequent $z'$ integration in Eq.~(\ref{eq:foota}) is
straightforward, giving
\begin{equation}
\label{eq:eightgraph}
 2\,{\int_0^\infty}dz' \,G_{\text{D}}(\bm{x}',\bm{x}')\;
\raisebox{-1.4em}{%
\begin{texdraw}  \drawdim pt \setunitscale 2 \linewd
    0.4 \lpatt(1 1)
\htext(-1 -4){$z$} \htext(-1.5 12){$z'$} \move(0 0)  \fcir f:0.0 r:0.7
\move(0 10) 
 \fcir f:0.0 r:0.7 \move(0 5) \lellip rx:3 ry:5 \move(5 0)
\end{texdraw}}
\,= - \bigg[ F_{m,\epsilon} 
\frac{\Gamma(2 - \epsilon)\, 
\sin({\epsilon\,\pi}/{2})}{\epsilon\,\pi\, \mathring{\sigma}^{m/4}}\bigg]^2
\frac{1}{1-\epsilon} 
\bigg\{\frac{1}{1-\epsilon}+\frac{\Gamma(\epsilon)\,
  \Gamma(\epsilon-1)}{\Gamma(2\, \epsilon -1)} {\bigg\}}\,
z^{2\epsilon-2}\;.
\end{equation}

Using Eqs.~(\ref{eq:zmtwoplusE}) and (\ref{eq:zpm2}), one can easily
compute the action of this distribution on the external lines. The
Laurent expansion of the result becomes
\begin{equation}
  \label{eq:Ginf112b}
   \raisebox{-0.75em}{$z_1$}\hspace{-0.8em}\raisebox{-0.3em}{\includegraphics[width=5em]{./ginf11.2b.eps}}
=\frac{\mathring{U}^2}{4}\,{\left(\frac{n+2}{3}\right)}^2\,
   {\left[\frac{-1}{4\epsilon^2}-
       \frac{1+4\,R_{\text{D}}(\bm{p},z_1)}{8\epsilon} 
 +O(\epsilon^0)\right]}\,
\raisebox{-0.75em}{$z_1$}\hspace{-0.8em}\raisebox{-0.3em}{\includegraphics[width=4em]{./ginf11.0.eps}} \;.
\end{equation}

The diagram 
\begin{eqnarray}
\raisebox{-0.75em}{$z_1$}\hspace{-0.8em}  \raisebox{-0.3em}{\includegraphics[width=6em]{./ginf11.2a.eps}}
&=& \bigg[\frac{\mathring{u} \,(n+2)}{6}\bigg]^2 {\int_0^\infty}dz \,
{\int_0^\infty}dz' \, \hat{G}_{\text{D}}(\bm{p};z_1,z)\,
\big({\hat{G}_{\text{D}}}\loarrow{\partial}_n\big)(\bm{p};z)\nonumber\\
&&\quad\times{\hat{G}}_{\text{D}}(\bm{p};z,z')
\,G_{\text{D}}(\bm{x},\bm{x})\, G_{\text{D}} (\bm{x}',\bm{x}')
\end{eqnarray}
is quite similar. Its resulting Laurent expansion reads
\begin{equation}
  \label{eq:Ginf112a}
   \raisebox{-0.75em}{$z_1$}\hspace{-0.8em}\raisebox{-0.3em}{\includegraphics[width=6em]{./ginf11.2a.eps}}
=\frac{\mathring{U}^2}{4}\,
    {\left(\frac{n+2}{3}\right)}^2\,
    {\left[\frac{1}{8\epsilon^2}+\frac{1+2\,
          R_{\text{D}}(\bm{p},z_1)}{8\epsilon} 
        +O(\epsilon^0)\right]}\,
\raisebox{-0.75em}{$z_1$}\hspace{-0.8em}\raisebox{-0.3em}{\includegraphics[width=4em]{./ginf11.0.eps}} \;.
\end{equation}
In deriving this equation we used the fact that the following integral
of the generalized function $z^{-2+\epsilon}
z'^{-2+\epsilon}\,e^{-\mathring{\kappa} |z-z'|}$ with a test function
$\varphi(z,z')$ has the Laurent expansion
\begin{eqnarray}
\label{eq:zzeLexp}
\lefteqn{
{\int_0^\infty}dz {\int_0^\infty}dz' \,\varphi(z,z')
\,z^{-2+\epsilon} z'^{-2+\epsilon}\,e^{-\mathring{\kappa} |z-z'|}
}&&\nonumber\\&=&
\frac{(\partial_z\partial_{z'}F)(0,0)}{2\epsilon^2} 
+\frac{1}{\epsilon}\bigg[\big({z'_+}^{-2}, (\partial_zF)(0,z')\big)
+\frac{(\partial_z^2F)(0,0)-(\partial_{z'}^2F)(0,0)}{4}\bigg]
+O(\epsilon^0)\;,\;\;\;\;
\end{eqnarray}
where $F$ is defined by
\begin{equation}
  \label{eq:F}
F(z,z')\equiv[\varphi(z,z')+\varphi(z',z)]\, e^{\mathring{\kappa}\,(z-z')}\;.
\end{equation}
In addition, Eqs.~(\ref{eq:GD}) and (\ref{eq:zmtwoplusE}) were
employed.

The computation of the diagram 
\raisebox{-0.75em}{\begin{texdraw} \drawdim pt 
\setunitscale 3 \linewd 0.3 \move(-6 0) \move(-3 0) \lcir r:1  
\htext(-4 -3.5){$$} \move(20 0) \lcir r:1
    \move(18 -1.5) 
    \rlvec(0 3)   
\move(19.5 -0.5)\rlvec(1 1) \move(19.5 0.5) \rlvec(1 -1)
\lpatt(0.75 0.75) \move(-2 0) \rlvec(5 0) \move(13.5 0) \rlvec(5 0)
\move(12 0) \move(8 0) \lellip rx:5
    ry:3  \move(3 0) \rlvec(10 0) \move(22 0)
\end{texdraw}}
is more cumbersome. In the position-space representation it can be
written as
\begin{equation}
  \label{eq:Ginf112c}
\raisebox{-0.85em}{\begin{texdraw} \drawdim pt 
\setunitscale 3 \linewd 0.3 \move(-6 0) \move(-3 0) \lcir r:1  
\htext(-4 -3.5){$z_1$} \move(20 0) \lcir r:1
    \move(18 -1.5) 
    \rlvec(0 3)   
\move(19.5 -0.5)\rlvec(1 1) \move(19.5 0.5) \rlvec(1 -1)
\lpatt(0.75 0.75) \move(-2 0) \rlvec(5 0) \move(13.5 0) \rlvec(5 0)
\move(12 0) \move(8 0) \lellip rx:5
    ry:3  \move(3 0) \rlvec(10 0) \move(22 0)
\end{texdraw}}
=\frac{\mathring{u}^2}{6}\,\frac{n+2}{3}\,(\psi, A\, \varphi)\;,
\end{equation}
where ${\mathcal A}$ is an integral operator defined through
\begin{equation}
(\psi, {\mathcal A}\, \varphi)\equiv\int_{{\mathbb R}^d_+}\!d^dx 
\, \int_{{\mathbb R}^d_+}\!d^dx'\, \psi(\bm{x})\,
G^3_{\text{D}}(\bm{x},\bm{x}') \,\varphi(\bm{x}')
\label{eq:Aform}
\end{equation}
with
\begin{equation}
\label{eq:psiphi}
\psi(\bm{x})\equiv G_{\text{D}} (\bm{x}_1,\bm{x}), \qquad
\varphi(\bm{x}')\equiv  ({G}_{\text{D}}
\loarrow{\partial}_n)(\bm{x}',\bm{r}_2)\;. 
\end{equation}

Since 
\begin{equation}
  \label{eq:A}
  (\psi,{\mathcal A}\,\varphi)=(\psi,{\mathcal A}\,\varphi)=
  \big\{\big(\varphi+\psi,{\mathcal A}\,[\varphi+\psi]\big) -(\varphi,
  {\mathcal A}\,\varphi)-(\psi,{\mathcal A}\,\psi)\}/2 \;,
\end{equation}
it is sufficient to calculate the diagonal element $(\varphi,{\mathcal
  A}\,\varphi)$ for a general test function $\varphi$. Upon
substituting Eq.~(\ref{eq:frD}) into Eq.~(\ref{eq:Aform}), one obtains
\begin{equation}
\label{eq:Abs}
(\varphi,{\mathcal A}\,\varphi)=I_0-3\,I_1+3\,I_2-I_3\;,
\end{equation}
where $I_j$ is defined through
\begin{equation}
\label{eq:Ij}
I_j\equiv (\varphi,{\mathcal A}_{j}\,\varphi)
=\int_{{\mathbb R}^d_+}\!d^dx \int_{{\mathbb R}^d_+}\!d^dx'\,
\varphi(\bm{x})\, {[G_{\text{b}}(\bm{x}-\bm{x}')]}^{3-j}\,
{[G_{\text{s}}(\bm{x},\bm{x}')]}^j\,\varphi(\bm{x}')\;.
\end{equation}
Let us calculate the terms on the right-hand side of
Eq.~(\ref{eq:Abs}) one by one, starting with the ``bulk'' graph
\begin{equation}
\label{eq:A0}
I_0={\int_0^\infty}\! dz\, {\int_0^\infty}\! dz' {\int}d^{d-1}r\,{\int}
d^{d-1}r'\, \varphi({\bm{r}},z)\,\varphi({\bm{r}}',z') 
\,G^3_{\text{b}}({\bm{r}}-{\bm{r}}',|z-z'|)\;.
\end{equation}
Exploiting the scaling property of the free bulk propagator,
\begin{equation}
\label{eq:Gscale}
G_{\text{b}}({\bm{r}};z) \equiv
G_{\text{b}}[(r_\alpha),(r_\beta);z]=|z|^{\epsilon-2} 
\, G_{\text{b}}{\big[\big(r_\alpha\,|z|^{-1/2}\big),(r_\beta/|z|;1\big]}\;,
\end{equation}
and making the changes of variables $r'_\alpha\to \rho_\alpha\equiv
|z-z'|^{-1/2}\,(r_\alpha'-r_\alpha)$ and $r'_\beta\to \rho_\beta\equiv
|z-z'|^{-1}\,(r_\beta'-r_\beta)$, we can recast Eq.~(\ref{eq:A0}) as
\begin{equation}
\label{eq:A0n}
I_0={\int}\!d^{d-1}r\,
{\int_0^\infty}\!dz\,{\int_0^\infty}\! dz' \,|z-z'|^{2 \epsilon-3}\,
\varphi(\bm{r},z) {\int}\!d^{d-1}\rho \;
\varphi{\big[\big(r_\alpha+\sqrt{|z-z'|}\,\rho_\alpha\big),
\big(r_\beta+|z-z'|\,\rho_\beta\big),z'\big]}
\,G^3_{\text{b}} ({\bm{\rho}};1)\;.
\end{equation}
This integral has a pole at $\epsilon=0$ caused by the ultraviolet
divergence on the line $|z'-z|=0$ which is due to the factor
$|z-z'|^{2 \epsilon-3}$. For our purposes here it is sufficient to
compute just the singular part of the integral (\ref{eq:A0n}). To this
end, we expand its integrand in powers of $\sqrt{|z-z'|}\,\rho_\alpha$
and $|z-z'|\,\rho_\beta$.  Subsequent termwise integration over $z'$
then yields
\begin{eqnarray}
\label{eq:Abbb}
I_0&=&\frac{1}{\epsilon}\,{\int}\!d^{d-1}r
{\int_0^\infty}\! dz \,\varphi(\bm{r},z)\bigg\{ \frac{f_0
  (1)}{2}\bigg[\frac{\partial^2 \varphi(\bm{r},z)}{\partial z^2}+
{\partial_\beta}{\partial_\beta}\varphi({\bm{r}},z)\bigg] + 
\frac{f_4(1)}{8m\,(m+2)}\,{(\partial_\alpha\partial_\alpha)^2} 
\varphi({\bm{r}},z)\bigg\}  \nonumber \\
&&\strut -\frac{f_2(1)}{4\,\epsilon\, m} {\int}\!d^{d-1}r\,
\varphi({\bm{r}},0)\, \partial_\alpha\partial_\alpha
\varphi({\bm{r}},0) +O(\epsilon^0)\;,
\end{eqnarray} 
where we have introduced
\begin{equation}
\label{eq:fk}
f_k (t) \equiv \left\{\int d^{d^*-1}\rho \,
\big(\rho_{\alpha'}\rho_{\alpha'}\big)^{k/2}
G^2_{\text{b}}({\bm{\rho}};1)\,
G_{\text{b}}{\big[\big(\rho_\alpha\,t^{1/2}\big),(\rho_\beta\,t);1\big]}
\right\}_{\epsilon=0} 
,\quad k=0,2,4\;,
\end{equation}
and utilized the identity
\begin{equation}
{\int_0^\infty}\! dz\,{\int_0^\infty}\! dz'\, |z-z'|^{2\epsilon -N}
\chi(z)\, \chi(z') 
=\frac{1}{\epsilon\, (N-1)!} \,{\int_0^\infty}\! dz \, \chi(z)
\frac{d^{N-1}}{dz^{N-1}}\chi(z) +O(\epsilon^0)\;, 
\end{equation}
valid for $N\in {\mathbb{N}}$ and test functions $\chi(z)$. 

Next, we turn to the terms $I_j$ with $j>0$. The associated Feynman
diagrams contain the image part
$G_{\text{s}}(\bm{x},\bm{x}')=G_{\text{b}}(\bm{r}-{\bm{r}}';|z+z'|)$,
which may induce additional ultraviolet singularities\cite{Die86a} at
$z= 0$ and $z'= 0$. To compute
\begin{equation}\label{eq:I1}
I_1=\int_{{\mathbb R}^d_+}\!d^dx \int_{{\mathbb R}^d_+}\!d^dx'\,
\varphi(\bm{x})\,G^2_{\text{b}}(\bm{x}-\bm{x}')\,
G_{\text{s}}(\bm{x},\bm{x}')\,\varphi(x')\;,
\end{equation}
we transform from the integration variables $\{z,z', {\bm{r}}'\}$ to
new ones, $\{v,t,\bm{\rho}\}$, defined by
\begin{equation}
  \label{eq:vtrho}
  z=v\frac{1+t}{2}, \quad z'=v\frac{1-t}{2}\;, \quad r'_\alpha=
  r_\alpha+\sqrt{v\,t}\,  \rho_\alpha , \quad
  r'_\beta= r_\beta+{v\, t}\,  {\rho}_\beta\;, 
\end{equation}
and then use again the  scaling property (\ref{eq:Gscale}) of the
propagator to rewrite Eq.~(\ref{eq:I1}) as
\begin{equation}
\label{eq:I1n}
I_1= {\int}d^{d-1}r {\int_0^\infty}\!dv \, v^{2\epsilon-2}
{\int_0^1}\!  dt\,
t^{\epsilon-1}\,{\mathcal G}(\bm{r}, v,t) 
\end{equation}
with
\begin{eqnarray}
\label{eq:Grvt}
\mathcal{G}(\bm{r},v,t)&=& \varphi[\bm{r},v\,(1+t)/2]\,
{\int}\!d^{d-1}\rho \;
 \varphi{\big[\big(r_\alpha+\sqrt{v\,t}\,\rho_\alpha\big),(r_\beta+
v \,t\,\rho_\beta),v\,(1-t)/2\big]}\, \times\nonumber\\
&&\strut \times G^2_{\text{b}}(\bm{\rho};1)\;
G_{\text{b}}{\big[\big(t^{1/2}\rho_\alpha\big),(t\,\rho_\beta);1\big]}\;.
\end{eqnarray}
Upon performing the integration over $t$ to obtain
\begin{equation}
{\int_0^1} dt\, t^{\epsilon-1}\,\mathcal{G}(\bm{r},v,t)
=\frac{\mathcal{G}(\bm{r}, v,0)}{\epsilon}+
{\int_0^1}\frac{dt}{t}\big[\mathcal{G}(\bm{r},v,t)
-\mathcal{G}(\bm{r}, v,0)\big]+O(\epsilon)\;, 
\end{equation}
we can do the integration over $v$ in Eq.~(\ref{eq:I1n}), using
Eqs.~(\ref{eq:zmtwoplusE}) and(\ref{eq:zpm2}). Straightforward
calculations then yield the result
\begin{eqnarray}
\label{eq:I1fin}
I_1&=&{\int}d^{d-1}r \,\Bigg\{\frac{f_0(0)}{2 \epsilon^2}
\varphi(\bm{r},0)\, \varphi'(\bm{r},0)+ \frac{\varphi(\bm{r},0)\,
 \varphi'(\bm{r},0)}{2\epsilon}\, 
{\int_0^1}\frac{dt}{t}\,[f_0(t)-f_0(0)] \nonumber\\ 
&&\strut + \frac{1}{4m\epsilon}\, \varphi(\bm{r},0)\,
\partial_\alpha\partial_\alpha \varphi(\bm{r}, 0)\, {\int_0^1}dt\,
f_2(t)+ \frac{f_0(0)}{\epsilon} \,
{\Big(v_+^{-2},\varphi^2(\bm{r},v/2) \Big)} 
 \Bigg\}+O(\epsilon^0)\;.\;\;\;
\end{eqnarray}

The Laurent expansions of the integrals $I_2$ and $I_3$ can be worked
out in a similar fashion. They read
\begin{eqnarray}
\label{eq:I2fin}
I_2 &=&{\int}d^{d-1}r \,\bigg\{\frac{1}{2 \epsilon}\,
\varphi(\bm{r},0)\, \varphi'(\bm{r}, 0)\, {\int_1^\infty}\!dt \, f_0(t)
\nonumber\\&&\strut +
\frac{1}{4m \epsilon}\, \varphi(\bm{r},0)\,
\partial_\alpha\partial_\alpha\varphi(\bm{r}, 0)\, {\int_1^\infty}\!dt \,
f_2(t)  \bigg\}+ O(\epsilon^0)
\end{eqnarray}
and
\begin{eqnarray}
\label{eq:I3fin}
I_3 &=&{\int}d^{d-1}r \,\bigg\{\frac{f_0 (1)}{2 \epsilon}
\,\varphi(\bm{r}, 0)\,\varphi'(\bm{r}, 0) +\frac{f_2 (1)}{4
  m\epsilon}\, \varphi(\bm{r},0)\, \partial_\alpha\partial_\alpha
\varphi(\bm{r}, 0)  \bigg\} +O(\epsilon^0)\;.
\end{eqnarray}

Equations (\ref{eq:Abbb}), (\ref{eq:I1fin}), (\ref{eq:I2fin}) and
(\ref{eq:I3fin}) give us the diagonal elements $(\varphi, {\mathcal
  A}_j\varphi)$ for a general test function $\varphi$. Utilizing these
results in conjunction with Eq.~(\ref{eq:A}), one can readily
determine the required off-diagonal elements $(\psi,{\mathcal
  A}_j\,\varphi)$ corresponding to the special choices
(\ref{eq:psiphi}) of $\psi(\bm{x})$ and $\varphi(\bm{x})$. The terms
involving $f_2$ do not contribute to the pole terms because of the
Dirichlet boundary condition satisfied by the test function $\psi$
from Eq.~(\ref{eq:psiphi}). This entails that there are no such
contributions from the diagonal elements $(\psi,{\mathcal A}_j\psi)$,
and that there analogs from the difference $(\psi+\phi,{\mathcal
  A}_j[\psi+\phi])-(\phi,{\mathcal A}_j\phi)$ cancel.
\end{widetext}

The coefficients $f_0(1)$ and $f_4(1)$ can be expressed in terms of
the single integrals $j_\phi(m)$ and $j_\sigma(m)$ defined by
Eqs.~(\ref{eq:jphidef}) and (\ref{eq:jsigmadef}) as
\begin{equation}
  \label{eq:f01}
f_0(1)=F_{m,0}^{2}\,\mathring{\sigma}^{-m/2}\;\frac{j_\phi(m)}{8-m}
\end{equation}
and
\begin{equation}
   \label{eq:f41}
f_4(1)=F_{m,0}^{2}\,\mathring{\sigma}^{1-m/2}\;\frac{j_\sigma(m)}{2}\;.
\end{equation}

The two integrals involving $f_0(t)$,
\begin{equation}
  \label{eq:intfJu}
  {\mathcal I}_1\equiv {\int_0^1}\frac{dt}{t} \bigg(f_0 (t)-f_0 (0)\bigg)
\end{equation}
and
\begin{equation}
  \label{eq:intf01inf}
{\mathcal I}_2\equiv{\int_1^\infty} dt \,f_0(t)=
{\int_0^1}\frac{dt}{t^2}\,f_0(1/t)\,\;, 
\end{equation}
both yield contributions $\sim\epsilon^{-1}$. Let us rewrite them in
terms of the integrals $J_u(m)$ and $j_1(m)$ defined by
Eqs.~(\ref{eq:Ju}), (\ref{eq:judef}), and (\ref{eq:j1m}).

To this end, we rewrite ${\mathcal I}_1$ as
\begin{equation}
  \label{eq:I1calc}
  {\mathcal I}_1=\lim_{\delta\to 0+}(B_\delta+f_0(0)\,\ln\delta)\;,
  \;\; B_\delta\equiv{\int_\delta^1}\frac{dt}{t}\,f_0(t)\,.
\end{equation}
\begin{widetext}
Using the scaling form (\ref{eq:Gbx}) of the free bulk propagator and
performing the angular integrations, one obtains
\begin{eqnarray}
  \label{eq:Bdelta}
  \frac{B_\delta\,\mathring{\sigma}^{m/2}}{S_m\,S_{3-m/2}}&=&
  {\int_\delta^1}dt\, 
  {\int^\infty_0}dr\,r^{2-m/2}{\int_0^\infty}dw\,
  (t^2+r^2)^{m/4}\, \frac{\Phi_{m,d^*}^2{\big(w^{1/4}\big)}\,
    \Phi_{m,d^*}{\big[(y(t,r)\,w)^{1/4}\big]}}{4 w^{1-m/4}\,
    (1+r^2)(t^2+r^2)^2}\nonumber\\
&=&{\int_0^\infty}dw\,
\frac{\Phi_{m,d^*}^2{\big(w^{1/4}\big)}}{8 w^{1-m/4}}\,
{\int^\infty_0}dr\,\frac{r^{2-m/2}}{(1+r^2)^{2-m/4}}\,
{\int_{y(\delta,r)}^1}dy\, 
\frac{\Phi_{m,d^*}{\big[(yw)^{1/4}\big]}}{y^{2-m/4}\,
  \sqrt{y\,(1+r^2)-r^2}} \;,\nonumber\\
\end{eqnarray}
where 
\begin{equation}
  \label{eq:ytrdel}
  y(t,r)\equiv \frac{t^2+r^2}{1+r^2}
\end{equation}
is the new integration variable to which we transformed to get the
second line of Eq.~(\ref{eq:Bdelta}). To extract the singularity of
the integral for small $\delta$, we subtract and add $\Phi_{m,d^*}(0)$
in the numerator of the $y$ integral. In the part of the integral
produced by the term
$\Phi_{m,d^*}{\big[(yw)^{1/4}\big]}-\Phi_{m,d^*}(0)$, we can safely
replace the lower integration limit by its $\delta\to 0$ limit
$y(0,r)$. In the double integral
${\int_0^\infty}dr\,{\int_{y(0,r)}^\infty}dy$ we reverse the order of
integrations whereby it becomes ${\int_0^1}dy\,{\int_0^{r_y}}dr$ with
$r_y=\sqrt{y/(1-y)}$. Upon changing from $r$ to the integration
variable $x=r/r_y$, we can perform the $x$ integration. In the $w$
integral of the resulting expression, one easily recognizes the
scaling function $\Theta_m(w^{1/4})$ given in Eq.(\ref{eq:Theta}) upon
recalling Eq.~(D.1) of Ref.~\onlinecite{SD01}. The result of the
remaining $w$~integration therefore is proportional to the integral
$j_u(m)$ whose definition is recalled in Eq.~(\ref{eq:judef}).

The part of the integral associated with the constant
$\Phi_{m,d^*}(0)$ in the numerator of Eq.~(\ref{eq:Bdelta}) yields
terms that behave as $\sim\ln\delta$ and approach a finite limit as
$\delta\to 0$, respectively. They can be computed in a straightforward
fashion. Upon gathering all contributions to $B_\delta$, substituting
them into Eq.~(\ref{eq:I1calc}), and taking into account that
$f_0(0)=\mathring{\sigma}^{-m/2}\,F_{m,0}^2$, we can perform the
required limit $\delta\to 0$. The result is
\begin{equation}
  \label{eq:I1res}
  {\mathcal I}_1/F_{m,0}^2\,\mathring{\sigma}^{-m/2}=
  {\int_0^1}\frac{dt}{t}\,{\bigg[\frac{f_0(t)}{f_0(0)}-1\bigg]} =
  J_u(m)-1-\ln 2\;.
\end{equation}

The integral ${\mathcal I}_2$ can be transformed in a similar manner
to show that it can be expressed in terms of the integral $j_1(m)$
introduced in Eq.~(\ref{eq:j1m}) as
\begin{equation}
  \label{eq:I2res}
  {\mathcal I}_2/F_{m,0}^2\,\mathring{\sigma}^{-m/2}
  ={\int_1^\infty}dt\,\frac{f_0(t)}{f_0(0)}
  =j_1(m)\;. 
\end{equation}

Combining the above results finally yields
\begin{equation}
\label{eq:decD}
\raisebox{-0.8em}{\begin{texdraw} \drawdim pt \setunitscale 3 \linewd
    0.3 \move(-6 0) \move(-3 0) \lcir r:1   \move(20 0) \lcir r:1
    \move(18 -1.5) 
    \rlvec(0 3)   
\move(19.5 -0.5)\rlvec(1 1) \move(19.5 0.5) \rlvec(1 -1)
\lpatt(0.75 0.75) \move(-2 0) \rlvec(5 0) \move(13.5 0) \rlvec(5 0)
\move(12 0) \move(8 0) \lellip rx:5
    ry:3  \move(3 0) \rlvec(10 0) \move(22 0)
\end{texdraw}}
 =
\raisebox{-0.8em}{\begin{texdraw} \drawdim pt \setunitscale 3 \linewd
    0.3 \move(-6 0) \move(-3 0) \lcir r:1  \move(8 0) \lellip rx:5
    ry:3  \move(3 0) \rlvec(10 0) \move(20 0) \lcir r:1 \move(18 -1.5)
    \rlvec(0 3)  
\move(19.5 -0.5)\rlvec(1 1) \move(19.5 0.5) \rlvec(1 -1)
\lpatt(0.75 0.75) \move(-2 0) \rlvec(5 0) \move(13.5 0) \rlvec(5 0)
\move(12 0)
\htext(7.4 0.5){\footnotesize{b}} \htext(7.4 3.5){\footnotesize b}
\htext(7.4 -2.5){\footnotesize{b}} \move(22 0)
\end{texdraw}}
 -
\raisebox{-0.8em}{\begin{texdraw} \drawdim pt \setunitscale 3 \linewd
    0.3 \move(-6 0) \move(-3 0) \lcir r:1  \move(8 0) \lellip rx:5
    ry:3  \move(3 0) \rlvec(10 0) \move(20 0) \lcir r:1 \move(18 -1.5)
    \rlvec(0 3)  
\move(19.5 -0.5)\rlvec(1 1) \move(19.5 0.5) \rlvec(1 -1)
\lpatt(0.75 0.75) \move(-2 0) \rlvec(5 0) \move(13.5 0) \rlvec(5 0)
\htext(7.4 0.5){\footnotesize b} \htext(7.4 3.5){\footnotesize b}
\htext(7.65 -2.5){\footnotesize s} \move(22 0)
\end{texdraw}}
 +
\raisebox{-0.8em}{\begin{texdraw} \drawdim pt \setunitscale 3 \linewd
    0.3 \move(-6 0) \move(-3 0) \lcir r:1  \move(8 0) \lellip rx:5
    ry:3  \move(3 0) \rlvec(10 0) \move(20 0) \lcir r:1 \move(18 -1.5)
    \rlvec(0 3)  
\move(19.5 -0.5)\rlvec(1 1) \move(19.5 0.5) \rlvec(1 -1)
\lpatt(0.75 0.75) \move(-2 0) \rlvec(5 0) \move(13.5 0) \rlvec(5 0)
\htext(7.65 0.5){\footnotesize s} \htext(7.4 3.5){\footnotesize b}
\htext(7.65 -2.5){\footnotesize s} \move(22 0)
\end{texdraw}}
-
\raisebox{-0.8em}{\begin{texdraw} \drawdim pt \setunitscale 3 \linewd
    0.3 \move(-6 0) \move(-3 0) \lcir r:1  \move(8 0) \lellip rx:5
    ry:3  \move(3 0) \rlvec(10 0) \move(20 0) \lcir r:1 \move(18 -1.5)
    \rlvec(0 3)  
\move(19.5 -0.5)\rlvec(1 1) \move(19.5 0.5) \rlvec(1 -1)
\lpatt(0.75 0.75) \move(-2 0) \rlvec(5 0) \move(13.5 0) \rlvec(5 0)
\htext(7.65 0.5){\footnotesize s} \htext(7.65 3.5){\footnotesize s}
\htext(7.65 -2.5){\footnotesize s} \move(22 0)
\end{texdraw}}
\end{equation}
with
\begin{eqnarray}
\label{eq:bbb}
\raisebox{-0.8em}{\begin{texdraw} \drawdim pt \setunitscale 3 \linewd
    0.3 \move(-6 0) \move(-3 0) \lcir r:1  \move(8 0) \lellip rx:5
    ry:3  \move(3 0) \rlvec(10 0) \move(20 0) \lcir r:1 \move(18 -1.5)
    \rlvec(0 3)  
\move(19.5 -0.5)\rlvec(1 1) \move(19.5 0.5) \rlvec(1 -1)
\lpatt(0.75 0.75) \move(-2 0) \rlvec(5 0) \move(13.5 0) \rlvec(5 0)
\htext(-4.1 -3.8){$z_1$}
\htext(7.4 0.5){\footnotesize b} \htext(7.4 3.5){\footnotesize b}
\htext(7.4 -2.5){\footnotesize b} \move(22 0)
\end{texdraw}}
 &=&\frac{n+2}{3}\, \mathring{U}^2 \bigg\{ \frac{\mathring{\sigma}\,z_1
   \,(p_\alpha 
   p_\alpha)^2 }{2 
   \mathring{\kappa}_{\bm{p}}\,\epsilon}\bigg[ \frac{j_\phi (m)}{12\,
   (8-m)}+\frac{j_\sigma (m)}{96 m\,(m+2)}\bigg]
    \nonumber\\&&\strut
    -\frac{j_\phi(m)}{24\epsilon\,(8-m)}+O(\epsilon^0)\bigg\} 
\raisebox{-0.75em}{$z_1$}\hspace{-0.8em}\raisebox{-0.3em}{\includegraphics[width=4em]{./ginf11.0.eps}}\,,
\end{eqnarray}
\begin{equation}
\label{eq:bbs}
\raisebox{-0.8em}{\begin{texdraw} \drawdim pt \setunitscale 3 \linewd
    0.3 \move(-6 0) \move(-3 0) \lcir r:1  \move(8 0) \lellip rx:5
    ry:3  \move(3 0) \rlvec(10 0) \move(20 0) \lcir r:1 \move(18 -1.5)
    \rlvec(0 3)  
\move(19.5 -0.5)\rlvec(1 1) \move(19.5 0.5) \rlvec(1 -1)
\lpatt(0.75 0.75) \move(-2 0) \rlvec(5 0) \move(13.5 0) \rlvec(5 0)
\htext(-4.1 -3.8){$z_1$}
\htext(7.4 0.5){\footnotesize b} \htext(7.4 3.5){\footnotesize b}
\htext(7.65 -2.5){\footnotesize s} \move(22 0)
\end{texdraw}}
=\frac{n+2}{24}\, \mathring{U}^2 \bigg\{
\frac{1}{\epsilon^2}+\frac{1}{\epsilon}\,{\big[J_u(m) +2
  R_{\text{D}}(\bm{p},z_1) 
\big]} +O(\epsilon^0)\bigg\}\,
\raisebox{-0.8em}{$z_1$}\hspace{-0.8em}\raisebox{-0.3em}{\includegraphics[width=4em]{./ginf11.0.eps}}\,,\qquad
\end{equation}
\begin{equation}
\label{eq:bss}
\raisebox{-0.8em}{\begin{texdraw} \drawdim pt \setunitscale 3 \linewd
    0.3 \move(-6 0) \move(-3 0) \lcir r:1  \move(8 0) \lellip rx:5
    ry:3  \move(3 0) \rlvec(10 0) \move(20 0) \lcir r:1 \move(18 -1.5)
    \rlvec(0 3)  
\move(19.5 -0.5)\rlvec(1 1) \move(19.5 0.5) \rlvec(1 -1)
\lpatt(0.75 0.75) \move(-2 0) \rlvec(5 0) \move(13.5 0) \rlvec(5 0)
\htext(-4.1 -3.8){$z_1$}
\htext(7.65 0.5){\footnotesize s} \htext(7.4 3.5){\footnotesize b}
\htext(7.65 -2.5){\footnotesize s} \move(22 0)
\end{texdraw}}
=\frac{n+2}{24}\, \mathring{U}^2  \bigg[\frac{1}{\epsilon}\,j_1(m)
+O(\epsilon^0) 
\bigg]
\raisebox{-0.75em}{$z_1$}\hspace{-0.8em}\raisebox{-0.3em}{\includegraphics[width=4em]{./ginf11.0.eps}}\,,
\end{equation}
and
\begin{equation}
\label{eq:sss}
\raisebox{-0.8em}{\begin{texdraw} \drawdim pt \setunitscale 3 \linewd
    0.3 \move(-6 0) \move(-3 0) \lcir r:1  \move(8 0) \lellip rx:5
    ry:3  \move(3 0) \rlvec(10 0) \move(20 0) \lcir r:1 \move(18 -1.5)
    \rlvec(0 3)  
\move(19.5 -0.5)\rlvec(1 1) \move(19.5 0.5) \rlvec(1 -1)
\lpatt(0.75 0.75) \move(-2 0) \rlvec(5 0) \move(13.5 0) \rlvec(5 0)
\htext(-4.1 -3.8){$z_1$}
\htext(7.65 0.5){\footnotesize s} \htext(7.65 3.5){\footnotesize s}
\htext(7.65 -2.5){\footnotesize s} \move(22 0)
\end{texdraw}}
=\frac{n+2}{72}\, \mathring{U}^2
\bigg[\frac{j_\phi(m)}{\epsilon\,(8-m)}+O(\epsilon^0)\bigg]\, 
\raisebox{-0.75em}{$z_1$}\hspace{-0.8em}\raisebox{-0.3em}{\includegraphics[width=4em]{./ginf11.0.eps}}\,,
\end{equation}
where it should be noticed that the line-symmetry factors of the
graphs on the right-hand side of Eq.~(\ref{eq:decD}) are $1/2$ (second
and third graph) and $1/3{!}$ (all others), depending on whether they
have two or three equivalent lines.

We can now substitute the above results into Eq.~(\ref{eq:Ginfren})
for $G_\infty^{(0,2)}(\bm{p};z)$ and express the bare variables
$\mathring{U}$ and $\mathring{\sigma}$ in terms of renormalized ones
using the expressions for the bulk renormalization factors $Z_\phi$,
$Z_\sigma$, and $Z_u$ given in Eqs.~(40), (41), and (48) of
Ref.~\onlinecite{SD01}. Upon determining the renormalization factor
$Z_{1,\infty}$ such that the dimensional poles are minimally
subtracted, one easily arrives at the series expansion
(\ref{eq:Z1inf}).

\section{Renormalization of $\hat{G}_\infty^{(0,2)}(\bm{p})$ to
  two-loop order}
\label{app:Ginf02}

In this appendix we consider the renormalization of the two-point
function ${G}_\infty^{(0,2)}(\bm{p})$ for $\rho=\tau=0$, which
requires the additive surface counterterm $\propto C_2(u,\epsilon)$
according to Eq.~(\ref{eq:Ginfren}). At two-loop order the sole
contribution to this counterterm comes from the diagram
{\raisebox{-0.4em}{\includegraphics[width=5em]{./ginf02.2c.eps}}\,}.
To determine its pole terms one can proceed as in
Appendix~\ref{app:Ginfgraphs}, choosing the test functions as
\begin{equation}
\label{eq:psi2phi}
\psi(\bm{x})\equiv  ({G}_{\text{D}}
\loarrow{\partial}_n)(\bm{x},\bm{r}_2)\;, \qquad
\varphi(\bm{x}')\equiv  ({G}_{\text{D}}
\loarrow{\partial}_n)(\bm{x}',\bm{r}_2)\;. 
\end{equation}

A straightforward calculation shows that the graph's pole terms in
question --- i.~e., those with no dependence on $p_\beta$ and a
$p_\alpha$~dependence of the form $\epsilon^{-1}\, p_\alpha p_\alpha$
--- can be written as
\begin{equation}
  \label{eq:G2ren}
   \frac{\mathring{U}^2}{8\,
    m\,\epsilon}\,\frac{n+2}{9}\,  
\,\frac{\mathring{\sigma}^{1/2}\,p_{\alpha}p_\alpha}{F_{m,0}^2}
\,\bigg[2\, f_2(1)
+3{\int_0^1}dt \, f_2(t)\,  
-3\,{\int_1^\infty}dt \, f_2(t)\, \bigg]_{\mathring{\sigma}=1}\;,
\end{equation}
where $\mathring{U}$ and $f_2(t)$ are defined through Eqs.~(\ref{eq:U0})
and (\ref{eq:fk}), respectively. Upon introducing the coefficient
\begin{equation}
  \label{eq:bm}
   b_m\equiv \frac{S_m S_{3-\frac{m}{2}}}{4\,m\, F_{m,0}^2}= 
\frac{2^{5+m}\,\pi^{(22+3m)/4}\,\Gamma{\big(1+\frac{m}{2}\big)}}%
{\Gamma{\big(\frac{3}{2}-\frac{m}{4}\big)}
  \Gamma^2{\big(1+\frac{m}{4}\big)}}\;,
\end{equation}
the two integrals can be written as
\begin{equation}
  \label{eq:f2Is}
  \frac{F_{m,0}^{-2}}{8m}\, {\int_1^\infty}dt \, 
  f_2(t)=b_m\,I^{(1)}(m)\;,\qquad
 \frac{F_{m,0}^{-2}}{8m}\,{\int_0^1}dt
  \,\left.f_2(t)\right|_{\mathring{\sigma}=1}=b_m\,I^{(2)}(m)\;,
\end{equation}
with
\begin{equation}
  \label{eq:Ione}
       I^{(1)}(m)=
       {\int_0^\infty}d\upsilon\,\upsilon^{m+1}\,\Phi_{m,d^*}(\upsilon)  
    {\int_0^1}\!dy\,
       y^{m+1}\,\Phi_{m,d^*}^2(\upsilon y)\,\Psi^{(1)}_m(y^4)
\end{equation}
and
\begin{equation}
  \label{eq:Itwo}
       I^{(2)}(m)=
       {\int_0^\infty}d\upsilon\,\upsilon^{m+1}\,\Phi_{m,d^*}(\upsilon)  
    {\int_1^\infty}\!dy\,
       y^{m+1}\,\Phi_{m,d^*}^2(\upsilon y)\,\Psi^{(2)}_m(y^4)\;,
\end{equation}
where 
\begin{equation}
  \label{eq:Psi1def}
  \Psi^{(1)}_m(y)\equiv \frac{1}{2\sqrt{y}}\,
  {\bigg(\frac{y}{1-y}\bigg)}^{(6-m)/4} \,{\int_0^1}\,dt\,
  \frac{t^{(2-m)/4}}{\sqrt{1-t}}\,
  {\bigg(1+\frac{y\,t}{1-y}\bigg)}^{(m-6)/4}%\quad y\le 1
\end{equation}
and
\begin{equation}
  \label{eq:Psi2def}
  \Psi^{(2)}_m(y)\equiv \frac{1}{2\sqrt{y}}\,
  {\bigg(\frac{y}{y-1}\bigg)}^{(6-m)/4} \,{\int_0^\infty}\,dt\,
  \frac{t^{(2-m)/4}}{\sqrt{1+t}}\,
  {\bigg(1+\frac{y\,t}{y-1}\bigg)}^{(m-6)/4}\;.
\end{equation}
The latter functions are expressible in terms of hypergeometric
functions as
\begin{equation}
 \label{eq:Psi1}
 \Psi^{(1)}_m(y)=\frac{\sqrt{\pi}\,y^{1-m/4}\, (1-y)^{(m-6)/4}}{2\,
   \Gamma(2-m/4)} \; \Gamma{\Big(\frac{3}{2}-\frac{m}{4}\Big)}\; {}_2
 F_1{\Big(\frac{3}{2}-\frac{m}{4},\frac{3}{2}-\frac{m}{4};2-\frac{m}{4};
\frac{y}{y-1}\Big)}
\end{equation}
and
\begin{eqnarray}
 \label{eq:Psi2}
 \Psi^{(2)}_m(y)&=&\frac{\sqrt{\pi} \,
   \Gamma(1-m/4)}{2 \,(y-1)^{1/2}\, 
\Gamma[(6-m)/4)]} \;
{}_2F_1{\Big(\frac{1}{2},\frac{1}{2};\frac{m}{4},\frac{y}{y-1}\Big)}
 \nonumber\\ &&\strut
+ \frac{y^{1-m/4} \, 
(y-1)^{(m-6)/4} }{2 \,\sqrt{\pi}} \,
\Gamma{\Big(\frac{3}{2}-\frac{m}{4}\Big)}\,  
 \Gamma{\Big(\frac{m}{4}-1\Big)}\; 
\times\nonumber\\&&\strut\times
 {}_2F_1{\Big(\frac{3}{2}-\frac{m}{4},\frac{3}{2}-\frac{m}{4};
   2-\frac{m}{4};\frac{y}{y-1}\Big)}\;.
\end{eqnarray}

From the above results, Eq.~(\ref{eq:C2}) for $C_2(u,\epsilon)$
follows at once.

As a consistency check we have explicitly verified to two-loop order
that all poles of $G^{(0,2)}_\infty(\bm{p};\sigma=\tau=0)$ get
canceled in the renormalized function
$G^{(0,2)}_{\infty,\text{ren}}(\bm{p};\sigma=\tau=0)$ for our choice
of bulk and surface renormalization functions. For interested readers
wishing to check this, we here give the Laurent expansions of the
corresponding Feynman diagrams to the required orders in $\epsilon$
and $\mathring{U}$. They read
\begin{equation}
  \label{eq:one}
  \raisebox{-0.5em}{\includegraphics[width=5em]{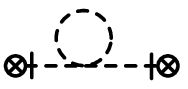}}= 
\frac{n+2}{3}\, \frac{\mathring{U}}{2}\,\frac{-\mathring{\kappa}_{\bm{p}}}{\epsilon}\,  
[1-\epsilon\ln\mathring{\kappa}_{\bm{p}}+O(\epsilon^2)]\;,
\end{equation}
\begin{equation}
  \label{eq:twoa}
  \raisebox{-0.5em}{\includegraphics[width=5em]{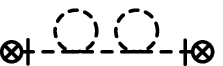}}= 
  \left(\frac{n+2}{3}\right)^2\,
  \frac{\mathring{U}^2}{4} \,(-\mathring{\kappa}_{\bm{p}})
  \Big[\frac{1}{2\epsilon^2}
  +\frac{1-4\ln(2\mathring{\kappa}_{\bm{p}})}{4\epsilon}
  +O(\epsilon^0)\Big]\;, 
\end{equation}
\begin{equation}
  \label{eq:twob}
  \raisebox{-0.5em}{\includegraphics[width=5em]{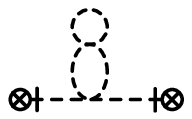}}= 
 \left(\frac{n+2}{3}\right)^2\,
  \frac{\mathring{U}^2}{4}
  \,(-\mathring{\kappa}_{\bm{p}})\,\frac{-1}{2\epsilon^2}\Big\{1
  +\epsilon\Big[\frac{1}{2}-2\ln(2\mathring{\kappa}_{\bm{p}})
  +O(\epsilon^3)\Big]\Big\}\;,
\end{equation}
and
\begin{eqnarray}
  \label{eq:twoc}
  \raisebox{-0.45em}{\includegraphics[width=5em]{./ginf02.2c.eps}} &=& 
\frac{\mathring{U}^2}{6}\,\frac{n+2}{3}\,\bigg\{
\frac{3\,\mathring{\kappa}^2_{\bm{p}}-p_\beta
  p_\beta}{4\mathring{\kappa}_{\bm{p}}\epsilon}\,\frac{j_\phi(m)}{8-m}
+\frac{\mathring{\sigma}\, (p_\alpha
  p_\alpha)^2}{32m(m+2)\mathring{\kappa}_{\bm{p}}}\, 
\frac{j_\sigma(m)}{\epsilon}\nonumber\\&&\strut
+6\,\frac{\sqrt{\mathring{\sigma}}\,p_\alpha 
  p_\alpha}{\epsilon}\,b_m\,{\bigg[\frac{2}{3}\frac{J_{2,3}(m)}{6-m}
  -I^{(1)}(m)+I^{(2)}(m)\bigg]}\nonumber\\&&\strut
 +\frac{3\,\mathring{\kappa}_{\bm{p}}}{2\epsilon}\bigg[\frac{1}{\epsilon} 
+J_u(m) -j_1(m)  - 2\,\ln (2\mathring{\kappa}_{\bm{p}})  
   \bigg]+O(\epsilon^0) \bigg\}\;,
\end{eqnarray}
where $J_{2,3}(m)$ denotes a particular one of the integrals
(\ref{eq:Jps}).

\section{One-loop calculation for general values of $\lambda$}
\label{app:Plam}
In this appendix we describe the calculation of the renormalization
functions $Z_1(u,\lambda,\epsilon)$, $Z_c(u,\lambda,\epsilon)$, and
$P_\lambda (u,\lambda,\epsilon)$, defined by Eqs.~(\ref{eq:surfrep}),
to one-loop order. For this purpose, it is sufficient to consider the
cumulants $\hat{G}^{(1,1)}(\bm{p};z_1)$ and $\partial
\hat{G}^{(2,0)}(\bm{p};z_1)/\partial \mathring{c}$ for
$\mathring{c}=0$ and generic $\mathring{\lambda} >0$. Let us indicate
that we have set $\mathring{c}=0$ by a subscript zero. The free
propagator (\ref{eq:Ghatgen}) in this case reduces to
\begin{equation}
  \label{eq:frN1}
\hat{G}_0(\bm{p};z_1,z_2) =
\frac{1}{2\mathring{\kappa}_{\bm{p}}}
  \bigg[e^{-\mathring{\kappa}_{\bm{p}}|z_1-z_2|}
+ \frac{\mathring{\kappa}_{\bm{p}}-\mathring{\lambda} \,
  p_\alpha p_\alpha}{\mathring{\kappa}_{\bm{p}}+\mathring{\lambda} \, p_\alpha
  p_\alpha}\, 
e^{-\mathring{\kappa}_{\bm{p}}(z_1+z_2)}\bigg]\;.
\end{equation}

Consider first $\hat{G}_0^{(1,1)}(\bm{p};z_1)$. Up to one-loop order,
we have
\begin{equation}
\label{eq:G011lam}
\hat{G}_0^{(1,1)}(\bm{p};z_1)=
\frac{e^{-\mathring{\kappa}_{\bm{p}}\, z_1}}{\mathring{\kappa}_{\bm{p}}+\mathring{\lambda}\,
  p_\alpha p_\alpha} 
 -\frac{\mathring{u}}{2}\,\frac{n+2}{3} \,{\int_0^\infty}dz \,
 \hat{G}_0(\bm{p};z_1,z)\,  
\hat{G}_0(\bm{p};z,0)\, G_0(\bm{x},\bm{x})
+O(\mathring{u}^2)
\end{equation}
with
\begin{equation}
\label{eq:Gxx}
G_0(\bm{x},\bm{x})={\int}\frac{d^{d-1}p}{(2 \pi)^{d-1}}\, 
\frac{\mathring{\kappa}_{\bm{p}}-\mathring{\lambda} \,p_\alpha
  p_\alpha}{\mathring{\kappa}_{\bm{p}} 
  +\mathring{\lambda} \, p_\alpha p_\alpha}\, \frac{e^{-2
    \mathring{\kappa}_{\bm{p}}z}}{2\mathring{\kappa}_{\bm{p}}} \, 
=i_1{\big(\mathring{\lambda}\,\mathring{\sigma}^{-1/2},\epsilon;m\big)}\, 
F_{m,\epsilon}\,\frac{\Gamma(2 - \epsilon)\, 
      \sin ({\epsilon\,\pi }/{2})}
    {\epsilon\,\pi\,\mathring{\sigma}^{m/4}}\,z^{\epsilon-2}\;.
\end{equation}
Here $i_1(\lambda,\epsilon;m)$ is defined by
\begin{equation}
\label{eq:i1def}
i_1{\big(\mathring{\lambda} \mathring{\sigma}^{-1/2},\epsilon;m\big)}
\equiv 
C_1(\epsilon)\,\mathring{\sigma}^{m/4}\, 
{\int}\frac{d^{d-1}p}{(2 \pi)^{d-1}}\,
\frac{\mathring{\kappa}_{\bm{p}}-\mathring{\lambda} 
  \, p_\alpha 
  p_\alpha}{\mathring{\kappa}_{\bm{p}}+\mathring{\lambda} \, p_\alpha
  p_\alpha}\,\frac{e^{-\mathring{\kappa}_{\bm{p}}}}{2\mathring{\kappa}_{\bm{p}}}\;, \quad
i_1(0,\epsilon;m)=1\;. 
\end{equation}
where the normalization constant $C_1$ is fixed by the specified value
of $i_1(0,\epsilon;m)$.  Thus the $\mathring{\lambda}$~dependence of the
generalized function $G_0(\bm{x},\bm{x})$ is entirely contained in the
prefactor $i_1(\lambda\,\mathring{\sigma}^{-1/2},\epsilon;m)$.

After performing the angular integrations in Eq.~(\ref{eq:i1def}), we
transform from the radial integration variables $P\equiv
(p_\beta\,p_\beta)^{1/2}$ and
$k=(\mathring{\sigma}\,p_\alpha\,p_\alpha)^{1/2}$ to
$\kappa\equiv\mathring{\kappa}_{\bm{p}}$ and $k$, using
${\int_0^\infty}dk{\int_0^\infty}dp\,g(k,p)= {\int_0^\infty}d\kappa
{\int_0^{\sqrt{\kappa}}}dk\,\kappa \, (\kappa^2-k^4)^{-1/2}\,
g{\big(k,\sqrt{\kappa^2-k^4}\big)}$. The $k$~integration then can be
performed, and one easily convinces oneself that
$i_1(\lambda,\epsilon;m)$ can be written in the form~(\ref{eq:i1}).

We can now employ Eq.~(\ref{eq:zmtwoplusE}) to perform the
$z$~integration in Eq.~(\ref{eq:G011lam}). This yields
\begin{eqnarray}
\label{eq:G011}
\hat{G}_0^{(1,1)}(\bm{p};z_1)&=&
{\bigg\{1+\mathring{U} \frac{n+2}{12}\, 
i_1{\big(\mathring{\lambda}\,\mathring{\sigma}^{-1/2};m\big)}\,{\bigg[  
\frac{1}{\epsilon}\,{\bigg( 1-\frac{2 \,\mathring{\lambda}\, p_\alpha
    p_\alpha}{\mathring{\kappa}_{\bm{p}}+\mathring{\lambda}\, p_\alpha
    p_\alpha} 
\bigg)}
+O(\epsilon^0)\,\bigg]}\bigg\}}\,
\frac{e^{-\mathring{\kappa}_{\bm{p}}\,
    z_1}}{\mathring{\kappa}_{\bm{p}}+\mathring{\lambda}\,  
p_\alpha p_\alpha}
\nonumber\\&&\strut
 +O(\mathring{u}^2)\;.
\end{eqnarray}
To determine $Z_1$, we can set $p_\alpha=0$ and determine the $O(u)$
term of $Z_1(u,\lambda,\epsilon)$ from the condition that the pole of
$\hat{G}_0^{(1,1)}(\bm{p};z_1)$ cancels in
$\hat{G}_{0,\text{ren}}^{(1,1)}(\bm{p};z_1)$. This yields the
result~(\ref{eq:Z1onel}). For $p_\alpha\ne 0$, the pole implied by the
first term of Eq.~(\ref{eq:G011}) in parentheses cancels upon
multiplication of $Z_1^{-1/2}$. Requiring that the remaining one
implied by the second term in parentheses gets absorbed by the
renormalization of $\mathring{\lambda}$ according to
Eq.~(\ref{eq:surfrep}) yields Eq.~(\ref{eq:Plambdares}).

Turning to the function
$[\partial\hat{G}^{(2,0)}(\bm{p};z_1,z_2)/\partial\mathring{c}]_0$, we
note that the derivative $-\partial/\partial\mathring{c}$ generates an
insertion of the surface operator
${\int}_{\mathfrak{B}}dA\,\bm{\phi}^2/2$, which we depict as
\raisebox{-0.0em}{\begin{texdraw} \drawdim pt \setunitscale
    2.5 \linewd 0.4 \move(-4 0)\move(-2 0) \rlvec(1 0) \move(1 0)
    \rlvec(1 0) \move(0 0) \lcir r:1 \move(-0.5 -0.5)\rlvec(1 1)
    \move(-0.5 0.5) \rlvec(1 -1) \move(0 1) \lpatt(0.4 1) \rlvec(0 3)
    \move(3 0)
\end{texdraw}}
. Hence we have
\begin{eqnarray}
  \label{eq:G20partc}
  \lefteqn{\frac{-\partial}{\partial\mathring{c}} \,\hat{G}^{(2,0)}(\bm{p};z_1,z_2)
  \bigg|_{\mathring{c}=0}}&&
\nonumber\\&=&
\raisebox{-0.9em}{\begin{texdraw} \drawdim pt \setunitscale 2.5 \linewd
    0.4 \move(-6 0) \move(-3 0) \lcir r:1   \move(6 0) \lcir r:1
 \move(-2 0) \rlvec(7 0)
\move(5.5 -0.5)\rlvec(1 1) \move(5.5 0.5) \rlvec(1 -1)
\htext(-4 -4.5){$z_1$} \move(7 0)
\rlvec(7 0) \move(15 0) \lcir r:1 \htext(13.75 -4.5){$z_2$}
\move(6.0 1) \lpatt(0.4 1) \rlvec(0 3) 
\move(20 0)
\end{texdraw}}
+
\left[
\raisebox{-0.9em}{\begin{texdraw} \drawdim pt \setunitscale 2.5 \linewd
    0.4 \move(-6 0) \move(-3 0) \lcir r:1   \move(6 0) \lcir r:1
 \move(-2 0) \rlvec(7 0)
\move(5.5 -0.5)\rlvec(1 1) \move(5.5 0.5) \rlvec(1 -1)
\htext(-4 -4.5){$z_1$} \move(7 0)
\rlvec(7 0) \move(14.5 4) \lellip rx:3 ry:4 
\move(14.5 0) \rlvec(7.5 0) \move(23 0) \lcir r:1
\htext(21.75 -4.5){$z_2$}
\move(6 1) \lpatt(0.4 1) \rlvec(0 3)
\move(28 0)
\end{texdraw}}
+ (z_1\leftrightarrow z_2)
\right]
+
\raisebox{-0.9em}{\begin{texdraw} \drawdim pt \setunitscale 2.5 
\linewd 0.4 \move(-6 0) \move(-3 0) \lcir r:1   \move(13 0) \lcir r:1
 \move(-2 0) \rlvec(14 0)
\htext(-4 -4.5){$z_1$} \move(8 0)
 \move(5 4) \lellip rx:3 ry:4 
\move(5 8) \fcir f:1 r:1 \lcir r:1 
\move(4.5 7.5)\rlvec(1 1) \move(4.5 8.5) \rlvec(1 -1)
\move(5 9) \lpatt(0.4 1) \rlvec(0 3)
\htext(11.75 -4.5){$z_2$}
\move(15 0)
\end{texdraw}}
+O(\mathring{u}^2)\;.
\end{eqnarray}
The Feynman integral of the graph inside the brackets is nothing but
$\hat{G}(\bm{p};z_1,0)$ times the term $\propto \mathring{u}$ of
Eq.~(\ref{eq:G011lam}), with the right external point taken at $z_2$.
The remaining last graph's loop involves the integral
\begin{equation}
  \label{eq:partcbG20}
  {\int}\frac{d^{d-1}p}{(2\pi)^{d-1}}\,
  \frac{e^{-2\mathring{\kappa}_{\bm{p}}z}}{{(\mathring{\kappa}_{\bm{p}}+
    \mathring{\lambda}\,p_\alpha\,p_\alpha)}^2}
=F_{m,\epsilon}\,\frac{4\,\Gamma(1-\epsilon)\,
  \sin(\epsilon\,\pi/2)}{\epsilon\,\pi\,\mathring{\sigma}^{m/4}}\, 
i_2{\big(\mathring{\lambda}\,\sigma^{-1/2};m\big)}\,z^{\epsilon-1}\;,
\end{equation}
where $i_2$ is the counterpart of $i_1$ defined by
\begin{equation}
  \label{eq:i2def}
  i_2{\big(\mathring{\lambda}
    \mathring{\sigma}^{-1/2},\epsilon;m\big)} \equiv 
C_2(\epsilon)\,\mathring{\sigma}^{m/4}\, 
{\int}\frac{d^{d-1}p}{(2 \pi)^{d-1}}\,
\frac{e^{-\mathring{\kappa}_{\bm{p}}}}{{(\mathring{\kappa}_{\bm{p}} +\mathring{\lambda} \, 
    p_\alpha 
  p_\alpha)}^2}\;, \quad
i_2(0,\epsilon;m)=1\;.
\end{equation}
Proceeding as in the case of $i_1$, one arrives at the form
(\ref{eq:i2}).

A combination of the above results yields the Laurent expansion
\begin{eqnarray}
  \label{eq:G20partcLaurent}
  \frac{-\partial}{\partial\mathring{c}} \,\hat{G}^{(2,0)}(\bm{p};z_1,z_2)
  \bigg|_{\mathring{c}=0}&=&
{\Bigg\{1+\frac{\mathring{U}}{\epsilon} \frac{n+2}{6}\, 
{\bigg[i_1{\big(\mathring{\lambda}\, \mathring{\sigma}^{-1/2}\big)}\,
  {\bigg( 1-\frac{2 
      \,\mathring{\lambda}\, p_\alpha p_\alpha}{\mathring{\kappa}_{\bm{p}}+
      \mathring{\lambda}\, 
      p_\alpha p_\alpha} \bigg)}}}\nonumber\\&&\mbox{}{{-
2\,i_2{\big(\mathring{\lambda}\,
  \mathring{\sigma}^{-1/2};m\big)}+O(\epsilon^0)\bigg]}
\Bigg\}} \,
\frac{e^{-\mathring{\kappa}_{\bm{p}}\,(z_1+z_2)}}{{(\mathring{\kappa}_{\bm{p}}+
    \mathring{\lambda}\,  
p_\alpha p_\alpha)}^2}+O(\mathring{u}^2)\;,
\end{eqnarray}
from which $Z_c$ can be determined in a straightforward fashion by
requiring that the poles of the function (\ref{eq:G20partcLaurent})
are minimally absorbed through $Z_c$ so that the renormalized quantity
$[\partial \hat{G}^{(2,0)}_{\text{ren}}(\bm{p};z_1,z_2)/\partial
c]_{c=0} = Z_c\,Z_\phi^{-1}\,[\partial
\hat{G}^{(2,0)}(\bm{p};z_1,z_2)/\partial\mathring{c}]_0= [\partial
\hat{G}^{(2,0)}_{\text{ren}}(\bm{p};z_1,z_2)/\partial c]_0$ becomes uv
finite. The result is given in Eq.~(\ref{eq:Zconel}).

Using {\sc Mathematica},\cite{rem:MAT} the integrals
$i_1(\lambda;m)\equiv i_1(\lambda,\epsilon=0;m)$ and
$i_2(\lambda;m)\equiv i_2(\lambda,\epsilon=0;m)$ introduced in
Eqs.~(\ref{eq:i1}) and (\ref{eq:i2}) can be computed for $0<m<6$ and
expressed in terms of hypergeometric functions. One obtains
\begin{equation}
  \label{eq:i1hyp}
  i_1(\lambda;m)=\frac{m}{6}\,\lambda^2\,
  {\mathop{_{2\!}{F}_{1}}\nolimits}{\Big(1,1+\frac{m}{4};
    \frac{5}{2};\lambda^2\Big)} +
 {\mathop{_{2\!}{F}_{1}}\nolimits}{\Big(1,\frac{m}{4};
    \frac{3}{2};\lambda^2\Big)}
-\frac{2\pi\,\big[1-(1-\lambda^2)^{(2-m)/4}\big]}{\lambda\,
  \cos(m\pi/4)\,B[m/4,(6-m)/4]} 
\end{equation}
and
\begin{equation}
  \label{eq:i2hyp}
 i_2(\lambda;m)= {\mathop{_{2\!}{F}_{1}}\nolimits}{\Big(1,\frac{m}{4};
   \frac{1}{2};{\lambda }^2\Big)} - 
  \frac{\pi \,\lambda \,
     {\left( 1 - {\lambda }^2 \right) }^
      {-(2 +m)/4}}
     {\cos(m\pi/4)\,B[(2-m)/4,m/4]}\;.
\end{equation}

For $m\to 0$, $m=2$, $m=4$, and $m\to 6$, these expressions simplify to
\begin{eqnarray}
  \label{eq:i1spec}
i_1(0+,m)&=& 1\;,\\
  i_1(\lambda;2)&=&-1+\frac{2}{\lambda}\,\ln(1+\lambda)\;,\\ 
i_1(\lambda;4)&=&\frac{\pi-\lambda}{\lambda}-
\frac{2\arccos\lambda}{\lambda\,\sqrt{1-\lambda^2}} \;,\\
i_1(\lambda;6-)&=&\frac{1-\lambda}{1+\lambda}\;,
\end{eqnarray}
and
\begin{eqnarray}
  \label{eq:i2spec}
i_2(0+;m)&=&1\;,\\
   i_2(\lambda;2)&=&\frac{1}{1+\lambda}\;,\\
   i_2(\lambda;4)&=&\frac{1-\lambda\,\arccos\lambda}{1-\lambda^2}\;,\\
   i_2(\lambda;6-)&=&\frac{1}{(1+\lambda)^2}\;,
\end{eqnarray}
respectively.

\section{Calculation of $P^{(2,-1)}_\lambda(0)$}
\label{app:Plam2}

In this appendix we wish to determine the renormalization
(\ref{eq:surfrep}) of $\mathring{\lambda}$ to order $u^2$ for
$\lambda=0$ and compute the residuum $P^{(2,-1)}_\lambda(0)$ introduced
by Eq.~(\ref{eq:Plambda}).

To this end, we consider the function
$\hat{G}^{(1,1)}_0(\bm{p};z_1)\equiv
\hat{G}^{(1,1)}_{\mathring{c}=0}(\bm{p};z_1)$ for $c =\mathring{c}=0$
and $\lambda=0$.  As claimed, the bare variable $\mathring{\lambda}$
does not vanish at order $u^2$ if $\lambda=0$. Hence we must keep a
nonzero $\mathring{\lambda}$ in our calculation.  Up to two-loop
order, the Feynman graph expansion of $\hat{G}^{(1,1)}_0(\bm{p};z_1)$
reads
\begin{eqnarray}
\label{eq:loop2N}
\hat{G}^{(1,1)}(\bm{p};z_1)&=&
\raisebox{-0.8em}{$z_1$}\hspace{-0.7em}\raisebox{-0.3em}{\includegraphics[width=4em]{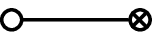}}
+\raisebox{-0.8em}{$z_1$}\hspace{-0.7em}\raisebox{-0.3em}{\includegraphics[width=5em]{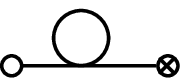}} +
\raisebox{-0.8em}{$z_1$}\hspace{-0.7em}\raisebox{-0.3em}{\includegraphics[width=6em]{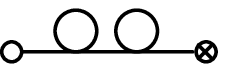}} +
\raisebox{-0.8em}{$z_1$}\hspace{-0.7em}\raisebox{-0.3em}{\includegraphics[width=5em]{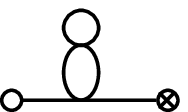}}
\nonumber\\&&\strut
+{}
\raisebox{-0.75em}{$z_1$}\hspace{-1.4em}\raisebox{-0.65em}{\begin{texdraw} \drawdim pt
      \setunitscale 3 \linewd 
    0.3 \move(-6 0) \move(-3 0) \lcir r:1  \move(8 0) \lellip rx:5
    ry:3  \move(3 0) \rlvec(10 0) 
\move(19.5 -0.5)\rlvec(1 1) \move(19.5 0.5) \rlvec(1 -1) \move(20 0)
\lcir r:1  \move(-2 0) \rlvec(5 0) \move(13.5 0) \rlvec(5 0) \move(22 0)
\end{texdraw}} +O(\mathring{u}^3)\;,
\end{eqnarray} 
where \raisebox{-0.7em}{\begin{texdraw} \drawdim pt \setunitscale 2.5 \linewd
    0.4 \move(-6 0) \move(-3 0) \lcir r:1   \move(9 0) \lcir r:1
\move(-2 0) \rlvec(10 0)
\htext(-4 -4.3){$z$} \htext(7.75 -4.3){$z'$}\move(12 0)
\end{texdraw}}
here represents the free propagator (\ref{eq:Ghatgen}) for
$\mathring{c}=\mathring{\tau}=0$. As before, the crossed circle
denotes a point on the surface. Thus the first graph simply becomes
\begin{equation}
  \raisebox{-0.8em}{$z_1$}\hspace{-0.7em}\label{eq:frN}
\raisebox{-0.3em}{\includegraphics[width=4em]{./ginf11.00.eps}} =\frac{\exp(-\mathring{\kappa}_{\bm{p}}z)}{\mathring{\kappa}_{\bm{p}}+
  \mathring{\lambda} \, 
  p_\alpha p_\alpha}\;. 
\end{equation}

In Appendix~\ref{app:Plam} we saw that at one-loop order the
renormalization of $\mathring{\lambda}$ remains multiplicative. Unless
$\lambda\ne 0$, the first two two-loop graphs in Eq.~(\ref{eq:loop2N})
obviously cannot produce primitive momentum dependent uv singularities
of a form corresponding to the renormalization of the boundary
operator $\sum_\alpha(\partial_\alpha\bm{\phi})^2$ because the
(tadpole) graphs one obtains through amputation of the external lines
are independent of the momentum $\bm{p}$. This momentum independence
holds, of course, also if $\lambda\ne 0$. In this case, uv
singularities requiring a multiplicative renormalization of
$\mathring{\lambda}$ are produced by the pole $\sim\delta'(z)$ of the
amputated one-loop tadpole graph. As we explicitly showed in
Appendix~\ref{app:Plam}, this follows quite simply by computing the
action of this distribution on the external legs. This does \emph{not}
happen for $\lambda=0$ as long as $\mathring{\lambda}$ may be set to
zero in the free propagator so that it obeys \emph{Neumann boundary
  conditions}. The upshot of these considerations is that the first
graph contributing to the (non-multiplicative) renormalization of
$\mathring{\lambda}$ is the right-most two-loop graph in
Eq.~(\ref{eq:loop2N}), involving the momentum dependent subgraph
\raisebox{-0.45em}{\begin{texdraw} \drawdim pt \setunitscale 3 \linewd
    0.3 \move(-1 0)  \move(8 0) \lellip rx:5
    ry:2.5  \move(1 0) \rlvec(14 0) 
\end{texdraw}}
.

Since we know that $\mathring{\lambda}=O(u^2)$ if $\lambda=0$, we can
set $\mathring{\lambda}=0$ in all but the first graph of
Eq.~(\ref{eq:loop2N}). Thus the lines of all those graphs correspond
to the free Neumann propagator, which in analogy to Eq.~(\ref{eq:frD})
can be written as
$\hat{G}_{\text{N}}=\hat{G}_{\text{b}}+\hat{G}_{\text{s}}$. Using this
decomposition and the graphical conventions introduced in
Eqs.~(\ref{eq:GDgraph}) and (\ref{eq:decD}), we may split the two-loop
graph in question as
\begin{eqnarray}
\label{eq:loopcbs}
\raisebox{-0.8em}{\begin{texdraw} \drawdim pt
      \setunitscale 3 \linewd 
    0.3 \move(-6 0) \move(-3 0) \lcir r:1  \move(8 0) \lellip rx:5
    ry:3  \move(3 0) \rlvec(10 0) 
\move(19.5 -0.5)\rlvec(1 1) \move(19.5 0.5) \rlvec(1 -1) \move(20 0)
\lcir r:1  \move(-2 0) \rlvec(5 0) \move(13.5 0) \rlvec(5 0) \move(22 0)
\end{texdraw}} 
&=&
 \raisebox{-0.8em}{\begin{texdraw} \drawdim pt
      \setunitscale 3 \linewd 
    0.3 \move(-6 0) \move(-3 0) \lcir r:1  \move(8 0) \lellip rx:5
    ry:3  \move(3 0) \rlvec(10 0) 
\move(19.5 -0.5)\rlvec(1 1) \move(19.5 0.5) \rlvec(1 -1) \move(20 0)
\lcir r:1  
 \move(-2 0) \rlvec(5 0) \move(13.5 0) \rlvec(5 0)
\htext(7.4 0.5){\footnotesize b} \htext(7.4 3.5){\footnotesize b}
\htext(7.4 -2.5){\footnotesize b} \move(22 0)
\end{texdraw}}+\raisebox{-0.8em}{\begin{texdraw} \drawdim pt
      \setunitscale 3 \linewd 
    0.3 \move(-6 0) \move(-3 0) \lcir r:1  \move(8 0) \lellip rx:5
    ry:3  \move(3 0) \rlvec(10 0) 
\move(19.5 -0.5)\rlvec(1 1) \move(19.5 0.5) \rlvec(1 -1) \move(20 0)
\lcir r:1  
 \move(-2 0) \rlvec(5 0) \move(13.5 0) \rlvec(5 0)
\htext(7.4 0.5){\footnotesize b} \htext(7.4 3.5){\footnotesize b}
\htext(7.65 -2.5){\footnotesize s} \move(22 0)
\end{texdraw}}
+
\raisebox{-0.8em}{\begin{texdraw} \drawdim pt
      \setunitscale 3 \linewd 
    0.3 \move(-6 0) \move(-3 0) \lcir r:1  \move(8 0) \lellip rx:5
    ry:3  \move(3 0) \rlvec(10 0) 
\move(19.5 -0.5)\rlvec(1 1) \move(19.5 0.5) \rlvec(1 -1) \move(20 0)
\lcir r:1  
 \move(-2 0) \rlvec(5 0) \move(13.5 0) \rlvec(5 0)
\htext(7.65 0.5){\footnotesize s} \htext(7.4 3.5){\footnotesize b}
\htext(7.65 -2.5){\footnotesize s} \move(22 0)
\end{texdraw}}
\nonumber\\&&\strut
+
\raisebox{-0.8em}{\begin{texdraw} \drawdim pt
      \setunitscale 3 \linewd 
    0.3 \move(-6 0) \move(-3 0) \lcir r:1  \move(8 0) \lellip rx:5
    ry:3  \move(3 0) \rlvec(10 0) 
\move(19.5 -0.5)\rlvec(1 1) \move(19.5 0.5) \rlvec(1 -1) \move(20 0)
\lcir r:1  
 \move(-2 0) \rlvec(5 0) \move(13.5 0) \rlvec(5 0)
\htext(7.65 0.5){\footnotesize s} \htext(7.65 3.5){\footnotesize s}
\htext(7.65 -2.5){\footnotesize s} \move(22 0)
\end{texdraw}}\;,
\end{eqnarray}
Each term on the right-hand side can be computed in the manner
described in Appendix~\ref{app:Plam2}. One chooses
the functions $\phi$ and $\psi$ as\cite{com:testfct}
\begin{equation}
\label{eq:psiphiN}
\psi(\bm{x})=G_{\text{N}}(\bm{x}_1,\bm{x})\;, \qquad \varphi(\bm{x}') =
\left.G_{\text{N}}(\bm{x}',\bm{x}_2)\right|_{z_2=0}\;,
\end{equation}
utilizes Eqs.~(\ref{eq:A})--(\ref{eq:Ij}), (\ref{eq:Abbb}), and
(\ref{eq:I1fin})--(\ref{eq:I3fin}), and finally computes the Fourier
transform ${\int}d^{d-1}r_{12}\dots\,e^{i\bm{p}\cdot\bm{r}_{12}}$.

Only those parts of the quantities $I_j$ that are proportional to
${\int}d^{d-1}r \,\varphi(\bm{r},0)\,\partial_\alpha\partial_\alpha
\varphi(\bm{r}, 0)$ contribute to the renormalization of
$\mathring{\lambda}$ at order $u^2$ if $\lambda=0$. They are given by
\begin{eqnarray}
I_0 &=&-\frac{f_2(1)}{4m\epsilon}\,{\int}d^{d-1}r \, \varphi(\bm{r},0)\,
\partial_\alpha\partial_\alpha \varphi(\bm{r}, 0)+\dots\;,
\nonumber \\
I_1 &=&\frac{1}{4m\epsilon}\,{\int_0^1}dt \,
f_2(t)\,{\int}d^{d-1}r \,\varphi(\bm{r},0)\,
\partial_\alpha\partial_\alpha \varphi(\bm{r},0)+\dots\;,
\nonumber\\
I_2 &=&\frac{1}{4m\epsilon}\, {\int_1^\infty} dt \, f_2(t)\,
{\int}d^{d-1}r \,\varphi(\bm{r},0)\, \partial_\alpha\partial_\alpha
\varphi(\bm{r},0)+\dots\;, \nonumber\\
I_3 &=&\frac{f_2(1)}{4m\epsilon}\, {\int} d^{d-1}r \,
\varphi(\bm{r},0)\,\partial_\alpha\partial_\alpha \varphi(\bm{r},0)
+\dots\;,  
\end{eqnarray}
where the ellipses stand for other types of terms. Upon going over to
the $\bm{p}z$~representation, we obtain the result
\begin{equation}
  \label{eq:lamren}
  \raisebox{-0.8em}{\begin{texdraw} \drawdim pt
      \setunitscale 3 \linewd 
    0.3 \move(-6 0) \move(-3 0) \lcir r:1  \move(8 0) \lellip rx:5
    ry:3  \move(3 0) \rlvec(10 0) 
\move(19.5 -0.5)\rlvec(1 1) \move(19.5 0.5) \rlvec(1 -1) \move(20 0)
\lcir r:1  \move(-2 0) \rlvec(5 0) \move(13.5 0) \rlvec(5 0) \move(22 0)
\end{texdraw}}=- \frac{\mathring{U}^2}{8\,
    m\,\epsilon}\,\frac{n+2}{3}\, 
\,\frac{\mathring{\sigma}^{m/2}}{F_{m,0}^2}\,{\int_0^\infty}dt \, f_2(t)\,
\frac{p_{\alpha}^2}{\mathring{\kappa}_{\bm{p}}}\;\raisebox{-0.3em}{\includegraphics[width=4em]{./ginf11.00.eps}}  +\dots\;. 
\end{equation}
The explicitly displayed pole term must be absorbed by the
renormalization of $\mathring{\lambda}$ according to
Eq.~(\ref{eq:surfrep}). The expansion of the zero-loop graph
\raisebox{-0.7em}{$z_1$}\hspace{-1.2em}
\raisebox{-0.2em}{\includegraphics[width=4em]{./ginf11.00.eps}}\ 
to first order in $\mathring{\lambda}$ produces a contribution of the
form $P_\lambda(u,\lambda{=}0,\epsilon)\,\mathring{\sigma}^{1/2}\,
[\partial\hat{G}^{(1,1)}/
\partial\mathring{\lambda}]_{\mathring{\lambda}=0}$.  Its
$u^2/\epsilon$ part must cancel the above pole. This is the case if we
make the choice
\begin{equation}
\label{eq:P11fin}
P^{(2,-1)}_\lambda(0)=-\frac{n+2}{3}\,
\frac{\mathring{\sigma}^{(m-1)/2}}{8\,m\,F_{m,0}^2}\,{\int_0^\infty}dt
\, f_2(t)=-
\frac{n+2}{3}\,b_m\,{\big[I^{(1)}(m)+I^{(2)}(m)\big]}\;,
\end{equation}
where $I^{(1)}(m)$ and $I^{(2)}(m)$ are the functions introduced in
Appendix~\ref{app:Ginf02} [cf.\
Eqs.~(\ref{eq:f2Is})--(\ref{eq:Itwo})].

\section{Numerical evaluation of integrals}
\label{app:numeval}

In this appendix we briefly explain how the numerical values of the
integrals $J_{2,3}(m)$, $j_1(m)$, $I_1^{(1)}(m)$, and $I^{(2)}(m)$
presented in Table~\ref{tab:jis} were obtained.

\subsection{Numerical calculation of the integrals $J_{2,3}(m)$ and
  $j_1(m)$}
\label{app:J23j1}

From its definition (\ref{eq:Jps}) it is clear that $J_{2,3}(m)$ is a
single integral of a similar form $\int_0^\infty d\upsilon\,
f(\upsilon)$ as the quantities $j_\phi(m)$, $j_\sigma(m)$,
$j_\rho(m)$, and $j_u(m)$, which were previously introduced and
numerically evaluated in Ref.~\onlinecite{SD01}. The same applies to
the integral $j_1(m)$ if we utilize its representation
(\ref{eq:j1Omegaform}) in terms of $\Omega_{m,d^*}$. Unless $m$ takes
the special values $2$ or $6$, the integrands $f(\upsilon)$ involve
differences of hypergeometric functions that grow exponentially as
$\upsilon\to\infty$, while the $f(\upsilon)$ themselves decay as
inverse powers of $\upsilon$ (cf.\ Appendix~E of
Ref.~\onlinecite{SD01}).

To cope with these difficulties, we proceeded as follows. Summing up
the Taylor series expansions of the scaling functions
$\Phi_{m,d^*}(\upsilon)$ and $\Omega_{m,d^*}(\upsilon)$ appearing in
the integrands $f(\upsilon)$ gave reliable results as long as
$\upsilon$ was not too large. Typically, this worked for all
$\upsilon\le\upsilon_0$ up to $\upsilon_0\simeq 10.7$. For larger
values of $m$, even larger choices of $\upsilon_0$ were possible. We
therefore worked with $\upsilon_0=10$ for both functions
$\Phi_{m,d^*}(\upsilon)$ and $\Omega_{m,d^*}(\upsilon)$, and all
$m$. Depending on whether $\upsilon$ was smaller or larger than this
value of $\upsilon_0$, we relied on the Taylor summation and the
asymptotic expansions of the functions $\Phi_{m,d^*}(\upsilon)$ and
$\Omega_{m,d^*}(\upsilon)$ (see below).

In order to gain precision and speed, it proved useful to compute the
required Taylor-series expansion coefficients in a recursive fashion.
Starting from Eq.~(\ref{eq:Phi}), one can easily show that the
function $\Phi_{m,d^*}$ can be written as
\begin{equation}
 \Phi_{m,d^*}(v) = 2^{-5-m} \pi^{-\frac{6+m}{4}} \Bigg[
  \frac{8}{\Gamma(\frac{1}{2}+\frac{m}{4})} \,\sum_{k=0}^\infty 
           s_k{\left({v^4}/{64}\right)}
    - \frac{\sqrt{\pi}}{\Gamma(1+\frac{m}{4})}\,
        v^2 \,\sum_{k=0}^\infty \check{s}_k{\left({v^4}/{64}\right)} \Bigg]\;,
\end{equation}
where $s_k(x)$ and $\check{s}(x)$ satisfy the recursion relations
\begin{equation}
    s_{k+1}(x) =
    \frac{8x\,s_k(x)}{(1+2k)(2+m+4k)} \;,
 \quad
    \check{s}_{k+1}(x) = \frac{4x\,\check{s}_k(x)}{(1+k)(4+m+4k)} \;,
   \quad s_0(x)=\check{s}_0(x)=1 \;. 
\end{equation}

From Eq.~(\ref{eq:Omegacf}) one easily derives the following analogous
representation of $\Omega_{m,d^*}(\upsilon)$:
\begin{equation}
 \Omega_{m,d^*}(\upsilon) =\frac{\sqrt{\pi}}{8\Gamma(1+{m}/{4})}\,
 \upsilon^{m-2}\, 
               \sum_{k=1}^\infty 
               \frac{S_k({\upsilon^4}/{64})}%
                    {m-2+4k}
          - \frac{1}{\Gamma(\frac{1}{2}+\frac{m}{4})}\, \upsilon^{m-4}\,
               \sum_{k=1}^\infty 
               \frac{ \check{S}_k({\upsilon^4}/{64})}%
                    {m-4+4k}
          + R(\upsilon)
\end{equation}
with
\begin{equation}
   \label{RekSnSnQuerZwei}
    S_{k+1}(x)\frac{4x\,S_k(x)}{(1+k)(4+m+4k)}\;,
\quad
    \check{S}_{k+1} (x) =
   \frac{8x\check{S}_k(x)}{(1+2k)(2+m+4k)}\;,\quad
  S_0(x)=\check{S}_0(x)= 1\;,
\end{equation}
and
\begin{equation}
 R (\upsilon) = 
   \frac{\sqrt{\pi}\,
     \upsilon^{m-2}}{8(m-2)\,\Gamma{\left(1+\frac{m}{4}\right)}} 
-\frac{8^{{m}/{2}}\,
   \Gamma\left(\frac{1}{2}+\frac{m}{4}\right)}% 
        {32\,(m-2)\,\sin\left({m\pi}/{4}\right)}
   - \frac{\upsilon^{m-4}}{(m-4)\,
     \Gamma\left(\frac{1}{2}+\frac{m}{4}\right)}\;.
\end{equation}

The apparent singularities at $m=2$ and $m=4$ actually cancel, so that
\begin{equation}
 R(\upsilon) = \left\{
  \begin{array}{l@{\qquad\text{for }}l}
   \frac{1}{2}\,\upsilon^{-2} + \frac{1}{8}\,(C_E - 1) + \frac{1}{4} \ln
   ({\upsilon}/{2}) 
   & m=2\;.\\[\smallskipamount]
   \frac{1}{16}\,\pi^{1/2}\, \upsilon^2 
   + \pi^{-1/2}\,{\left[1 - C_E + \ln({2}{\upsilon^{-2}})\right]}
   & m=4\;.
  \end{array}\right.
\end{equation}

In order to determine the values of the scaling functions for
$\upsilon\ge \upsilon_0$, we utilized their asymptotic expansions.
Those of the functions $\Phi_{m,d^*}$ can be found in Eqs.~(A5)
and (A6) of Ref.~\onlinecite{SD01}; those of the
$\Omega_{m,d^*}(\upsilon)$ can be shown to read
\begin{equation}
   \label{eq:AsymptOmega}
 \Omega_{m,d^*}(\upsilon) = -\frac{16}{\sqrt{\pi}} \,\upsilon^{m-8}\, 
             \sum_{k=0}^\infty \left(-\frac{64}{\upsilon^4}\right)^k
                \frac{\Gamma\big(\frac{3}{2} + k\big)}%
        {\big(\frac{m}{4}-k-2\big)\,\Gamma{\big(\frac{m}{4} -
            \frac{1}{2}- k\big)}}\;.
\end{equation}
Note that if $m=2$ and $m=6$, this asymptotic series truncates at
zeroth and first order, respectively, just as its analog for
$\Phi_{m,d^*}(v)$ does for these choices of $m$. 

Keeping the first three terms of the asymptotic expansions turned out
to be sufficient. At $\upsilon=\upsilon_0=10$, the Taylor series of
both $\Omega_{m,d^*}(\upsilon)$ and $\Phi_{m,d^*}(\upsilon)$ agreed
with their respective (truncated) asymptotic series to within
$10^{-4}$ to $10^{-7}$ percent.

Employing the above methods to compute the values of the integrands
$f(\upsilon)$, the required integrals could be determined by
straightforward numerical integration, giving the results for
$J_{2,3}(m)$ and $j_1(m)$ presented in Table~\ref{tab:jis}. The
numerical values found for $m=2$ and $m=6$ in this manner agree with
the analytical results to the number of decimal digits retained or
better. We also computed $j_1(m)$ by numerical evaluation of the
double integral (\ref{eq:j1m}), obtaining results in conformity with
those obtained by the above method.
\end{widetext}

\subsection{Calculation of $I^{(1)}(m)$, $I^{(2)}(m)$, and related
  quantities}
\label{app:I12}

The renormalization functions $C_2(u,\epsilon)$ and
$P_\lambda^{(2,-1)}(0)$ involve the double integrals $I^{(1)}(m)$ and
$I^{(2)}(m)$ defined by Eqs.~(\ref{eq:Ione}) and (\ref{eq:Itwo}).  In
the special cases $m=2$ and $m\to 6$, the required integrations can again
be done analytically. The functions $\Psi_{m}^{(1)}$ and
$\Psi_{m}^{(2)}$ [cf.\ Eqs.~(\ref{eq:Psi1}) and (\ref{eq:Psi2})]
appearing in the integrands of $I^{(1)}(m)$ and $I^{(2)}(m)$
reduce to the simple expressions
\begin{equation}
  \label{eq:Psi12m2}
 \Psi_2^{(1)}(y)=\Psi_2^{(2)}(y)\ = \frac{1}{2} \ln
 \frac{1+y^{1/2}}{\big|1-y^{1/2}\big|} 
\end{equation}
and
\begin{equation}
  \label{eq:Psi12m6}
 \Psi_m^{(1)}(y) =\Psi_m^{(2)}(y)= \frac{2}{6-m}\,[y^{-1/2}+O(6-m)]\;,
\end{equation}
respectively. Upon substituting these results together with the
simplified expressions for the scaling functions $\Phi_{2,5}(\upsilon)$
and $\Phi_{6,7}(\upsilon)$ into Eqs.~(\ref{eq:Ione}) and
(\ref{eq:Itwo}), the integrations can be performed with the aid of
{\sc Mathematica}\cite{rem:MAT} to obtain
the results (\ref{eq:I1m2res})-(\ref{eq:I2m6res}). From them
Eqs.~(\ref{eq:P21zerom2}) and (\ref{eq:P21zerom6}) for
$P_\lambda^{(2,-1)}(0)$ follow by insertion into
Eq.~(\ref{eq:P11fin}).

For other values of $m$, we performed the double integrals
(\ref{eq:Ione}) and (\ref{eq:Itwo}) numerically, employing {\sc
  Mathematica}.\cite{rem:MAT} In contrast to $\Phi_{m,d^*}(y)$, the
functions $\Psi_m^{(1)}(y)$ and $\Psi_m^{(2)}(y)$ are not differences
of functions that grow exponentially as $y\to\infty$.  Therefore,
their calculation poses no problem for {\sc Mathematica} even for
large $y$.  In the case of $\Phi_{m,d^*}(y)$ (and its powers appearing
in the integrands), we simply let {\sc Mathematica} evaluate its
representation (\ref{eq:Phi}) in terms of hypergeometric functions if
$y\le y_0\simeq 8.5$, and utilized its asymptotic large-$y$ expansion
for $y>y_0$.
In each case we made sure that the values of the integrands the
two methods yielded at the matching point $y=y_0$ were sufficiently
close so that their difference could safely be neglected, given the
limited number of decimal digits of our final numerical results.

The integrands of $I^{(1)}(m)$ and $I^{(2)}(m)$ have integrable
singularities at $y=1$ and vary considerably in some parts of the
two-dimensional integration regime. However, {\sc Mathematica}'s
algorithm for numerical integration requires that the integrand does
not vary too much. Letting us guide by two-dimensional plots of the
integrands, we therefore divided the integration regimes into suitable
rectangular subregions such that the algorithm was able to perform the
integration over each of them without running into precision problems.
We then added up the contributions from these rectangles to determine
the integrals $I^{(1)}(m)$ and $I^{(2)}(m)$. The resulting numbers can
be found in Table~\ref{tab:jis}.

The numerical values {\sc Mathematica} produces for $I^{(1)}(m)$ and
$I^{(2)}(m)$ in the special cases $m=2$ and $m=6$ agree with the
analytical results (\ref{eq:I1m2res})--(\ref{eq:I2m6res}) to within
$10^{-6}$ percent. However, since {\sc Mathematica} automatically
makes use of the fact that scaling functions such as $\Phi_{m,d^*}$
simplify considerably if $m=2$ and $m=6$, we must caution not to
conclude that the same high numerical precision applies to our
numerical results for other choices of $m$.
 
Let us also note that in the case $m=4$, 
$\Psi_m^{(1)}(y)$ and $\Psi_m^{(2)}(y)$ simplify to
\begin{equation}
\label{eq:Psi1m4}
 \Psi_4^{(1)}(y)   = (1-y)^{-1/2}\, K[y/(y-1)]
\end{equation}
and
\begin{equation}
\label{eq:Psi2m4}
 \Psi_4^{(2)}(y) = \frac{2}{(y-1)^{1/2} + y^{1/2}}\,
               K{\Bigg(\bigg[\frac{(y-1)^{1/2} -y^{1/2}}{%
                       (y-1)^{1/2} + y^{1/2}}\bigg]^2\Bigg)}\,,
\end{equation}
where $K(y)$ denotes a complete elliptic integral of the first
kind.

A final remark concerns the sum $I^{(1)}(m)+I^{(2)}(m)$ appearing in
$P_\lambda^{(2,-1)}(0)$. From Eq.~(\ref{eq:f2Is}) we see that it is
proportional to ${\int_0^\infty}dt\,f_2(t)$. The integration over $t$
is straightforward,  producing a free propagator in $d^*-1$
dimensions. The result translates into 
\begin{eqnarray}
\label{eq:I12sum}
\lefteqn{I^{(1)}(m)+I^{(2)}(m)}&&\nonumber\\&=&
\frac{1}{4} \,{\int_0^\infty}dR\,R^{1-m/2}\,
{\int_0^\infty}d\varrho \,\varrho^{m+1}\,
\Phi_{m,d^*-1}{\big(\varrho\,R^{-1/2}\big)} \times
\nonumber\\&&\strut\times
(R^2+1)^{-2}\,\Phi_{m,d^*}^2{\big[\varrho\,(R^2+1)^{-1/4}\big]}\,.
\end{eqnarray}
Since the integrand on the right-hand side has less structure than
those of the individual integrals $I^{(1)}(m)$ and $I^{(2)}(m)$, the
form (\ref{eq:I12sum}) of the sum lends itself more easily to
numerical evaluation via {\sc Mathematica}. Our results for
$P_\lambda^{(2,-1)}(0)$ presented in Eq.~(\ref{eq:P2min1numval}) were
obtained in this fashion.

%\bibliography{/home/hwd/papers/bank,/home/hwd/papers/bankLP,/home/hwd/papers/Sergej/dgrcom,/home/hwd/Ldb/PrivFB}

\begin{thebibliography}{61}
\bibitem[{com({\natexlab{a}})}]{com:ASIgen} \bibinfo{note}{In general,
    there may even be several different anisotropy exponents. If the
    isotropy is broken in both subspaces ${\mathbb R}^m$ and ${\mathbb
      R}^{d-m}$, then up to $d-1$ anisotropy exponents $\theta_i\ne 1$
    are conceivable.}
\bibitem[{DRG()}]{DRGcom}
\bibinfo{note}{A brief account of parts of this work has been given in
  Ref.~\onlinecite{DRG03}. Note that Eq.~(12) of this Letter contains a sign error, which is corrected by Eq.~(\ref{eq:ylambda}) of the present paper.}
\bibitem[{com({\natexlab{b}})}]{com:mres}
\bibinfo{note}{Since $d^*(m)$ varies between $d^*(0)=4$ and $d^*(8)=8$, and $m$
  can at most become equal to the surface dimension $d-1$ if $z$ is a
  $\beta$-direction, we have $m\le d^*(7)-1=13/2$. Thus the largest permissible
  integer value of $m$ is $m=6$.}
\bibitem[{com({\natexlab{d}})}]{com:comp} \bibinfo{note}{In order to
    compare with the $m=0$ results of Refs.~\onlinecite{DD81b},
    \onlinecite{DD83a}, and \onlinecite{Die86a}, one should take into
    account that a different normalization of the coupling constant
    was employed in these references, which implies the following
    correspondence between their $Z$-factors $Z_{\text{DD}}(u)$ and
    ours here: $Z_{\text{DD}}(u/2)=Z(u)$.}
\bibitem[{rem({\natexlab{a}})}]{rem:Xi} \bibinfo{note}{In
    Ref.~\onlinecite{DS00a}, explicit results for the scaling function
    $\Xi_{m,d}(\upsilon)$ in terms of generalized hypergeometric
    functions are given which generalize Eq.~(\ref{eq:Xidstar}) to the
    case $d\ne d^*(m)$; cf.\ its equation (A3). Integral
    representations of $\Xi_{m,d}(\upsilon)$ can be found in this
    reference as well.}
\bibitem[{com({\natexlab{e}})}]{com:GSSing} \bibinfo{note}{In the
    continuous symmetry case, a surface phase with homogeneous
    long-range order above the bulk transition temperature is not
    expected unless the surface dimension $d-1$ exceeds the usual
    lower bulk critical dimension $d_*=2$ of the $O(n)$ vector model.}
\bibitem[{com({\natexlab{g}})}]{com:amplitudes} \bibinfo{note}{These
    are analogs of the nonuniversal amplitudes $E_c^*$ and $E_{h_1}^*$
    utilized in Eqs.~(3.105b) and (3.105d) of
    Ref.~\onlinecite{Die86a}. In the present $m>0$ case, they do not
    only depend on $u$, the initial value of the running coupling
    constant $\bar{u}$, but also on $\lambda$.}
\bibitem[{com({\natexlab{h}})}]{com:smalleps} \bibinfo{note}{Strictly
    speaking, this is guaranteed at this stage only for sufficiently
    small $\epsilon>0$ because the $\epsilon$ dependent terms of the
    exponent $1+O(\epsilon)$ appearing in
    $(1/\bar{c})^{M\,(1+O(\epsilon))}$ can then be trusted to be
    positive. However, from the well-studied $m=0$ case we know that
    this exponent is proportional to the difference
    $\beta_1^{\text{ord}}-\beta_1^{\text{sp}}$ of the surface critical
    exponents $\beta^{\text{ord/sp}}_1$ of the ordinary and special CP
    transitions, respectively. This difference is positive and
    expected to be so even in the $m>0$ case.}
\bibitem[{rem({\natexlab{b}})}]{rem:corrl} \bibinfo{note}{In the
    special case $\rho=0$, $\xi_{l2}$ and $\xi_{l4}$ correspond to
    correlation lengths for the $d-m$ ``hard'' and $m$ ``easy''
    directions. In the general case $\rho\ne0$ and $g_\tau\ne 0$, this
    is no longer true, irrespective of whether such correlation
    lengths are defined via second moments of the two-point function
    $G^{(2,0)}$ or in terms of its exponential decay along these
    directions (``true correlation lengths''). Unlike our $\xi_{l2}$
    and $\xi_{l4}$, which are independent of the scaling field
    $\sigma$, such correlation lengths depend on both relevant scaling
    fields $g_\tau$ and $\sigma$.}
\bibitem[{com({\natexlab{i}})}]{com:bulknc} \bibinfo{note}{In other
    words, the analogous bulk model with homogeneous values of
    $\mathring{\tau}=\mathring{\tau}_2$ and
    $\mathring{\rho}=\mathring{\rho}_2$ and the Hamiltonian ${\mathcal
      H}={\int_{\mathbb{R}^d}} {\mathcal
      L}_{\text{b}}(\bm{x};\mathring{\tau}, \mathring{\rho})\,dV$
    should have these properties.}
\bibitem[{com({\natexlab{j}})}]{com:etaperp} \bibinfo{note}{A natural
    way to define the above-mentioned surface index $\eta_{l4,\perp}$
    would be to write the exponent $(2-\eta_\perp)/\theta$ in the
    second line on the right-hand side of Eq.~(\ref{eq:chi1psing}) as
    $2-\eta_{l4,\perp}$.}
\bibitem[{com({\natexlab{k}})}]{com:testfct} \bibinfo{note}{Test
    functions are normally required to have strong smoothness
    properties, e.g., to have continuous partial derivatives of
    arbitrary order and to decay faster than any power as any of its
    arguments approaches $\pm\infty$. The amputated legs we are
    concerned with do not possess such strong properties. For example,
    the amputated leg $G_{\text{D}}(\bm{x}_1,\bm{x})$ and its Fourier
    transform $\hat{G}_{\text{D}}(\bm{p};z_1,z)$ depend on $|z_1-z|$
    and $|z_1+z|$, so that sufficiently high $z$~derivatives produce
    (derivatives of) $\delta(z_1\pm z)$. However, one can easily see
    that the actions of the distributions in question on the external
    legs are well defined. The formulae derived for genuine test
    functions $\varphi$ and $\psi$ apply, if we consider the cumulants
    in the $\bm{p}z$~representation as functions of the
    $z$~coordinates $z_1$ and $z_2$ of their external points off the
    surface. Amputated legs whose external points are located on the
    surface, such as
    \;
    %
    \raisebox{-0.70em}{$z$}\hspace{-0.5em}\raisebox{-0.3em}{\includegraphics[width=4em]{./ginf11.0.eps}}
    %
    \,,
    are exponentials in $z$ and hence good test functions of $z$.\\
    With regard to the coordinates $\bm{r}$ parallel to the surface
    the situation is even simpler and completely analogous to the bulk
    case.  Owing to the translational invariance we have parallel to
    the surface, the required integrals over $\bm{r}$~variables can be
    rewritten as convolutions. Such convolutions of well-defined
    generalized functions of $\bm{r}$~variables (i.e., of the
    amputated graph with the two amputated legs) are well-defined,
    becoming simple products in $\bm{p}$~space.}
\end{thebibliography}

\end{document}